\documentclass[sn-mathphys,Numbered]{sn-jnl}
\usepackage{graphicx}%
\usepackage{multirow}%
\usepackage{amsmath,amssymb,amsfonts}%
\usepackage{amsthm}%
\usepackage{mathrsfs}%
\usepackage[title]{appendix}%
\usepackage{xcolor}%
\usepackage{textcomp}%
\usepackage{manyfoot}%
\usepackage{booktabs}%
\usepackage{algorithm}%
\usepackage{algorithmicx}%
\usepackage{algpseudocode}%
\usepackage{listings}%
\usepackage{lineno}
\usepackage{url}

\theoremstyle{thmstyleone}%
\theoremstyle{thmstyletwo}%

\theoremstyle{thmstylethree}%

\begin{document}
\graphicspath{ {./images_revise/} }

\title{Bangladesh’s Amplified Coastal Storm Tide Hazard
from Tropical Cyclones and Rising Sea Levels in a Warming Climate}
\author*[1]{\fnm{Jiangchao} \sur{Qiu}}\email{qiujch24@mit.edu}
\author*[1]{\fnm{Sai} \sur{Ravela}}\email{ravela@mit.edu}
\author[1]{\fnm{Kerry} \sur{Emanuel}}\email{emanuel@mit.edu}

\affil[1]{\orgdiv{Department of Earth, Atmospheric and Planetary Sciences}, \orgname{Massachusetts Institute of Technology}, \orgaddress{\street{77 Massachusetts Avenue}, \city{Cambridge}, \postcode{01239}, \state{MA}, \country{USA}}}

\abstract{The risk of extreme storm tides to Bangladesh's low-lying and densely populated coastal regions, already vulnerable to tropical cyclones, remains poorly quantified under a warming climate. Here, using a statistical-physical downscaling approach, our multimodel large-ensemble projections under the IPCC6 SSP2-4.5, SSP3-7.0, and SSP5-8.5 scenarios show that Bangladesh's 100-year storm tide will likely intensify from 3.5 m to between 4.9 m and 5.4 m by the end of the 21st century. The Meghna-North Chattogram region is the most vulnerable, and the storm tide season will broaden significantly, amplifying the strongest during the late monsoon and late post-monsoon seasons. We project substantial increases in seasonal storm tide frequencies, with a four-fold increase in back-to-back extremes in the post-monsoon season. Across the SSP2-4.5, SSP3-7.0, and SSP5-8.5 scenarios assessed using multiple climate models, the frequency of storm tide from destructive cyclones like Bhola and Gorky will significantly increase by 7-18 times and 6-23 times, respectively. Our study indicates a need to re-examine the ongoing coastal improvement and heighten the urgency to enhance coastal resilience in Bangladesh.}


\maketitle

\newpage
\textbf{{Science for society}}
\\
Bangladesh is highly susceptible to severe tropical cyclones and the associated storm tides. The absence of timely assessments of climate change's impact on storm tides hinders progress toward achieving sustainable development goals. Our analyses reveal an intensification of storm tides in Bangladesh, even under middle-of-the-road climate scenarios, emphasizing the need for renewed attention.

This work provides critical climate-relevant information for significant ongoing investments, such as the Coastal Embankment Improvement Project (CEIP) and the Multipurpose Disaster Shelter Project (MDSP). These projects may underestimate future storm tide risks due to their reliance on deterministic methods that do not explicitly account for changes in tropical cyclone activity under warming climates. For example, underestimating new design crest levels will fail to effectively reduce the population's vulnerability, necessitating a revision of these significant investments.

Our findings offer key insights to help Bangladesh address the growing impacts of climate change on its coastal communities. Furthermore, the methodology presented here, particularly the statistical-physical tropical cyclone downscaling approach and the joint sampling of cyclone parameters, sea-level rise probabilities, and tides, can inform workflows for other low-lying coastal nations facing similar challenges.
\\
\\
\textbf{Highlights}
\begin{itemize}
    \item TC climatology change and SLR drive Bangladesh's exacerbating storm tides in both IPCC AR5 and AR6 multimodel large-ensemble projections.
    
    \item The storm tide season significantly broadens, with the most substantial intensification occurring during the late monsoon and late post-monsoon seasons. 
    
    \item Substantial increases in seasonal storm tide frequencies are projected, including a four-fold rise in back-to-back extremes during the post-monsoon season.
    
    \item A joint sampling methodology probabilistically integrates tropical cyclones, sea level rise, and tides into storm tide simulations.
\end{itemize}

\newpage
\section{Introduction}
\label{introduction}

Tropical cyclone (TC)-induced coastal floods rank among the deadliest and costliest worldwide catastrophes~\cite{needham2015review}. The Bay of Bengal (BoB), located in the northeastern part of the Indian Ocean, has consistently experienced some of the most destructive coastal floods in history. Although it accounts for only 5--6 percent of global TC activity, approximately 80--90 percent of global TC fatalities occur in this basin~\cite{chowdhury2002cyclone,paul2009relatively}. The BoB's funnel-shaped and shallow northern region naturally amplifies the water level, raising it to 10 meters above mean sea level when strong TCs strike~\cite{islam2009climatology,dasgupta2010vulnerability}. Six TCs in the BoB have each caused more than $140,000$ fatalities~\cite{needham2015review}, primarily due to coastal flood inundation of the low-lying (less than 5 m above mean sea level), densely populated mega-delta (with a population density of $6,734$ per $km^{2}$)~\cite{streatfield2008population}.

Bangladesh is a downstream riparian state for three major trans-Himalayan rivers, namely the Ganges, the Brahmaputra, and the Meghna, and is fringed by the BoB. Due to its location, the country is prone to coastal floods caused by intense TCs. It has a history of devastation caused by TCs, with 14 events each resulting in the loss of over 10,000 lives from 1760 to 2020~\cite{emanuel_tropical_2021}. The latest of these was Cyclone Gorky in April 1991, which claimed at least 140,000 lives. Sitting on the frontline in the battle against coastal floods, Bangladesh has since emerged as an international champion~\cite{noauthor_worlds_2020}, implementing proactive policies to improve its resilience significantly. The government has improved its early warning system, increasing access to a network of cyclone shelters and evacuation roads, improving polders\footnote{Polders are low-lying areas surrounded by embankments.}, and implementing community-based cyclone preparedness programs~\cite{kazi_bangladesh_2022}. These efforts have yielded impressive results, reducing mortality from TCs nearly a hundred-fold. In May 2020, Super Cyclone Amphan hit the western coast of Bangladesh, resulting in a limited death toll of 128 despite inducing a 5-meter storm surge~\cite{khan2021towards}.

However, a warming climate likely poses a significant threat to Bangladesh. The polder embankment systems, which consist of 139 polders covering 1.2 million hectares of land, were built to protect about 8 million people from flooding and ensure their safety and livelihoods~\cite{kazi_bangladesh_2022}. If Sea-Level Rise (SLR) and extreme TCs become more frequent and destructive due to climate change, the exposed infrastructure and vulnerable populations will be at greater risk. Increased sedimentation, elevating riverbeds, and land subsidence within polders increase embankment stress exposure to even low-intensity TCs~\cite{becker2020water}. Furthermore, rising coastal water levels, reduced upstream river discharge, and polder-induced tidal amplification may worsen salinity intrusion~\cite{chen2018coastal}. If the cyclone season expands into the monsoon season, the combined impact of coastal and inland flooding could devastate agriculture and water supply, further straining the vulnerable population.

The consequences faced by coastal Bangladesh are severe and often appear irreversible. In light of the emerging climate hazard, Bangladesh's coastal architecture seems fragile and unsustainable. Risk-informed solutions are needed, so it is necessary to quantify climate change impacts accurately. In the context of TC-induced storm surge risk, downscaling TCs to establish climatologies in future climate scenarios holds promise for quantifying the hazard. However, it is challenging to do so using the limited observational record of cyclones passing through Bangladesh or bearing the enormous computational expense of running high-resolution numerical climate models.

To address these limitations, synthetic TC models have become a valuable tool for risk assessment~\cite{emanuel2006statistical,emanuel2008hurricanes,lee2018environmentally,bloemendaal2020generation,jing2020environment,leijnse2022generating}. Emanuel et al.'s statistical-physical downscaling framework, in particular, has proven effective in regions with sparse observational data and under future climate scenarios~\cite{emanuel2006statistical,emanuel2008hurricanes}. A multimodel intercomparison study~\cite{meiler2022intercomparison} demonstrated the superior performance of this approach in the BoB, in contrast to purely statistical models like STORM (Synthetic Tropical cyclOne geneRation Model)~\cite{bloemendaal2020generation}, which tend to overestimate high-intensity TCs due to data limitations.

Recent studies have applied Emanuel et al.'s framework to assess storm surge risks in Bangladesh~\cite{khan2022storm} but confine the assessments to the current climate. Earlier projections by Emanuel~\cite{emanuel_tropical_2021} indicate that under a high greenhouse gas emissions scenario, the likelihood of extreme TC winds (exceeding 150 knots) could increase tenfold by the end of the century. Despite these alarming findings, few studies have examined TC-induced storm surge risks for Bangladesh under changing climate scenarios.

In this study, we explicitly and efficiently downscale TCs and simulate storm tides across coastal Bangladesh under projected climate scenarios by applying a statistical-physical framework~\cite{emanuel2006statistical,emanuel2008hurricanes}. This approach allows us to assess future risks and address a critical knowledge gap comprehensively. Although researchers have used similar frameworks in other regions~\cite{lin2014heightened,neumann2015risks,garner2017impact,yin2021hazard}, few have extended these methods to the BoB, which limits Bangladesh's ability to adapt effectively to escalating climate risks.

\begin{figure*}[htb!]
\centering
\includegraphics[width=1\textwidth]{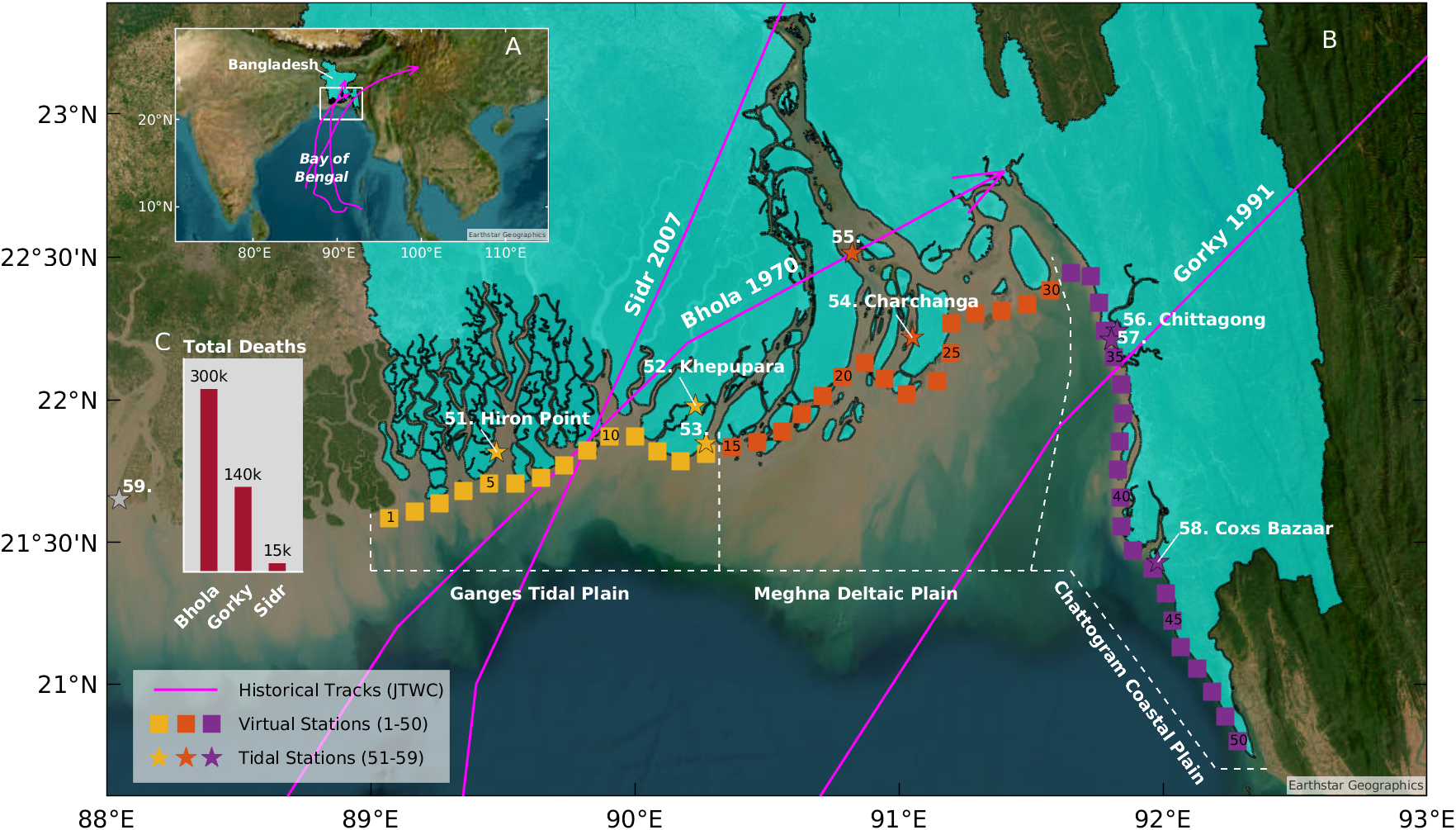}
\caption{\textbf{Maps of historically deadliest TCs that made landfall in Bangladesh.} \textbf{A}, The Joint Typhoon Warning Center (JTWC) tracks Bhola (1970), Gorky (1991), and Sidr (2007) TCs, marking them in magenta with arrows. These cyclones originate in the BoB and move northward to strike Bangladesh, highlighted in cyan. \textbf{B}, The map zooms into coastal Bangladesh to show where these three TCs made landfall and identifies 58 water level stations across Southwest Bangladesh (Ganges Tidal Plain, marked yellow), Middle Bangladesh (Meghna Deltaic Plain, marked orange), and East Bangladesh (Chattogram Coastal Plain, marked purple). The equidistant sampling divides Bangladesh's 300 km coastline into 50 points (defined as ``Virtual Stations [VS]") spaced 6 km apart. These points supplement the limited existing tidal stations. The map labels VS with names of MIT/Earth Signals and Systems Group alums, representing them as squares with IDs from 1 to 50, while pentagrams represent tidal stations with IDs from 51 to 59. Table~\ref{tab:tabs1} provides detailed information on these regional stations. \textbf{C}, The histogram associates each reported total fatality count with the corresponding TC (data from reference~\cite{needham2015review}). The base map integrates data from BDP 2100 (Baseline Volume 1, pg. 403), Humanitarian Data Exchange, World Bank, ESRI ArcGIS, Maxar, Earthstar Geographics, USDA FSA, USGS, Aerogrid, IGN, IGP, and the GIS User Community.}
\label{fig:fig1}
\end{figure*} 
\subsection{Structure of Paper}

The coupled downscaling and hydrodynamic framework involves three primary components: synthetic TC downscaling, hydrodynamic simulation of storm tides, and statistical analysis incorporating bias correction. A brief description of the workflow using these components is as follows. We use a downscaled synthetic TC track set~\cite{emanuel_tropical_2021} to investigate the effects of climate change (including TC climatology change and probabilistic SLR) on Bangladesh's storm tides. Simulated synthetic TCs drive a verified hydrodynamic model~\cite{luettich1992adcirc} to simulate storm tides, dynamically incorporating astronomic tides and SLR using updated higher-accuracy regional bathymetry~\cite{krien2016improved,khan2022storm}.

The simulated hydrodynamic storm-tide ensemble provides the results in this paper. Specifically, we estimate storm-tide return periods in present climate and future climate scenarios, including the combined effect of future TC climatology change and SLR. We then investigate storm tide distributions across seasons for severity and frequency. We additionally analyze the frequency that storms similar to the historically deadliest TC-induced storm tides (Bhola and Gorky, tracks and fatalities can be found in Figure~\ref{fig:fig1}) might have in a warming climate.

The paper begins by presenting the main results first in Section~\ref{sec:results}, followed by an in-depth discussion in Section~\ref{sec:discussion}. Section~\ref{sec:methods} presents a detailed methodology for the workflow and its components. The supplementary material contains additional information supporting the paper.

\section{Results}
\label{sec:results}

\subsection{Bangladesh's Storm Tides under Current and Future Climate Across Scales}

\begin{figure*}[htb!]%
\centering
\includegraphics[width=1\textwidth]{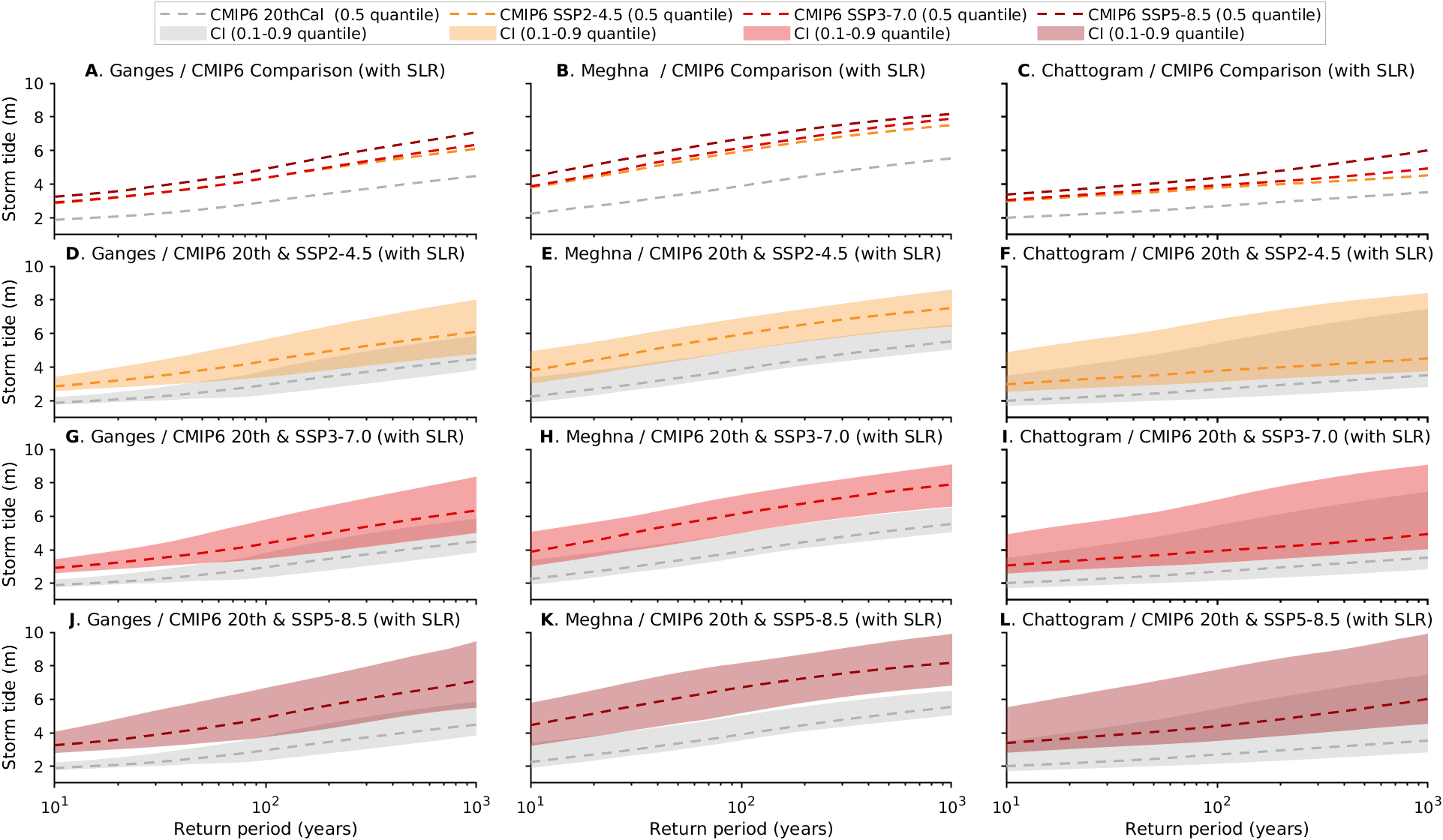}
\caption{\textbf{Bangladesh's storm tides versus return periods, as projected by CMIP6 models at the regional scale.} Storm tides are the total water levels (combined components of astronomic tide, storm surge, and mean sea-level state) relative to the mean sea level of the 1995-2014 baseline. \textbf{A, D, G, J,} Projections for the Ganges. \textbf{B, E, H, K,} Projections for the Meghna. \textbf{C, F, I, L,} Projections for the Chattogram. \textbf{A, B, C,} Comparison between the current climate and the future CMIP6 SSP2-4.5, SSP3-7.0, and SSP5-8.5 climates. \textbf{D, E, F,} CMIP6 model ensembles under the SSP2-4.5 scenario. \textbf{G, H, I,} CMIP6 model ensembles under the SSP3-7.0 scenario. \textbf{J, K, L,} CMIP6 model ensembles under the SSP5-8.5 scenario. Dashed lines indicate the ensemble median (0.5 quantiles), while shaded areas indicate each estimate's confidence interval (CI, 0.1-0.9 quantiles). The current climate period spans from 1981 to 2000, while the future climate period spans from 2081 to 2100. The confidence intervals account for variability in Tide, SLR, TCs, climate models, multi-station data, and Kernel-GPD parameters.}
\label{fig:fig2}
\end{figure*}  

\begin{figure*}[htb!]%
\centering
\includegraphics[width=1\textwidth]{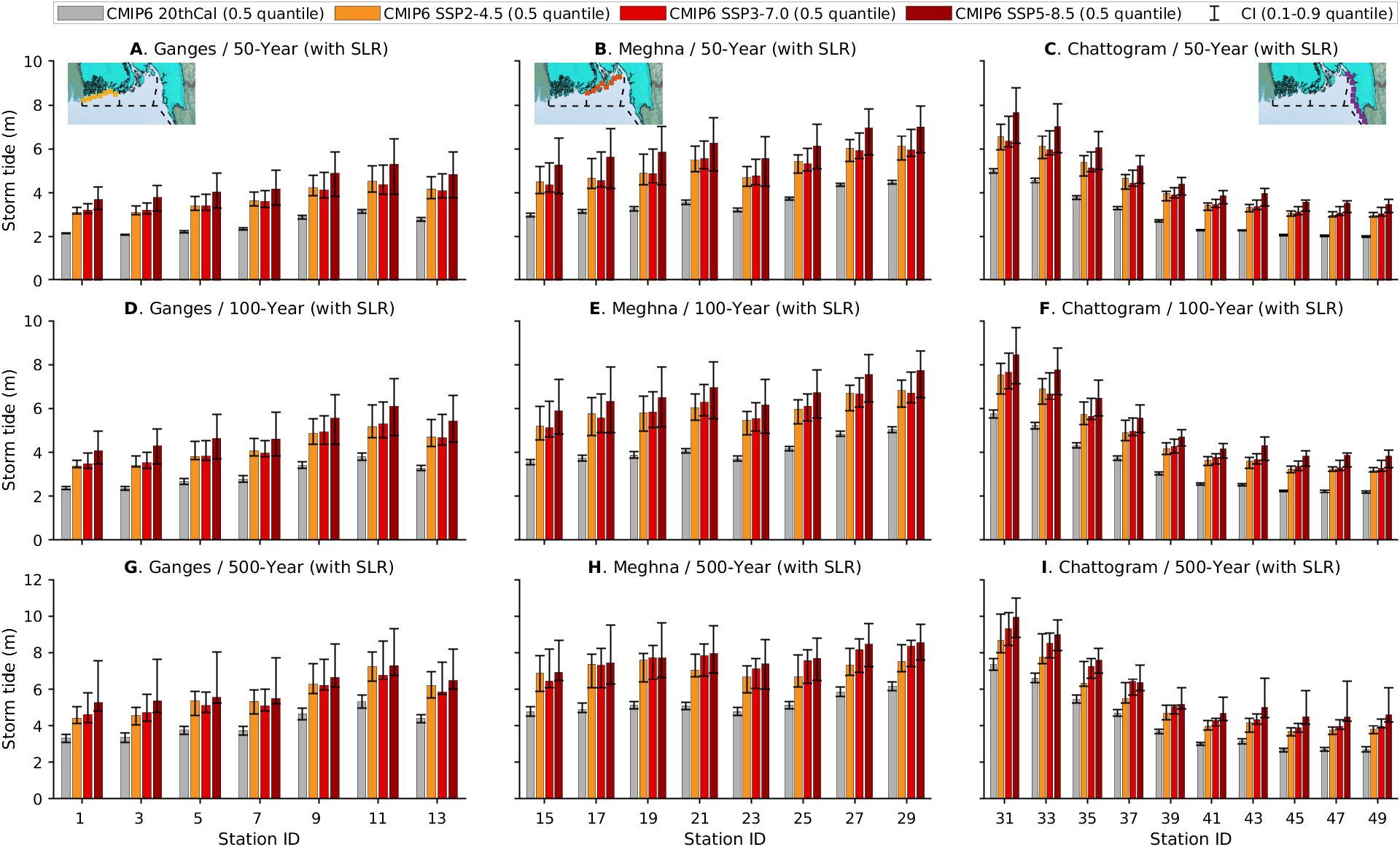}
\caption{\textbf{Bangladesh's 50-, 100-, 500-year storm tides projected by CMIP6 models at the station scale.} Storm tides are the total water levels (combined components of the astronomic tide, storm surge, and mean sea-level state) relative to the mean sea level of the 1995-2014 baseline. \textbf{A, B, C,} 50-year return period. \textbf{D, E, F,} 100-year return period. \textbf{G, H, I,} 500-year return period. \textbf{A, D, G,} Stations located in the Ganges (southwest Bangladesh). \textbf{B, E, H,} Stations located in the Meghna (middle Bangladesh). \textbf{C, F, I,} Stations located in the Chattogram (east Bangladesh). Projections at all 58 stations are available, but the graph displays only every other station (VS). Histogram heights indicate the ensemble median (0.5 quantiles) for the current climate, CMIP6 SSP2-4.5, SSP3-7.0, and SSP5-8.5 climates. Vertical error bars indicate each estimate's confidence interval (CI, 0.1-0.9 quantiles). The current climate period spans from 1981 to 2000, while the future climate period spans from 2081 to 2100. The confidence intervals account for variability in Tide, SLR, TCs, climate models, and Kernel-GPD parameters. Base map sourced from the BDP 2100 (Baseline Volume 1, pg. 403), Humanitarian Data Exchange, World Bank, ESRI ArcGIS, Maxar, Earthstar Geographics, USDA FSA, USGS, Aerogrid, IGN, IGP, and the GIS User Community.}
\label{fig:fig3}
\end{figure*} 

We conducted storm tide hazard assessments in Bangladesh to evaluate the impacts of climate change on storm tides and identify the most vulnerable regions. These assessments span national (aggregate projections from all 50 ``Virtual Stations [VS]"), regional (projections aggregated for VS in the Ganges, Meghna, and Chattogram regions), and local (station-specific) scales.

At the \textbf{national scale}, we find that climate change will significantly increase Bangladesh's storm tides by the end of the $21^{\texttt{st}}$ century, even under moderate scenarios (see Figure~\ref{fig:figs1}). The projected increases in storm tide risk under the CMIP6 SSP2-4.5 and SSP3-7.0 scenarios are similar in magnitude, regardless of whether sea-level rise (SLR) is included, but remain lower than those under the SSP5-8.5 scenario. Statistical uncertainty also grows at extended return periods. Specifically, the 100-year storm tide, currently at 3.5 m (CI: 2.2 m to 5.0 m), is projected to rise to:
\begin{itemize}
    \item 4.9 m (CI: 3.2 m to 6.7 m) under SSP2-4.5,
    \item 5.0 m (CI: 3.4 m to 6.9 m) under SSP3-7.0, and
    \item 5.4 m (CI: 3.7 m to 7.8 m) under SSP5-8.5.
\end{itemize}

Including TC climatology changes and SLR yields median increases of 1.4 m, 1.5 m, and 1.9 m under SSP2-4.5, SSP3-7.0, and SSP5-8.5, respectively. Excluding SLR, TC climatology changes alone will increase storm tides by 0.6 m, 0.6 m, and 0.9 m under the exact scenarios. Additionally, we assessed storm tide changes under CMIP5 RCP4.5 and RCP8.5 scenarios (see Figure~\ref{fig:figs4}), finding that CMIP6 projections indicate a higher risk than CMIP5 projections, as discussed in Section~\ref{sec:discussion}.

At the \textbf{regional scale}, storm tide impacts are uneven across Bangladesh. Northern Chattogram is the most vulnerable region, followed by Meghna and Ganges, with southern Chattogram ranking fourth. Under CMIP6 SSP2-4.5, the 100-year storm tides for:
\begin{itemize}
    \item Ganges, Meghna, and Chattogram are projected to increase by 1.4 m, 2.0 m, and 1.1 m (medians), respectively.
    \item Under SSP3-7.0, the increases are 1.4 m, 2.3 m, and 1.2 m, respectively.
    \item Under SSP5-8.5, the increases are 1.9 m, 2.8 m, and 1.7 m, respectively.
\end{itemize}

Further dividing Chattogram into northern (VS ID 31--37, Figure~\ref{fig:fig1}) and southern segments (VS ID 38--50, Figure~\ref{fig:fig1}), we find that:
\begin{itemize}
    \item Northern Chattogram experiences more significant increases, with median rises of 1.5 m, 1.5 m, and 2.3 m under SSP2-4.5, SSP3-7.0, and SSP5-8.5, respectively.
    \item In contrast, the southern Chattogram shows smaller median increases of 1.0 m, 1.2 m, and 1.6 m under the exact scenarios.
\end{itemize}
Similar regional assessments under CMIP5 RCP4.5 and RCP8.5 scenarios are presented in Figures~\ref{fig:figs5} and~\ref{fig:figs6}.

At the \textbf{local scale} (station-specific; see Figure~\ref{fig:fig3}), we observe that increases under SSP5-8.5 are more pronounced than under SSP2-4.5 and SSP3-7.0. The most substantial storm tide increases occur at stations in northern Chattogram. For example:
\begin{itemize}
    \item At VS Navi (ID=31) in northern Chattogram, the 100-year storm tide is projected to increase by 1.8 m, 1.9 m, and 2.7 m (medians) under SSP2-4.5, SSP3-7.0, and SSP5-8.5, respectively.
    \item In comparison, at VS Zhuchang (ID=49) in southern Chattogram, the increases are 1.0 m, 1.1 m, and 1.6 m (medians) under the exact scenarios.
\end{itemize}

Storm tides driven solely by TC climatology changes (see Figure~\ref{fig:figs3}) under SSP3-7.0 show slightly smaller increases than those under SSP2-4.5, as discussed further in Section~\ref{sec:discussion}.

\subsection{The Contributions of Changing SLR and TC Climatology to Storm Tide Risk}\label{TC_SLR_Contribution}

\begin{figure*}[htb!]%
\centering
\includegraphics[width=1\textwidth]{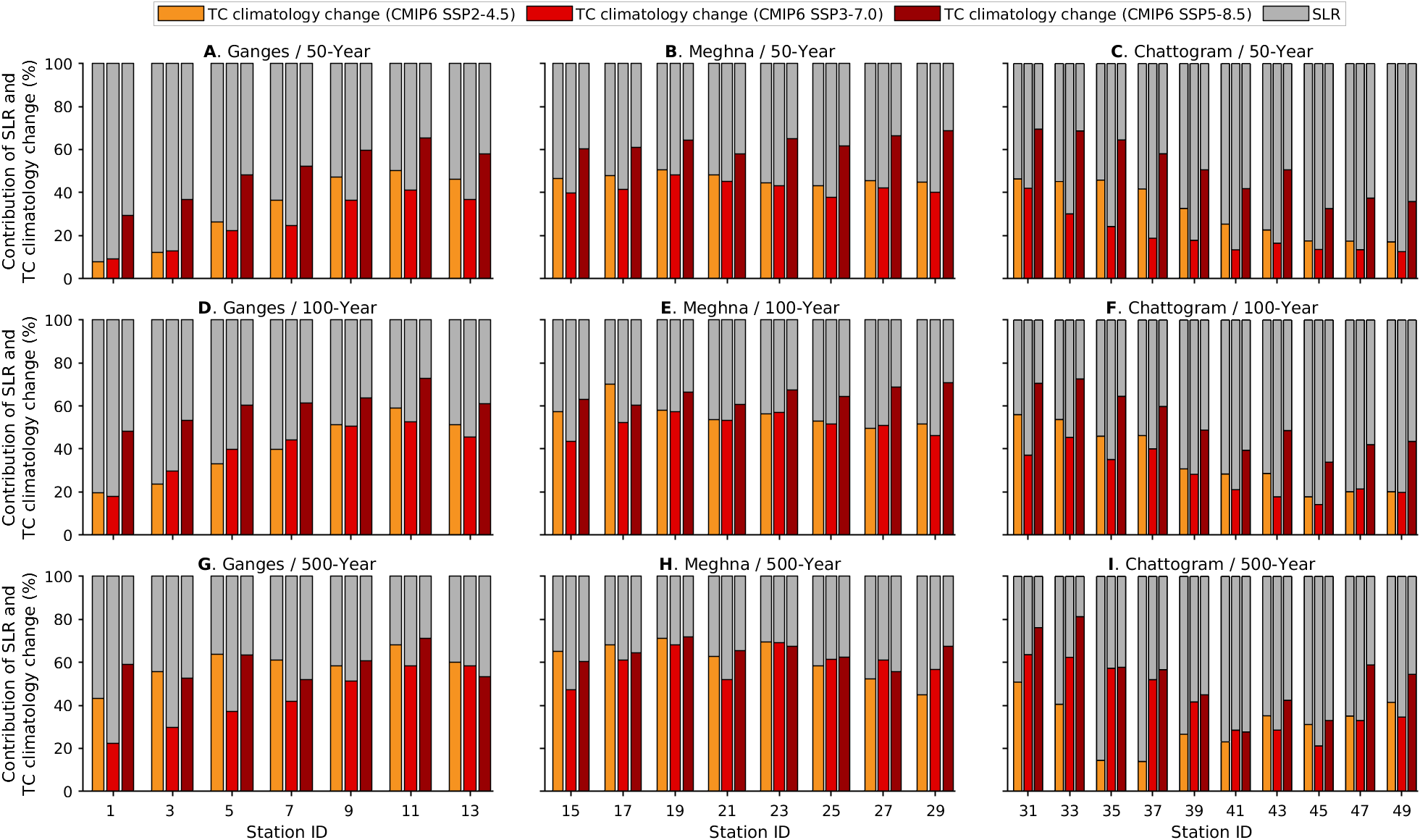}
\caption{\textbf{Contribution of TC climatology change and SLR to changes in Bangladesh's storm tides (50-, 100-, and 500-year return periods) at the regional scale.} \textbf{A, B, C,} 50-year return period. \textbf{D, E, F,} 100-year return period. \textbf{G, H, I,} 500-year return period. \textbf{A, D, G,} Stations located in the Ganges. \textbf{B, E, H,} Stations located in the Meghna. \textbf{C, F, I,} Stations located in the Chattogram. While projections for all 58 stations are available, the graph displays only every other VS. The stacked histogram heights indicate the contributions of TC climatology change and SLR to storm tide changes projected under the CMIP6 SSP2-4.5, SSP3-7.0, and SSP5-8.5 climate scenarios, respectively. The changes in storm tide are calculated based on the ensemble median (0.5 quantile). We assess the contributions of TC climatology change and SLR over the future period (2081–2100) compared to the baseline period (1981–2000).}
\label{fig:fig4}
\end{figure*}  

Two primary factors influence the changes in the storm tide: TC climatology changes and SLR. However, their roles differ across Bangladesh's coastline. As depicted in Figure~\ref{fig:fig4}, projections under CMIP6 SSP5-8.5 indicate that TC climatology change contributes more to storm tide than SLR, with TC contributions exceeding 50\%. Moreover, the proportion of the storm tide attributed to TC climatology change is more significant under SSP5-8.5 compared to SSP2-4.5 and SSP3-7.0. The assessments compare the ensemble medians of storm tide with SLR incorporated (Figure~\ref{fig:fig3}) and without SLR incorporated (Figure~\ref{fig:figs3}) for each return period. The SLR case injects stochastic time-dependent SLR in storm tide simulations (Section~\ref{SLR} and Figures~\ref{fig:figs22},~\ref{fig:figs23},~\ref{fig:figs24}). 

Specifically, using the 100-year storm tide as an example (Figure~\ref{fig:fig4}, panels \textbf{D, E, F}), under SSP2-4.5, 44\% of all 50 VS experience an increase predominantly caused by TC climatology change rather than SLR. On average, TC climatology change contributes 41.8\% (in the Ganges), 54.7\% (in the Meghna), and 33\% (in the Chattogram) to the increase in storm tides.

Similarly, under SSP3-7.0, TC climatology change contributes more than SLR to the rise in 26\% of all 50 VS. On average, TC climatology change contributes 41.6\% (in the Ganges), 50.4\% (in the Meghna), and 27\% (in the Chattogram) to the increase in storm tides.

In comparison, projections under SSP5-8.5 indicate that TC climatology changes dominate the growth in 74\% of all 50 VS. On average, TC climatology changes contribute 61.1\% (in the Ganges), 64.6\% (in the Meghna), and 51.1\% (in the Chattogram) to the increase in storm tides.


\begin{figure*}[htb!]%
\centering
\includegraphics[width=1\textwidth]{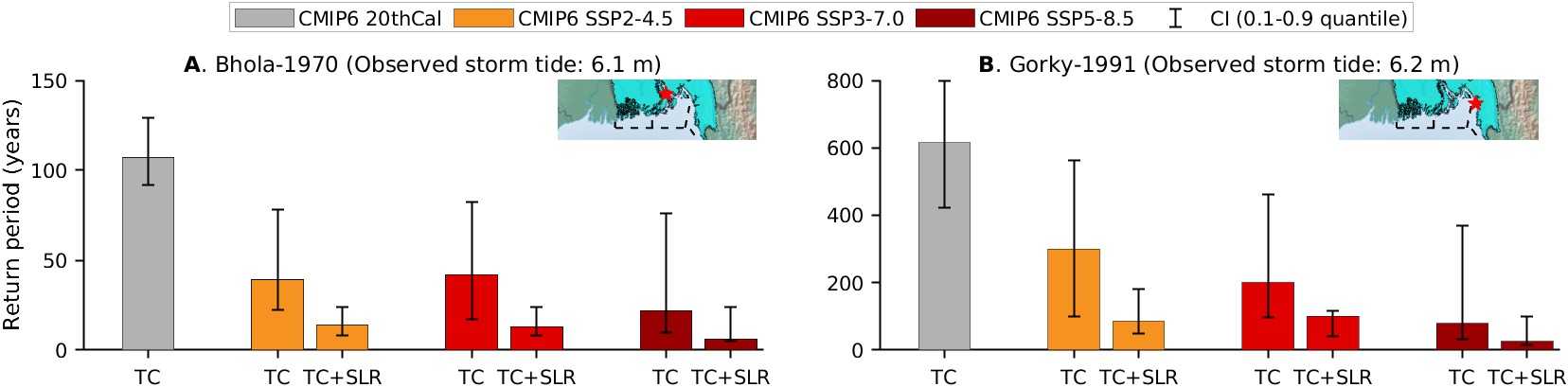}
\caption{\textbf{Changing annual frequency of storm tides similar to historically deadliest TCs in a warming climate.} \textbf{A}, Assessment for TC Bhola at station Northern Bhola Island (ID=55). \textbf{B}, Assessment for TC Gorky at station Anwara (ID=57, near the Karnaphuli River mouth). The histogram height represents the ensemble median for the estimated return period of storm tide corresponding to the observed peak storm tide (based on local mean sea level, sourced from previous studies~\cite{frank1971deadliest,as1998coastal}). The whiskers indicate the estimated confidence interval (CI, 0.1–0.9 quantiles). The red pentagram icon in the top-right corner of each subplot marks the location of the maximum measured storm tide during the two landfall TCs. The confidence intervals account for variability in Tide, SLR, TCs, climate models, and Kernel-GPD parameters. The effects of wave setup and river discharge on observed historical extreme water levels are poorly documented and assumed to be non-significant for this hazard assessment. Base map sourced from the BDP 2100 (Baseline Volume 1, pg. 403), Humanitarian Data Exchange, World Bank, ESRI ArcGIS, Maxar, Earthstar Geographics, USDA FSA, USGS, Aerogrid, IGN, IGP, and the GIS User Community.}
\label{fig:fig5}
\end{figure*} 

\subsection{Assessing Future Annual Exceedance Frequency of Deadliest TC-Induced Storm Tides}\label{Historical_event_assessment}

Estimating the annual exceedance frequency (the reciprocal of the return period) of storm tides in future scenarios similar to the deadliest TCs from the past, such as TC Bhola (1970) and TC Gorky (1991), is crucial for communicating the potential dangers of climate change to the public and developing effective climate adaptation strategies. Figure~\ref{fig:fig5} illustrates the annual frequency of these two deadliest TC-induced storm tides in the current climate and their potential change in a warming climate, considering the joint effect of TC climatology change and SLR. The projected annual frequency increases significantly: 7–18 times for the Bhola-like TC and 6–23 times for a Gorky-like TC, respectively. We note that the Meghna estuary region is more susceptible to extreme storm tides, as the Bhola and Gorky cyclones produced maximum storm tides of similar magnitudes but significantly different return periods at two different locations. This fact underscores the need for heightened attention and preparedness in this region.

Assessing the risk of historic events across the entire coastline is challenging due to poor monitoring of water level records. Only limited observations document the maximum storm tide with explicit vertical datum information during Bhola (6.1 m above mean sea level, observed at Northern Bhola Island)~\cite{frank1971deadliest} and Gorky (6.2 m above mean sea level, observed at a location near the Karnaphuli River mouth)~\cite{as1998coastal}, respectively. We examine the $55^{\texttt{th}}$ station and $57^{\texttt{th}}$ station (which are nearest to the observations) to approximately represent Bhola-induced and Gorky-induced maximum storm tides, respectively. Table~\ref{tab:return_periods} summarizes the projected return periods.

\subsection{Changing Seasonal Severity, Frequency, and the Summer Monsoon Season} \label{Seasonalresult}

\begin{figure*}[htb!]%
\centering
\includegraphics[width=1\textwidth]{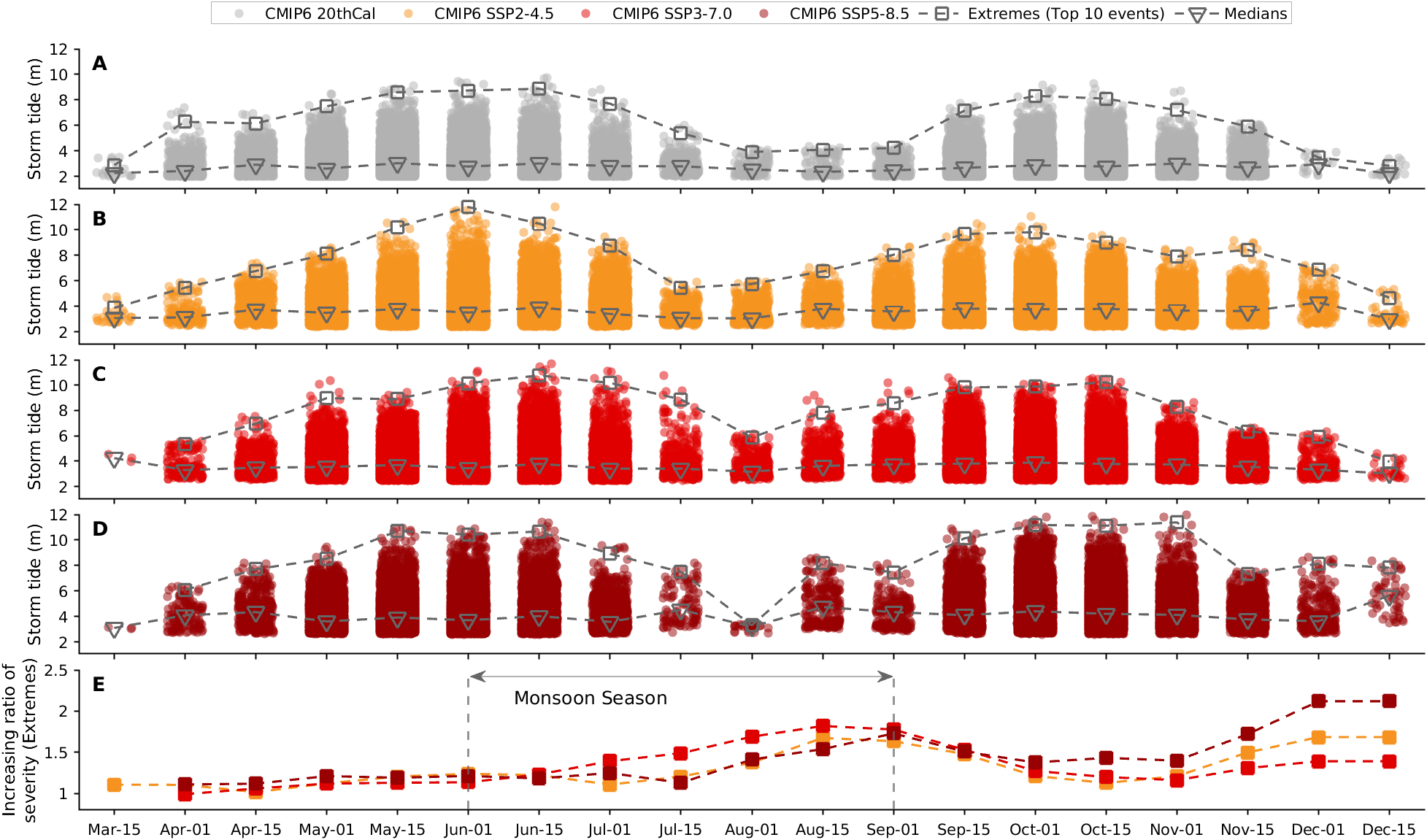}
\caption{\textbf{Shifted seasonal regimes of storm tide severity (greater than 2 m water level height) based on CMIP6 climate model ensembles.} \textbf{A, B, C, D,} Seasonal distribution of storm tide severity under the current climate and CMIP6 SSP2-4.5, SSP3-7.0, and SSP5-8.5 climates, respectively. As shown in panel \textbf{A}, two relatively inactive storm tide seasons are identified based on their seasonal behavior under the current climate: one during the late summer monsoon (August 01, August 15, and September 01) and the other during the late post-monsoon (December 01 and December 15). \textbf{E,} The increasing ratio of extreme storm tide intensities across the year, with a 3-point sliding average applied to obtain monthly ratios. The light square located at the top of each swarm in subplots \textbf{A, B, C, D} indicates the mean value of the top ten extremes, while the area between two vertical dashed lines in subplot \textbf{E} indicates the monsoon season of the BoB.}
\label{fig:fig6}
\end{figure*} 

\begin{figure*}[htb!]%
\centering
\includegraphics[width=1\textwidth]{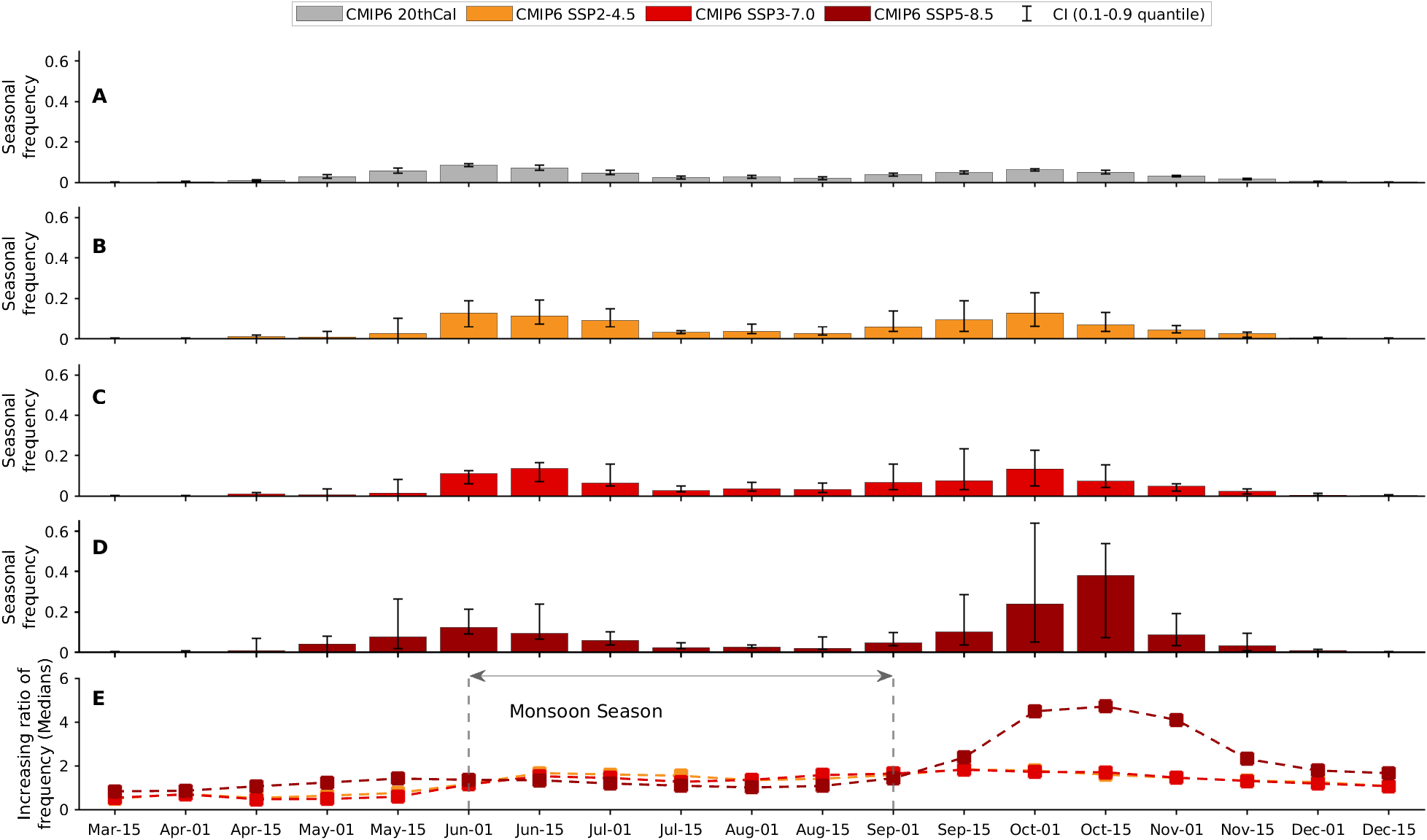}
\caption{\textbf{Shifted seasonal regimes of storm tide frequency (greater than 2 m water level height) based on CMIP6 climate model ensembles.} \textbf{A, B, C, D,} Seasonal distribution of storm tide frequency under current climate and CMIP6 SSP2-4.5, SSP3-7.0, and SSP5-8.5 climates, respectively. \textbf{E,} The increasing ratio of storm tide frequency for the medians throughout the year, with a 3-point sliding average applied to get monthly ratios. The error bar indicates the confidence interval from the 0.1 to 0.9 quantile. The area between two vertical dashed lines in subplot \textbf{E} indicates the monsoon season of the BoB. The confidence intervals account for variability in climate models and multi-station data.}
\label{fig:fig7}
\end{figure*} 

Bangladesh's landfalling TCs exhibit a clear bimodal seasonality, with activity peaks during the pre-monsoon period (April–May) and the post-monsoon period (October–December). These periods are separated by a relatively silent phase from June to August, attributed to the strong vertical wind shear caused by the South Asian summer monsoon~\cite{wu2023unraveling}.

To explore how climate change may impact these seasonal storm tide regimes, we analyze changes in storm tide severity and frequency. Figure~\ref{fig:fig6} highlights shifts in severity, while Figure~\ref{fig:fig7} focuses on frequency alterations across various climate scenarios.

Our findings indicate that climate change significantly amplifies Bangladesh's extreme storm tides. In Figure~\ref{fig:fig6} (panel \textbf{A, B, C, D}), ensemble projections suggest that the most dangerous storm tides still occur during the pre-monsoon and post-monsoon seasons. Climate change appears likely to shorten the interval of cyclone dormancy from 75 days ($\pm7$ days around August 01, August 15, September 01, and December 01, December 15) to a mere 15 days ($\pm7$ around August 01). Moreover, destructive storm tides expand seasonally in nearly all scenarios and significantly, with severity amplifying most within both the late summer monsoon ($\pm7$ days around August 01, August 15, and September 01) and late post-monsoon ($\pm7$ days around December 01 and December 15) seasons. In Figure~\ref{fig:fig6} (panel \textbf{E}), the severity ratios increase significantly during the late summer monsoon season and the late post-monsoon season compared to any other time.

Specifically, the storm tide severity in late summer monsoon season for the top ten extreme events under the current climate (Figure~\ref{fig:fig6} panel \textbf{A}) are 3.9 m, 4.1 m, and 4.2 m, respectively. The projections change significantly, reaching 5.8 m, 6.7 m, and 8.0 m under SSP2-4.5 (Figure~\ref{fig:fig6} panel \textbf{B}); 5.9 m, 7.9 m, and 8.6 m under SSP3-7.0 (Figure~\ref{fig:fig6} panel \textbf{C}); and 3.3 m, 8.2 m, and 7.5 m under SSP5-8.5 (Figure~\ref{fig:fig6} panel \textbf{D}).  Figure~\ref{fig:fig6} (panel \textbf{E}) shows increasing ratios:  1.4, 1.7, and 1.6 under SSP2-4.5; 1.7, 1.8, and 1.8 under SSP3-7.0; and 1.4, 1.5, and 1.7 under SSP5-8.5, respectively. 

In the late post-monsoon season, the storm tide severity for the top ten extreme events under the current climate (Figure~\ref{fig:fig6} panel \textbf{A}) is 3.5 m and 2.8 m, respectively. The extreme storm tides during the same periods change significantly, reaching 6.9 m and 4.7 m under SSP2-4.5 (Figure~\ref{fig:fig6} panel \textbf{B}), 5.9 m and 4.0 m under SSP3-7.0 (Figure~\ref{fig:fig6} panel \textbf{C}), 8.1 m and 7.8 m under SSP5-8.5 (Figure~\ref{fig:fig6} panel \textbf{D}). The ratios indicate increases again: 1.7 and 1.7 under SSP2-4.5, 1.4 and 1.4 under SSP3-7.0, and 2.1 and 2.1 under SSP5-8.5, respectively.

Although our analysis focuses solely on storm tides, the overlap between heightened water levels during the late monsoon season (mid-August) and potentially heavy monsoon rainfall (e.g., the devastating floods of August 2017~\cite{philip2019attributing} and July 2020~\cite{han2021grace}) can significantly amplify cascading inland-coastal flood risks in Bangladesh—the risk further compounds considering additional flooding from tropical cyclone (TC)-induced rainfall.

During the first cyclone dormancy period (July 25 to September 8), projections under SSP2-4.5 and SSP3-7.0 indicate that even the minimum storm tide (5.8 m) in warming climates exceeds the maximum storm tide (4.2 m) under the current climate. The second cyclone dormancy period shows a similar phenomenon. However, under the high-emission SSP5-8.5 scenario, a pronounced seesaw effect emerges: the highest storm tide during the first dormancy period decreases to just 3.3 m, while the second dormancy period (November 24 to December 22) experiences a substantial increase, reaching 7.8 m.

Notably, the second cyclone dormancy period disappears entirely under SSP5-8.5, suggesting that Bangladesh may face extreme storm tide hazards during winter—a situation unlikely under the current climate. This shift underscores the need for adaptive strategies to address the growing risk of compounded flooding events under future warming scenarios.

In contrast to severity, seasonal storm tide frequency changes are not unanimous across scenarios. While there are indications of a potentially many-fold increase in back-to-back flooding events during post-monsoon seasons under the SSP5-8.5 scenario, the frequency remains stable under the SSP2-4.5 and SSP3-7.0 scenarios. Figure~\ref{fig:fig7} illustrates the seasonal frequency shifts based on CMIP6 climate models for storm tides exceeding a 2 m threshold, which we attribute to the combined impact of TC climatology change and SLR. The most significant rise occurs during the post-monsoon season, particularly in October. The seasonal frequencies of storm tide during the three successive periods (October 1, October 15, and November 1) are 0.06 (CI: 0.05 to 0.07), 0.05 (CI: 0.04 to 0.06), and 0.03 (CI: 0.02 to 0.04), respectively, under the current climate (Figure~\ref{fig:fig7} panel \textbf{A}). However, under SSP5-8.5 (Figure~\ref{fig:fig7} panel \textbf{D}), these frequencies change to 0.24 (CI: 0.05 to 0.64), 0.38 (CI: 0.07 to 0.54), and 0.09 (CI: 0.03 to 0.19), with a significantly higher increase in frequency ratio of 4.5, 4.7, and 4.1 compared to any other seasonal period (Figure~\ref{fig:fig7} panel \textbf{E}). More frequent events during each period will decrease the interval between occurrences, leading to a higher likelihood of back-to-back coastal flooding.

\section{Discussion} \label{sec:discussion}
\subsection{Comparison With Previous Studies}

Several noteworthy findings from our results deserve further discussion. Reliably estimating the frequency of extreme storm tides is challenging, particularly in regions with insufficient observations, such as Bangladesh. Even studies evaluating Bangladesh's storm tide hazard under current climate conditions are limited, let alone those addressing future scenarios. Khan et al.~\cite{khan2022storm} applied the same downscaling-hydrodynamic method to assess storm tide hazards in coastal Bangladesh but confined their analysis to the current climate. By comparing their results with other existing return period estimates, they suggest that a biased extreme TC-event sampling strategy in previous studies may overestimate storm tide return periods. For example, Jakobsen et al.~\cite{jakobsen2006cyclone} estimate that the 100-year water level height is about 5 m at the mouth of Meghna and about 8–10 m at Sandwip.

In contrast, Khan et al.'s estimates are about 4 m at the mouth of Meghna and about 6 m at Sandwip. Unsurprisingly, our current climate estimates align with those of Khan et al. We estimate that the 100-year storm tide at these two locations (see estimation under the current climate in Figure~\ref{fig:fig3}) is about 4.2 m (Meghna region average) and 5.8 m (VS Navi with ID=31), respectively. 

Leijnse et al.~\cite{leijnse2022generating} estimate future climate storm-surge return periods in Bangladesh. Their estimate of the 100-year surge level (without incorporating astronomic tide) for Charchanga and Chittagong is over 60 cm lower than Khan et al. Several limitations in Leijnse et al.'s methodology may contribute to this discrepancy. Their hydrodynamic model lacked the updated bathymetric data in northern BoB. Further, they apply a purely statistical approach based on the historical TC dataset to generate synthetic TCs without an explicit representation addressing TC activity change under warming climates. As a result, significant uncertainties may remain in their return period estimation. So, we do not make the comparison here.

\subsection{Comparison Across Scales, Climate Models and Scenarios}
Our projections at regional and local scales reveal a pronounced vulnerability in the Meghna-North Chattogram region despite the widespread distribution of TC tracks across the Bay of Bengal (BoB). Simulations suggest that the funnel-shaped morphology of this region frequently amplifies surges from TCs traveling north and northeast. These storms often move along the coast and reflect off the Chattogram coastline, which runs parallel to the longitude, depositing significant surge energy at the mouth of the Meghna River. Similarly, surges from storms traveling west also follow the Chattogram coastline, culminating in high surge levels in the Meghna-North Chattogram area.

The interplay between the BoB's funnel-shaped morphology and the TC wind-field structure creates a unique and heightened risk for this segment, as highlighted in previous studies~\cite{as1998coastal}. While our assessment provides valuable insights for localized coastal climate adaptation planning and risk mitigation, it is crucial to recognize that the Meghna-North Chattogram region's vulnerability also poses significant risks to the substantial upstream population.

The elevated storm tide risk under CMIP6 SSP5-8.5 (Figure~\ref{fig:fig3}) compared to CMIP5 RCP8.5 (Figure~\ref{fig:figs7}) is primarily attributable to the greater frequency and intensity of TCs projected under SSP5-8.5, given that identical SLR samples in both RCP8.5 and SSP5-8.5 simulations. Equations~\ref{eu_eqn1} and~\ref{eu_eqn2} (see Section~\ref{StatisticalAnalysis}) illustrate an inverse relationship between annual TC frequency and storm tide severity, as captured in the storm tide-exceedance probability curve. Additionally, Figure~\ref{fig:figs26} demonstrates that storm tide levels and exceedance probabilities are inversely related.

An increase in TC annual frequency for a fixed return period $T$ reduces the return period for the same water level, leading to a higher storm tide for that return period. Specifically, as shown in Figure~\ref{fig:figs10}, TC frequency under SSP5-8.5 is significantly higher, with a median of 1.7 (CI: 0.7 to 2.5), compared to RCP8.5, which has a median of 0.8 (CI: 0.5 to 1.2).

Furthermore, Emanuel's~\cite{emanuel_tropical_2021} TC wind projections for Bangladesh indicate that TC wind intensity increases more substantially under SSP5-8.5 than under RCP8.5. The 100-year TC wind intensity rises from 123 knots under the current climate to 145 knots under RCP8.5 and further escalates to 168 knots under SSP5-8.5.

The differences in TC projections between CMIP5 and CMIP6 are partly attributable to the more extensive spread and generally higher average equilibrium climate sensitivity (ECS) within the CMIP6 model ensemble compared to CMIP5~\cite{zelinka2020causes,meehl2020context}. The choice of climate model also plays a significant role. For example, even within the same climate model families (e.g., IPSL and MIROC, as shown in Figure~\ref{fig:figs10}), TC frequency projections differ substantially between the fifth and sixth generations under both middle-pathway and high-pathway scenarios.

This disparity is likely due to the generally higher ECS values in CMIP6 compared to CMIP5, driven primarily by more substantial positive low-cloud feedback. Previous studies~\cite{zelinka2020causes,meehl2020context} emphasize that these feedback mechanisms amplify warming projections, which, in turn, influence TC activity projections under future climate scenarios.

The storm tide estimates under the SSP2-4.5 and SSP3-7.0 scenarios are similar in magnitude, regardless of whether SLR is considered, but remain lower than those projected under the SSP5-8.5 scenario, as shown in Figures~\ref{fig:fig2},\ref{fig:fig3},\ref{fig:figs1},\ref{fig:figs2}, and\ref{fig:figs3}. Figure~\ref{fig:figs29} provides an example of storm tide severity (empirical exceedance probability-storm tide curves) under the EC-Earth3 model at VS Navi without considering SLR.

Under SSP3-7.0, storm tide severity exceeds that under SSP2-4.5, driven by more intense TCs associated with higher emission pathways. However, the TC annual frequency is lower under SSP3-7.0 compared to SSP2-4.5 (see Figure~\ref{fig:figs10}). The combined effects of TC annual frequency and storm tide severity, as described by Equations~\ref{eu_eqn1} and~\ref{eu_eqn2}, counterbalance one another, resulting in annual storm tide levels of comparable magnitude under the SSP2-4.5 and SSP3-7.0 scenarios. Additionally, five out of seven CMIP6 models exhibit a higher TC annual frequency under SSP2-4.5 compared to SSP3-7.0, although the difference in magnitude remains relatively small.

Our observations indicate a higher storm tide risk (median values) for the historical cyclone Bhola under SSP2-4.5 compared to SSP3-7.0, as shown in Figure~\ref{fig:fig5} (panel \textbf{A}). Differences in storm tide return periods and associated dynamics between the two scenarios drive this disparity. Specifically, higher return periods observed under SSP3-7.0 for MRI-ESM2.0 and CanESM5.0.3 lead to higher median storm tide levels compared to SSP2-4.5. Conversely, the combination of higher storm tide severity (Figure~\ref{fig:figs30}, panel \textbf{B}) and greater annual TC frequency under SSP2-4.5 (Figure~\ref{fig:figs10}) results in lower return periods for the same storm tide level relative to SSP3-7.0.

Equations~\ref{eu_eqn1} and~\ref{eu_eqn2}, along with the inverse relationship between storm tides and exceedance probabilities, provide insights into these results. Among the seven CMIP6 models analyzed, two (MRI-ESM2.0 and CanESM5.0.3) exhibit higher return periods under SSP3-7.0 than under SSP2-4.5 for a storm tide level of 6.1 m at Northern Bhola Island. For illustration, MRI-ESM2.0 is highlighted as a representative example in Figure~\ref{fig:figs30}, panel \textbf{A} (results for the other six models are not shown). In contrast, the remaining five models demonstrate similar or lower return periods under SSP3-7.0 compared to SSP2-4.5 for the same storm tide level.

\subsection{Implications}
Our study has broad implications for understanding and mitigating the risks posed by future storm tides. By estimating the return periods of storm tides that account for the combined effects of future TC climatology changes and SLR and analyzing storm tide distributions across seasons in terms of severity and frequency, our findings provide critical insights. These results can inform policymakers and stakeholders about when and where to prioritize efforts to enhance coastal resilience and mitigate risks, leading to more targeted and practical solutions in a changing climate.

Our analysis underscores the need for future planning, rehabilitation, and improvement of infrastructure investments to focus on the Meghna region and northern Chattogram. Additionally,  the growing risks of cascading inland-coastal flooding during the late summer monsoon (August) and back-to-back coastal floods during the post-monsoon period (October and early November) deserve greater attention. Examples of such investments include the World Bank-supported CEIP and MDSP projects~\cite{kazi_bangladesh_2022}. CEIP Phase-I estimated that the annual risk from cyclone-induced flooding across coastal Bangladesh under the current climate amounts to US\$300 million, a figure projected to rise to US\$570 million (a 90\% increase) under future climate scenarios. These projections assumed a deterministic 0.5 m SLR and an 8\% increase in TC wind speeds, which might significantly underestimate the true risks~\cite{CEIP_I_2022}.

Accurately estimating the probabilities of extreme storm tides and their seasonal distributions is urgent. A well-designed risk mitigation plan can safeguard livelihoods, reduce financial burdens by minimizing disaster impacts, and avoid unnecessary and costly over-protection measures~\cite{calafat2020probabilistic}. Furthermore, our storm tide risk estimates provide critical ocean-side boundary information for hydrologists and engineers. By integrating coastal (including wave), fluvial, and pluvial processes, these findings can help create compound inundation maps for Bangladesh, enabling more comprehensive and effective climate adaptation strategies.
\section{Methods} \label{sec:methods}
\subsection{Synthetic TC Downscaling} \label{syntheticdownscaling}

We use a statistical-physical downscaling technique to create sets of synthetic TCs that affect Bangladesh~\cite{emanuel2006statistical,emanuel2008hurricanes}. The method uses thermodynamic and kinematic statistics from gridded global reanalyses or climate models to produce many synthetic TCs. Initially, we synthetically generate wind time series at 250hPa and 850hPa levels as a Fourier series of random phases and a geostrophic turbulence power-law distribution of the kinetic energy spectrum constrained to have accurate monthly means, variances, and covariance. We obtain these statistics and large-scale environmental factors such as potential intensity, wind shear, humidity, and ocean thermal stratification from gridded global reanalyses or climate models. 

The time-evolving environment is seeded randomly in space and time with warm-core vortices drawn from a Gaussian distribution of peak wind speeds centered at 12 m/s (25 knots). Seed vortices propagate forward with a weighted average of synthesized winds at the 250 and 850 hPa levels, according to the beta-and-advection model~\cite{holland1982tropical}. We then calculate the intensity of the vortices deterministically following the track using the Coupled Hurricane Intensity Prediction System (CHIPS) model~\cite{emanuel2004environmental}, which phrases the dynamics in angular momentum coordinates that allow for very high spatial resolution in the storm core. The thermodynamic input to the intensity model includes monthly mean potential intensity, along with 600 hPa temperature and specific humidity, derived from global climate models. The intensity model also accounts for salinity effects on density, affecting TC potential intensity, since the BoB has strong salinity gradients, especially in summer~\cite{emanuel_tropical_2021}. The storms used here are identical to the ones used in the cyclone study~\cite{emanuel_tropical_2021}, with additional storms generated to model the SSP2-4.5 and SSP3-7.0 scenarios.

Over 99\% of the seeded tracks dissipate quickly and are discarded. The remaining successfully grow to make up the downscaled TC climatology of a reanalysis or climate model. Only seeds that develop a maximum wind speed of at least 21 m/s (40 kt) during their lifetime become synthetic TCs. We represent each simulated synthetic TC as an hourly time series of storm parameters, including time, central position, maximum wind speed, pressure deficit, and radius to maximum wind. 

\subsubsection{Bangladesh TC Catalog}
We identify synthetic TCs affecting Bangladesh based on their passage over one or both of the two-line segments displayed in Figure~\ref{fig:figs9}. Overall, we generated a total of 100,200 physically consistent TCs, which is ample to resolve the tails, significantly reduces extrapolation uncertainties, and enhances the robustness of the risk assessment. 

The catalog includes ECMWF/ERA5 and GMAO/MERRA2 climate reanalyses, yielding 4,100 TCs for the current period (1980-2020), representing 5,860 synthetic years based on the annual TC frequency of 0.7 (see Figure~\ref{fig:figs9}). Please see Emanuel et al.~\cite{emanuel2008hurricanes} for comparisons of downscaled TC behavior with observations across basins and see~\cite{emanuel_tropical_2021} specifically for Bangladesh. 

We also generate 2,000 TCs using six CMIP5 and seven CMIP6 global climate models for two time periods: 1981-2000 for historical simulations and 2081-2100 for RCP4.5 and RCP8.5 (CMIP5) and SSP2-4.5, SSP3-7.0, and SSP5-8.5 (CMIP6) simulations. We select the climate models based on the availability of climate variables and in alignment with previous study~\cite{emanuel_tropical_2021}. Table~\ref{tab:tabs2} summarizes additional information about this study's reanalyses and climate models.

\subsection{Hydrodynamic Simulation of Storm Tide}
We used ADCIRC (ADvanced CIRCulation model, two-dimensional barotropic tides, Version 56.02)~\cite{luettich1992adcirc, westerink_adcirc_1994, pringle2021global} for storm tide simulations. Preparing this model requires several steps: Mesh Generation, Model Configuration, and Model Verification, as discussed in this section.

\subsubsection{Mesh Generation}

We use OceanMesh2D~\cite{roberts2019oceanmesh2d, roberts2019automatic} to generate a varying-resolution unstructured mesh for BoB (spanning latitudes from 9°N to 23°N and longitudes from 80°E to 100°E). OceanMesh2D generates a mesh based on several feature-driven geometric and topographic-bathymetric mesh size functions, providing adequate resolution to capture the intricate coastal characteristics. The final unstructured mesh consisted of 42,915 vertices and 77,385 triangular elements, with a resolution ranging from 20 kilometers over the deep ocean to 250 meters near the coastlines (Figure~\ref{fig:figs11}). Table~\ref{tab:tabs4} summarizes the mesh size functions and their corresponding parameters for spatially distributed resolution.

\paragraph{Bathymetry and Shoreline:} To represent bathymetry and shorelines in the computational mesh, we used data from different sources for the primary BoB and the Bengal Delta. 

For the primary BoB, the full-resolution Global Self-consistent Hierarchical High-resolution Shorelines (GSHHS, Version 2.3.7, the vertical datum is mean high water) dataset~\cite{wessel1996global} defines the shoreline boundaries, and the General Bathymetric Chart of the Oceans (GEBCO, Version 2024, the vertical datum is mean sea-level) on a 15 arc-seconds geographic latitude and longitude grid~\cite{weatherall2015new} is the source of bathymetry data. 

For the Bengal Delta, we utilize higher-resolution shoreline and bathymetry data. We derive the shoreline boundaries from the vectorized OpenStreetMap Water Layers (Version2)~\cite{yamazaki2017high}, while an updated bathymetric dataset~\cite{krien2016improved, khan2022storm} provides the bathymetry, which incorporates 77,000 newly digitized points from 34 updated Bangladesh Navigational Charts (see previous studies' supplementary material~\cite{krien2016improved, khan2022storm}). We interpolate the bathymetric measurement points for the Bengal Delta (originally referenced to Chart Datum and subsequently unified to mean sea-level) onto a structured grid with a 200-meter resolution to generate the Digital Elevation Model (DEM) using a simple Kriging method~\cite{qiu2022quantitative}. Bathymetry and bathymetric slope were interpolated onto inner (Bengal Delta) and outer (BoB) mesh vertices directly from the original DEM, then merged to ensure consistency across the connected areas (Figure~\ref{fig:figs12}). The function \textit{lim\_bathy\_slope} in OceanMesh2D kept the maximum bathymetric gradient below 0.1, which is essential to ensure numerical stability~\cite{roberts2019oceanmesh2d}. The constraint helps maintain the water depth variation between adjacent nodes within a defined slope, minimizing unrealistic or abrupt changes in the model.

\paragraph{Mesh Boundaries:} The mesh's outer node string boundary includes both oceanic and riverine boundaries. We classify segments as the mainland boundary if their distance from the shoreline is within 0.4 geographic degrees and their depth relative to sea level is less than 30 m. In contrast, the remaining segments become ocean boundaries. To account for the impact of upstream riverine flow, we implemented 29 normal flux boundaries within the mesh. The locations of river inflows were identified based on the availability of bathymetric data and the attributes of vectorized river reaches, such as average discharge, bankfull discharge, average river width, and length~\cite{lin2020global}. Details regarding the setup of riverine flow in OceanMesh2D are available in a previous study~\cite{qiu2022quantitative}, and Figure~\ref{fig:figs13} illustrates the final mesh boundary conditions. 

\subsubsection{Model Configuration} \label{ModelSetup}
The model setup involves configuring the inputs, forcings, and parameters, as discussed next.

\paragraph{Upstream Riverine Input:} Accurate upstream riverine discharge inputs are challenging in the complex deltaic model because long-term reliable observations for the Bengal Delta river network are lacking. A common approach is to assume an average climatological upstream discharge for astronomical tide and cyclone-induced storm tide periods, providing a constant hourly input to drive the model at the upstream boundary~\cite{tazkia2017seasonal,khan2022storm}. The global reach-level estimation supplied in a previous study~\cite{lin2020global} provides climatological discharge data in our study. We note that sensitivity experiments (driving the model with 10 discharge values ranging from the average discharge and bankfull discharge) demonstrate that both the astronomical tide and storm tide model responses are not sensitive to the upstream discharge forcing, as the locations of our tidal stations (and VS) are far from regions influenced by riverine dynamics. The same assumption also applies to the synthetic TC simulations.

\paragraph{Meteorological Forcing:} We use the analytical wind profile model derived by Chavas et al. (CLE15)~\cite{chavas2015model} to calculate the 1-minute average storm wind at the gradient level. CLE15 is a physics-based model that integrates two existing TC theories to represent wind variations in the inner and outer core regions~\cite{emanuel2004tropical, emanuel2011self}. These two solutions are seamlessly connected at a merging point, ensuring the mathematical continuity of both the angular momentum and its radial derivative. CLE15 utilizes the maximum wind speed and the radius of maximum wind to compute the complete wind profile. To account for the asymmetry of the wind field, we apply the asymmetrical wind model developed by Lin and Chavas (LC12)~\cite{lin2012hurricane} to add surface environmental wind to the storm wind. The gradient wind is converted to the 10-meter winds using a surface wind reduction factor (SWRF) of 0.8 (calibrated) and an empirical formula for inflow angles~\cite{bretschneider1972non}. A reduction factor of 0.893 converts the 1-minute average wind to a 10-minute average~\cite{powell1996hurricane} for the simulations. 

The parametric pressure model~\cite{holland1980analytic} calculates the radial pressure profile based on the pressure deficit. The Powell formulation~\cite{powell2003reduced}, with a capped wind drag coefficient ($C_d$) of 0.0016 (calibrated), converts surface wind velocity into wind stress. A previous study demonstrated that the CLE15-LC12 coupling wind model provides a superior method for simulating TC-induced storm tides, compared to the traditional Holland1980 wind model~\cite{holland1980analytic} with its translation-speed-based asymmetric approach~\cite{wang2022investigation}. Our testing experiments also indicate that the Holland1980 wind model tends to underestimate storm surges, whereas the CLE15 wind model performs more accurately, consistent with the conclusions of Wang et al.~\cite{wang2022investigation}.

\paragraph{Astronomical Tide Forcing:} We take fifteen astronomical tidal components into account to obtain more precise and comprehensive astronomical tide solutions, including eight major tides ($M_2$, $S_2$, $N_2$, $K_2$, $K_1$, $O_1$, $P_1$, $Q_1$), two long-period tides ($Mf$, $Mm$), three nonlinear tides ($M_4$, $MS_4$, $MN_4$) and two minor tides ($2N_2$, $S_1$). The self-attraction, loading (SAL), and internal tide terms are also considered~\cite{pringle2018modifications, shihora2022self}. The latest global satellite-assimilated tidal model TPXO10-Atlas-V2 with a resolution of 1/30°~\cite{egbert2002efficient} provides the equilibrium tide for the model domain and the periodic tidal signals along the oceanic open boundaries. We interpolate the amplitudes and phases of the SAL terms from FES2014 data-assimilated tidal solutions~\cite{lyard2021fes2014}. The internal tide drag coefficient ($C_{it}$) is set to 2.2, with a depth cutoff of 100 meters, based on globally optimized results for the Indian Ocean~\cite{blakely2022dissipation}. 

\paragraph{Bottom Friction:} Boundary layer dissipation in nearshore shallow areas primarily influences tidal solution accuracy. \textit{Manning's N} parameterizes quadratic bottom friction, whose depth dependency enables better redistribution of resistance. Here, we follow a similar approach to previous studies~\cite{krien2016improved,khan2021towards,khan2022storm}, but with further optimization for the spatial distribution to ensure consistency between simulations and observed tidal signals. Figure~\ref{fig:figs14} illustrates the spatially optimized \textit{Manning's N} coefficient. 

For water depths greater than 20 m, we set the \textit{Manning's N} coefficient to 0.02. In the Bengal Delta, where water depths are less than 20 m, we initially divide the region into five subdomains (each with a tidal gauge station) for parameter optimization. After conducting a series of sensitivity tests, we determined that using a \textit{Manning's N} coefficient of 0.013 for the left three regions (covering the Sargar Road, Hiron Point, and Dhulasar stations) and 0.01 for the right two regions (covering Chachanga and Chittagong stations) ensures that the tidal solutions at the five stations achieve an accuracy comparable to and consistent with that reported in previous studies~\cite{krien2016improved,khan2021towards,khan2022storm}. 

The lower bottom friction identified in the Meghna-Chattogram coastal plain is likely attributable to its fine-grained silty substrate~\cite{islam2021characteristics}. Pringle et al.~\cite{pringle2018finite} report a similar phenomenon in the Yellow Sea region of China. We recommend additional optimization of bottom friction dissipation using semi-data-informed approaches~\cite{pringle2018finite} when detailed and accurate sediment type information becomes available for the nearshore Bengal Delta.

\paragraph{Computational Setup and Performance:} To balance computational cost and ensure numerical stability for ensemble simulations, selecting an appropriate time step and domain decomposition strategy (CPU cores used for parallel simulation) is crucial. Based on previous experience~\cite{pringle2021global}, we determined a maximum time step of 40 seconds for the 250-meter mesh. Using Cyclone Sidr (2007), we tested the impact of domain decomposition on simulation wall-clock cost. We found that a parallel strategy with approximately 40 domains maximized computational efficiency, with limited gains observed beyond this point, providing critical insights into optimal computational resource allocation. Figure~\ref{fig:figs15} and Figure~\ref{fig:figs16} display the mesh decomposition and efficiency performance, respectively. Besides, the 40-second time step proved robust for synthetic TC-storm tide simulations, with only a tiny fraction (approximately 0.5\%) crashing. A reduced time step of 10 seconds corrected all the failed simulations without affecting the others. We apply a two-day model spin-up to all simulations. 

\subsubsection{Model Verification} \label{model verification}
Accurate simulation of astronomical tides is essential for reliably modeling storm tides. In this section, we verify astronomical and storm tide performance, collecting substantial data.

\paragraph{Astronomical Tide Verification:} We compare the model's output with harmonic constituents at five tide gauge stations: Sagar Roads, Hiron Point, Dhulasar, Charchanga, and Chittagong (see Figure~\ref{fig:fig1} and Table~\ref{tab:tabs1} for detailed locations). The model was initially driven by five leading astronomical tidal constituents ($M_2$, $S_2$, $N_2$, $K_1$, and $O_1$) for 30 days, with two days for spin-up and 28 days for harmonic analysis~\cite{ngodock2016improving,pringle2021global}. The co-tidal chart (Figure~\ref{fig:figs17}) illustrates tidal wave propagation patterns, which aligns with the findings of previous studies~\cite{sindhu2013characteristics, rose2022tidal}. The spatially distributed root-mean-square error (RMSE) discrepancies for each tidal constituent compared to the TPXO-10-Atlas-V2 are less than 2.5 cm across most of the ocean (Figure~\ref{fig:figs18}), indicating that our model performs well in the primary BoB basin. 

However, we note relatively higher discrepancies at the head of the Bengal Delta. That is primarily due to the model grid of TPXO-10-Atlas-V2 not extending into the river network and the lack of assimilation of accurate bathymetric data in these nearshore areas. Thus, the TPXO-10-Atlas-V2 product itself exhibits significant uncertainty in these shallow areas. By employing spatially optimized \textit{Manning's N} coefficient in the Bengal Delta (Section~\ref{ModelSetup}), the total complex error of all four available constituents ($M_2$, $S_2$, $K_1$, and $O_1$) is comparable to that of previous studies~\cite{krien2016improved,khan2021towards,khan2022storm}, the global satellite as well as site-assimilated TPXO-10-Atlas-V2 model, as shown in Table~\ref{tab:tabs5}. 

\paragraph{Storm tide Data Collection and Processing:} Accurately simulating storm tide requires high-quality water level measurements for model verification. The lack of water level observations during TC landfalls in Bangladesh poses a significant challenge. To address this, we assembled the most comprehensive water level observations available to date for storm tide verification. We collected long-term water level observations from two tidal stations, Hiron Point (sourced from Bangladesh Inland Water Transport Authority [BIWTA]) and Chittagong (sourced from the University of Hawaii Sea Level Center [UHSLC]), starting in 2007. We also collected water level observations from two other stations, Khepupara and Cox's Bazaar (sourced from BIWTA), during Cyclone Sidr's (2007) landfall. Since the Khepupara station stopped working after Sidr's landfall, we obtained an estimate of the peak water level from the ITJSCE report~\cite{ITJSCE2008, mamnun2020forcing}. 

Consequently, our dataset includes water level observations from eight recent TCs that made landfall in Bangladesh or India but significantly impacted Bangladesh. For four of these TCs, verification data from two stations were available. In contrast, for Sidr, data from four stations were available, with stations distributed on both sides of the TC's forward path. Thus, we assembled the most comprehensive water level observations for model verification. 

Some processing was further required to use the observations effectively. The observations were based initially on local Chart Datum and Bangladesh Standard Time (UTC+06:00). To ensure consistency in model verification, we converted them to local mean sea level and UTC. The datum conversion used were Hiron Point (-1.86 m), Chittagong (-3.50 m), Khepupara (-2.30 m), and Cox's Bazaar (-2.10 m), averaged over multiple years of observations. Previous studies~\cite{mamnun2020forcing,as1998coastal} report similar datum conversion values, and UHSLC's official website also documents datum information for several stations. 

\paragraph{Storm Tide Verification:} Having processed the observational data, we use statistical metrics from multiple historical TCs at various stations to evaluate the model's performance in simulating storm tides. As shown in Figure~\ref{fig:figs19}, the overall RMSE of 0.25 m with a 0.94 coefficient of determination ($R^2$) indicates that our model performed well in simulating peak storm tides. In addition, Figure~\ref{fig:figs20} shows strong agreement in both storm tide phase and amplitude for cyclone Sidr (which was the only cyclone for which observations at many stations were available). For visualization, Figure~\ref{fig:figs21} also displays Sidr's peak storm tide map. Thus, the hydrodynamic model's storm tide simulation performance is consistent with other results and reasonably verified.

\subsection{IPCC AR6 Relative Sea-Level Projection}\label{SLR}

Our probabilistic, localized relative sea-level projections are based on the Framework for Assessing Changes To Sea-level (FACTS)~\cite{kopp2023framework}, which emphasizes the role of the Antarctic and Greenland ice sheet as drivers of structural uncertainty in sea-level rise projections. FACTS can generate seven alternative probability distributions relative to a 1995-2014 baseline under multiple alternative emissions scenarios presented in the Intergovernmental Panel on Climate Change Sixth Assessment Report (IPCC AR6)~\cite{kopp2023communicating}. 

In this study, we apply the gauge-based Monte Carlo samples~\cite{kopp2023framework} (20,000 in total) of future relative sea-level under workflows 2-E and scenario CMIP6 SSP2-4.5 (Figure~\ref{fig:figs22}), SSP3-7.0 (Figure~\ref{fig:figs23}), and SSP5-8.5 (Figure~\ref{fig:figs24}), covering the period from 2080 to 2100. The workflow 2-E of FACTS employs Gaussian Process emulation for Greenland, glaciers, and Antarctica and forms the basis of the \textit{medium confidence} projections presented by IPCC AR6. The four gauge-based stations located in coastal Bangladesh are Hiron Point (Permanent Service for Mean sea-level [PSMSL] ID=1451), Khepupara (PSMSL ID=1454), Charchanga (PSMSL ID=1496), and Cox's Bazaar (PSMSL ID=1476). We assume that the sea level will rise monotonically within each decade. Thus, we represent the distributions using statistics at 2080, 2090, and 2100, linearly interpolated to the year of the simulated storm to get SLR samples. From a sampling perspective, we can thus draw SLR samples, with replacement, from a distribution for any scenario, location, and year of interest, which is essential to incorporate SLR, see Section~\ref{jointlysamplingTC_SLR}.

\subsubsection{Nonlinear Interactions} Rising sea levels increase regional water depth, causing the quadratic bottom stress to redistribute with depth. In turn, the total water level driven by both astronomical and meteorological forcing is affected, as shown in Figure~\ref{fig:figs25}). Neglecting the SLR-surge interaction underestimates the total water level. In our work, we account for the nonlinear interactions among sea level, tides, and storm surges, offsetting the initial sea surface with an SLR sample value paired with each TC sample. Section~\ref{jointlysamplingTC_SLR} provides details on the joint sampling, and Section~\ref{ReturnPeriodEst} provides details on the impacts of including SLR. 

\subsubsection{Storm Tide Catalogs: Jointly Sampling TCs, Tides, and SLR} \label{jointlysamplingTC_SLR}

We use a {sampling approach} to integrate tides and SLR with TCs in storm tide assessments. A critical feature of this approach is the {time association} of synthetic downscaled TC samples under specific climate models and scenarios, which connects these three components. The method utilizes the TC's origin and arrival times to synchronize tidal phases and constrain SLR samples to the associated year. By generating an extensive ensemble of synthetic TCs (100 per year) with random times distributed across the decades of interest, the approach effectively produces {Monte Carlo samples} from the joint distribution of TCs (for a given climate model, scenario, and period), tidal phases, and SLR (varying over the period). This results in an ensemble representing the marginal distribution of storm tides that incorporates the complex interactions between these effects. For a large ensemble, this strategy offers several advantages:
\begin{itemize}
    \item {Avoids Simplistic Assumptions}: By linking TC timing with tides and SLR, the approach avoids relying on an oversimplified linear relationship between these variables.
    \item {Scalability}: It circumvents the {combinatorial explosion} that would arise from explicitly rotating the tidal phase for each storm or pairing every SLR sample with each TC. 
\end{itemize}

Synchronizing storm timing with the tide implicitly accounts for tidal phase effects. For SLR, previous studies have used {convolutional} methods\cite{marsooli2019climate} or {copula-based} approaches\cite{moftakhari2017compounding} to derive the cumulative distribution function of storm tides. However, these methods face significant limitations:
\begin{itemize}
    \item {Copulas}: These are challenging to construct, particularly in scenarios involving complex dependencies and nonlinearities.
    \item {Convolution}: By definition, convolution assumes linearity and shift invariance and applies when the distributions are independent. This assumption does not hold for the nonlinear interactions between TCs, tides, and SLR. The latter two also depend on TC event properties.
\end{itemize}

Monte Carlo sampling overcomes these challenges by accommodating {nonlinear dependencies}. In cases where the relationships are linear and independent, it is easy to see that Monte Carlo converges to the convolutional case in the large sample limit. However, it remains the only generally viable method for capturing {nonlinear interactions} among tides, SLR, and TCs. TC characteristics influence SLR and tides and are not independent variables.

To be sure, hydrodynamic simulations do not inherit a single SLR (or a few) values representing the entire time interval (e.g., 2081-2100). The SLR distributions, which change with the year (Figure~\ref{fig:figs22},~\ref{fig:figs23},~\ref{fig:figs24}), are sampled to randomly associate with TCs generated in that year (100 pairs). We posit that this is reasonable (given typical annual TC frequencies) for sampling SLR.

\subsection{Statistical Analysis}\label{StatisticalAnalysis}
\subsubsection{Return Period Estimation} \label{ReturnPeriodEst}

The primary statistical analysis produces return periods corresponding to return levels with and without SLR. Further, we separately calculate the storm tide versus return period curves for each of the 58 coastal stations. For each storm tide simulation, we record the highest water elevation during the cyclone's lifetime and the corresponding time. The peak storm-tide arrival times help assess the seasonal distribution of extreme storm tides, as presented in Section~\ref{Seasonalresult}. The return period calculations use a vector of 2,000 peaks (per climate model) or 4,100 peaks (for climate reanalyses) and employ the following steps.

We assume that TCs arrive as a stationary Poisson process in a given climate~\cite{lin2012physically}. The storm-tide return period, incorporating storm surge and astronomical tide excluding SLR, exceeding a given return level \textit{h} uses the formula~\cite{marsooli2019climate}:

\begin{equation} \label{eu_eqn1}
T_{\eta_\mathrm{storm\,tide(noSLR)}}(h)=\frac{1}{\lambda \times EP\{\eta_{\mathrm{storm\,tide(noSLR)}} > h\}} 
\end{equation}

Where $\lambda$ is the TC annual frequency, ${EP\{\eta_{\mathrm{storm\,tide(noSLR)}} > h\}} =1-P\left\{\eta_{\mathrm{storm\,tide(noSLR)}} \leq h\right\}$ is the exceedance probability of maximum TC-induced storm tide, and $P\left\{\eta_{\mathrm{storm\,tide(noSLR)}} \leq h\right\}$ is the cumulative distribution function. Please note that for climate models and reanalyses, $\lambda$, the TC annual frequency, relates to the ratio of successful to unsuccessful seeds in a given year (see Section~\ref{syntheticdownscaling}), with a constant of proportionality established by comparing time averages to dependable historical records. For reference, Bangladesh's historical record suggests $\lambda = 0.7$ (see Section~\ref{BiasCorrection}) under a Poisson arrival model.

The storm-tide return period (incorporating storm surge, astronomical tide, and SLR) exceeding a given return level \textit{h} uses the same method~\cite{marsooli2019climate}:

\begin{equation} \label{eu_eqn2}
T_{\eta_\mathrm{storm\,tide(withSLR)}}(h)=\frac{1}{\lambda \times EP\{\eta_{\mathrm{storm\,tide(withSLR)}} > h\}}
\end{equation}

Where $EP\{\eta_\mathrm{storm\,tide(withSLR)} > h\} =1-P\{\eta_\mathrm{storm\,tide(withSLR)} \leq h\}$ is the exceedance probability (tail probability) of maximum storm tide denoted, and $P\{\eta_\mathrm{storm\,tide(withSLR)} \leq h\}$ is the cumulative distribution function. 

Extreme value theory suggests using the Peaks-Over-Threshold method with a Generalized Pareto Distribution (GPD) to model the upper tail~\cite{coles2001introduction} to estimate the storm tide probability. We fit GPD to the upper tail and a kernel density estimate~\cite{terrell1992variable} to the remaining portion (function {\it paretotails} in MATLAB), as shown in Figure~\ref{fig:figs26}. For a given return period, the inverse cumulative distribution function (function {\it icdf} in MATLAB) calculates the corresponding return levels. We characterize the upper tail as the $98^{\texttt{th}}$ percentile through a trial-and-error approach, testing from the $95^{\texttt{th}}$ to the $99^{\texttt{th}}$ percentile and to ensure the tail achieved the best fit with the smallest RMSE~\cite{marsooli2019climate}. 

These methods apply to each climate model in each scenario and at each station. We estimate the return periods with and without SLR in all scenarios after additionally correcting for bias (see Section~\ref{BiasCorrection} and quantifying uncertainty (see Section~\ref{UncertaintyQuantify})).

Additionally, Section~\ref{TC_SLR_Contribution} uses these methods to assess the relative impacts of TC climatology SLR on storm tides in future climate scenarios over the interval 2081-2100, relative to the present (1981-2000), each representing a stationary climate regime within which SLR continuously varies (see Section~\ref{SLR}). Section~\ref{Historical_event_assessment} assesses annual frequencies (the inverse of the return period) of historically destructive cyclones using return periods in different climate regimes, including the current climate (which offers better resolution than the observational record) and various warming scenarios.

\subsubsection{Bias Correction} \label{BiasCorrection}

Climate model projections are biased in various ways, including TC frequency, seasonal distributions, and storm tide return periods. These biases originate from distinct sources and require tailored correction methods, which we detail below.

\paragraph{TC Frequency:} 
As shown in Figure~\ref{fig:figs10}, the annual frequency of TCs varies significantly depending on the climate model used. The observed annual frequency of historical TCs (with peak wind speeds exceeding 33 knots) making landfall in Bangladesh is 0.7, based on 26 events recorded in the JTWC dataset from 1980 to 2020~\cite{bhardwaj2020climatological}. We assume that TC occurrences follow a Poisson process, making the arrival rate $\lambda=0.7$ a reasonable representation of the current climate's annual TC frequency. Additional details on landfall events and Poisson distribution fits are provided in Figure~\ref{fig:figs9}.

The observed annual frequency calibrates both reanalysis and climate model frequencies. Under the current climate, this calibration entails multiplying the projected annual frequencies in future climate scenarios with a correction ratio derived from the current climate. Table~\ref{tab:tabs3} presents the bias-corrected TC annual frequencies for future scenarios. The substantial variation among climate models reflects both systematic differences among models and the inherent uncertainties caused by internal climate variability~\cite{emanuel2022tropical}.

\paragraph{Storm Tide Return Period:} In addition to frequency, biases in storm-tide return period curves are evident (see Section~\ref{ReturnPeriodEst}). A quantile-quantile mapping method~\cite{cannon2015bias} is employed to align distributions and calibrate storm-tide return periods under the current climate. Specifically, climate model return levels are adjusted to match the average return levels derived from reanalyses (ECMWF/ERA5 and GMAO/MERRA2) across all quantiles in yearly return period increments. Bias corrections are computed for each station, return period, and climate model in the current climate and applied to the corresponding model and station for future scenarios. As illustrated in Figure~\ref{fig:figs27}, this approach ensures that the entire storm-tide return period curve is effectively corrected, providing consistency and reliability in the assessment of future storm-tide risks.

\paragraph{Storm Tide Seasonal Regimes:}

Time-indexed synthetic TCs enable the association of arrival times with peak storm tides. For seasonal assessments, we bin bias-corrected storm tides into 14-day windows, defined as $\pm 7$ days around the beginning and midpoint of each month. This results in a 24-bin distribution of fractional allocations of annual storm tide frequencies.

To address seasonal biases, we align the 24-bin distribution between the climate model projections and averaged reanalyses using a methodology similar to the one applied in previous bias corrections. This alignment yields a seasonal storm tide bias correction for each climate model. The per-model bias correction then applies to future seasonal frequency assessments for the corresponding models.

This seasonal bias correction allows for a detailed evaluation of shifts in storm-tide severity and frequency across different seasonal periods, as further discussed in Section~\ref{Seasonalresult}.

\subsubsection{Storm Tide Return Period Confidence Intervals} \label{UncertaintyQuantify}
Within any climate scenario, certain factors reduce the confidence in storm-tide return period estimates. While the joint sampling of SLR, tides, and TCs (see Section~\ref{jointlysamplingTC_SLR}) propagates essential uncertainties (see Section~\ref{Limitations} for additional discussion) into storm tides, additional sources remain. We account for two primary sources:

\paragraph{Model Error:}
Climate model ensembles implicitly represent model error, a significant source of uncertainty. The rapid TC downscaling process allows us to account for model error more comprehensively by incorporating a more extensive set of climate models rather than relying on just a few selected models.

\paragraph{Parameter Uncertainty in GPD Fitting:}
The fitting procedure for GPDs introduces parameter uncertainties. These uncertainties are quantified using a bootstrap approach, generating confidence intervals from 1,000 trials with replacement. Figure~\ref{fig:figs26} illustrates these bootstrap-derived confidence intervals from the piecewise Kernel-GPD fitting process. While GPD-induced uncertainty contributes to confidence loss, it is relatively minor compared to the more substantial impact of model error (see Figure~\ref{fig:figs28}).

We assess confidence intervals for storm-tide return periods across all spatial scales (station, regional, national) and climate scenarios. To do so, we:
\begin{itemize}
    \item Generate multiple bootstrap samples to fit storm-tide return period curves for each climate model.
    \item Accumulate the full mixed ensemble, accounting for both GPD parameter variability and climate model variability.
\end{itemize}
This approach allows us to calculate the mean (or median) return period curve and its upper and lower confidence bounds, providing a comprehensive assessment of return period uncertainty for each scenario.

\subsection{Limitations and Future Perspectives}
\label{Limitations}
There are a few limitations in this study that require further improvement.  Firstly, the hydrodynamic model does not account for wave setup, which refers to the increased water level at the coast caused by the breaking of waves in the surf zone. While wind waves naturally contribute to extreme water levels and are valuable in site-specific applications~\cite{leijnse2022generating,bakker2022estimating,CEIP_I_2022}, studies in Bangladesh suggest that wind waves typically account for 10\% to 15\% of the maximum total water level~\cite{krien2017towards,khan2021towards}. Since our primary focus is quantifying large-scale climate change impacts, using storm tides is a practical impact variable.  Simulating over 100,000 TCs downscaled from five CMIP5 and seven CMIP6 climate models would possibly add over two orders of magnitude of computational effort to our resources without changing our conclusions. We encourage future studies to explore more computationally efficient approaches for quantifying climate risk while considering the influence of waves, such as representative TC sampling~\cite{bakker2022estimating} or machine learning-based wave emulation~\cite{wang2024physics}.

Secondly, significant uncertainty exists at the nexus of climate scenarios, climate models~\cite{knutson2020tropical} and SLR~\cite{kopp2023communicating}. Although the newer CMIP6 has reduced these uncertainties compared to CMIP5, there is still a need for future research to tighten the scenarios and improve the accuracy of these models in projecting the atmospheric (e.g., cloud feedbacks and cloud-aerosol interactions) and oceanic variables that control TC activity and sea-level dynamics. 

Thirdly, it is essential to note that TCs induce storm surges but also produce extreme precipitation, especially when they stall during landfall. Emanuel~\cite{emanuel_tropical_2021} projects a twenty-fold increase in severe storm accumulated rainfall (exceeding 1,000 mm in Dhaka) due to climate change. The potentially growing overlap with the monsoon season and summer heat stress~\cite{matthews2019emerging} should motivate future research to consider these factors jointly in Bangladesh, integrating coastal (including wave), fluvial, and pluvial processes components~\cite{merz2021causes,bates2021combined,leijnse2021modeling}, in addition to the compound effects of TC winds and rainfall that is well underway in the community. 

Furthermore, in this paper, the monsoon season (June 1 to September 1) is defined based on the present climate and applied as-is in future projections. While our study does not directly focus on the monsoon, it provides a new perspective on potential future monsoon changes in Bangladesh from the perspective of cyclones. We recommend a study of interactions between TC and the monsoon in future studies. 

Lastly, this study assumes that bathymetry and morphology will remain unchanged by the end of this century under future climate scenarios. However, bathymetry and morphology will likely evolve over this time horizon, particularly in a dynamic deltaic system shaped by complex sediment transport processes and human disturbances~\cite{paszkowski2021geomorphic}. We recommend that future studies incorporate time-evolving bathymetry and morphology through interdisciplinary collaboration with estuarine geomorphologists to achieve more robust long-term projections~\cite{nienhuis2020global, becker2024coastal}.

\section{Resource availability}

\subsection{Data availability} \label{dataavailability}
Some public data sets used for this study are available at \url{https://www.soest.hawaii.edu/pwessel/gshhg/} (GSHHS), \url{http://hydro.iis.u-tokyo.ac.jp/\~yamadai/OSM_water/} (Rasterized OSM Water Layer Map), \url{https://www.gebco.net/data_and_products/gridded_bathymetry_data/} (GEBCO), \url{https://www.metoc.navy.mil/jtwc/jtwc.html?north-indian-ocean} (JTWC), \url{ftp://ftp.legos. obs-mip.fr/pub/FES2012-project/data/LSA/FES2014/} (FES2014 tidal database), \url{https://www.tpxo.net/global/tpxo10-atlas} (TPXO10-Atlas-V2), \url{https://github.com/rutgers-ESSP/ipCC-AR6-Sea-Level-Projections} (IPCC AR6 Sea-Level Projections), \url{https://uhslc.soest.hawaii.edu/stations/?stn=124#levels} (UHSLC tidal data), \url{https://zenodo.org/records/3552776} (Global reach-level hydrological attributes). WindRiskTech L.L.C. performs TC-induced risk assessments for clients worldwide and provides datasets free of charge for scientific research upon request (info@windrisktech.com), subject to a non-redistribution agreement. We will make our storm-tide return-period estimates incorporating SLR based on the IPCC AR6 multimodel ensemble publicly available upon the acceptance of our manuscript.

\subsection{Code availability}
The hydrodynamic model used in this study is ADCIRC (Version 56.02), which can be accessed freely at \url{https://github.com/adcirc/adcirc}. The OceanMesh2D toolbox generates the unstructured mesh. It is accessible at \url{https://github.com/CHLNDDEV/OceanMesh2D}. MATLAB (Version 2022a) {\it paretotails} function at \url{https://www.mathworks.com/help/stats/paretotails.html} provides the Peaks-Over-Threshold based Generalized Pareto Distribution fitting. pyTMD available at \url{https://github.com/tsutterley/pyTMD} provides astronomical tide preprocessing analyses. The codes used to create the unstructured mesh, ADCIRC inputs, and visualization are available from the corresponding authors upon reasonable request.

\section{Acknowledgments} \label{acknowledgements}
We acknowledge funding from the MIT Climate Resilience Early Warning Systems Climate Grand Challenges project, the Jameel Observatory JO-CREWSNet project, and the Weather Extremes Grand Challenges project. Schmidt Sciences, LLC provided support. 

The authors thank Yann Krien for offering the initial improved bathymetric dataset for the Northern Bay of Bengal and AKM Saiful Islam for sharing the updated bathymetric dataset for the Bengal Delta, as well as the long-term water level observations from multiple BIWTA gauge stations. We thank Michael Steckler for sharing the GPS coordinates and observed water level data for Hiron Point and Khepupara. We thank Md Wasif E Elahi for providing their model output for comparison. We thank Shuai Wang and Coleman P. Blakely for incorporating the CLE15-LC12 wind model into ADCIRC. We also express our gratitude to the developers of ADCIRC and OceanMesh2D, with special thanks to William J. Pringle and Keith J. Roberts for their help over the years. We thank Svetlana Erofeeva for discussing tidal solutions and sharing the latest TPXO10-Atlas-V2 dataset. 

Last but not least, we sincerely appreciate the editorial team and three anonymous reviewers for their precious time and valuable suggestions, which have immensely improved the manuscript.

\section{Author contributions}
Conceptualization, J.Q., S.C.R. and K.A.E; Methodology, J.Q., S.C.R. and K.A.E; Investigation, J.Q. and S.C.R.; Verification, J.Q.; Visualization, J.Q.; Writing – Original Draft, J.Q.; Writing – Drafting, Review \& Editing, S.C.R.; Writing - Review, K.A.E;  Resources, S.C.R.; Supervision, S.C.R.; Funding Acquisition, S.C.R. and K.A.E.;

\section{Declaration of interests}
The authors declare no competing interests.

\bibliography{mainbib}

\newpage
\renewcommand{\thefigure}{A\arabic{figure}} 
\renewcommand{\theHfigure}{A\arabic{figure}}
\setcounter{figure}{0}
\begin{appendices}
\section{Supplementary1}\label{secS1}
\begin{figure*}[htb!]%
\centering
\includegraphics[width=1\textwidth]{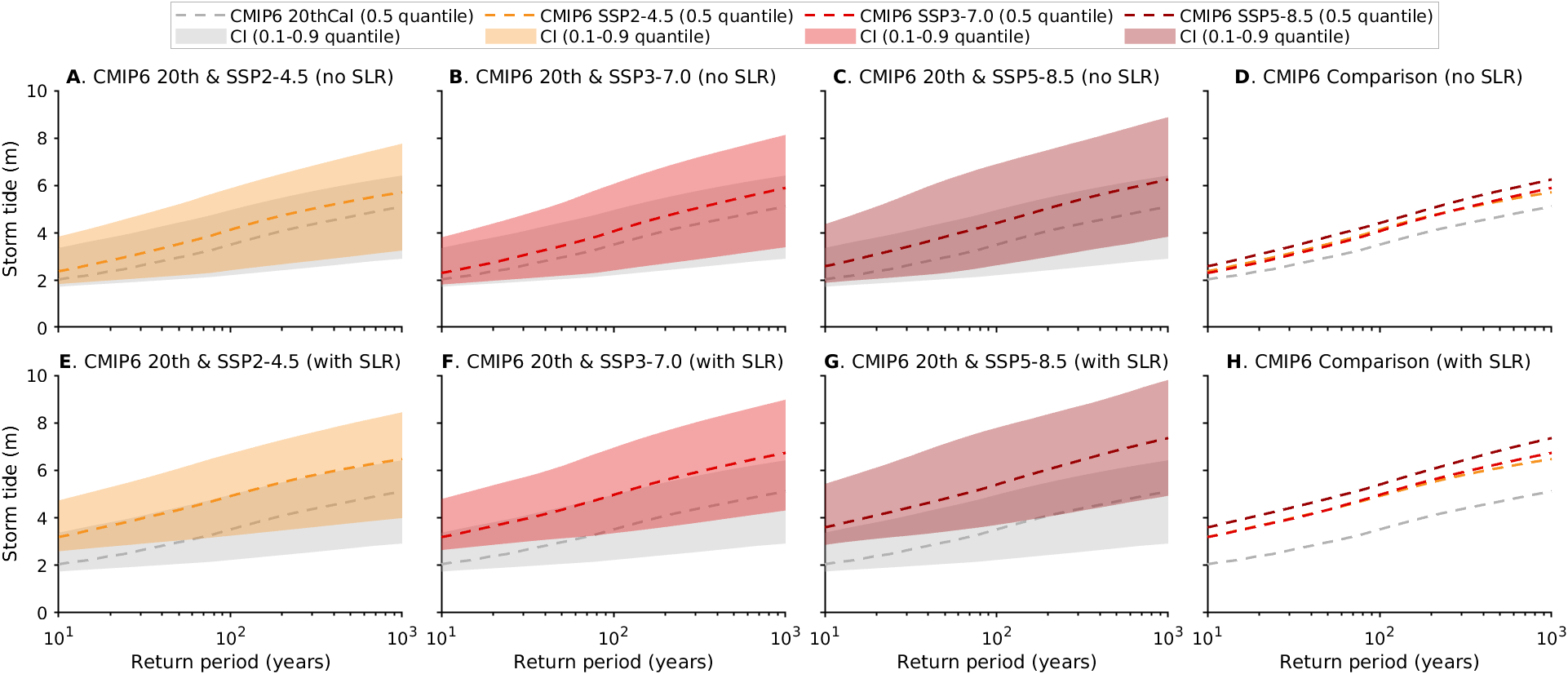}
\caption{\textbf{Bangladesh's storm tide versus return period curves projected by CMIP6 models at the national scale.} Storm tides are the water level elevations (combining components of astronomic tide and storm surge, and mean sea-level state) relative to the mean sea-level of the 1995-2014 baseline. \textbf{A, B, C, D,} Projections without considering SLR. \textbf{E, F, G, H,} Projections with incorporating SLR. \textbf{D, H,} Comparison between CMIP6 current, SSP2-4.5, SSP3-7.0 and SSP5-8.5 climates, respectively. Dashed lines indicate the ensemble median (0.5 quantiles) while shaded areas indicate each estimate's confidence interval (CI, 0.1-0.9 quantile). The current climate period spans from 1981 to 2000, while the future climate period spans from 2081 to 2100. The confidence intervals account for variability in Tide, SLR, TCs, climate models, multi-station data, and Kernel-GPD parameters.}
\label{fig:figs1}
\end{figure*}     

\begin{figure}[htb!]%
\centering
\includegraphics[width=1\textwidth]{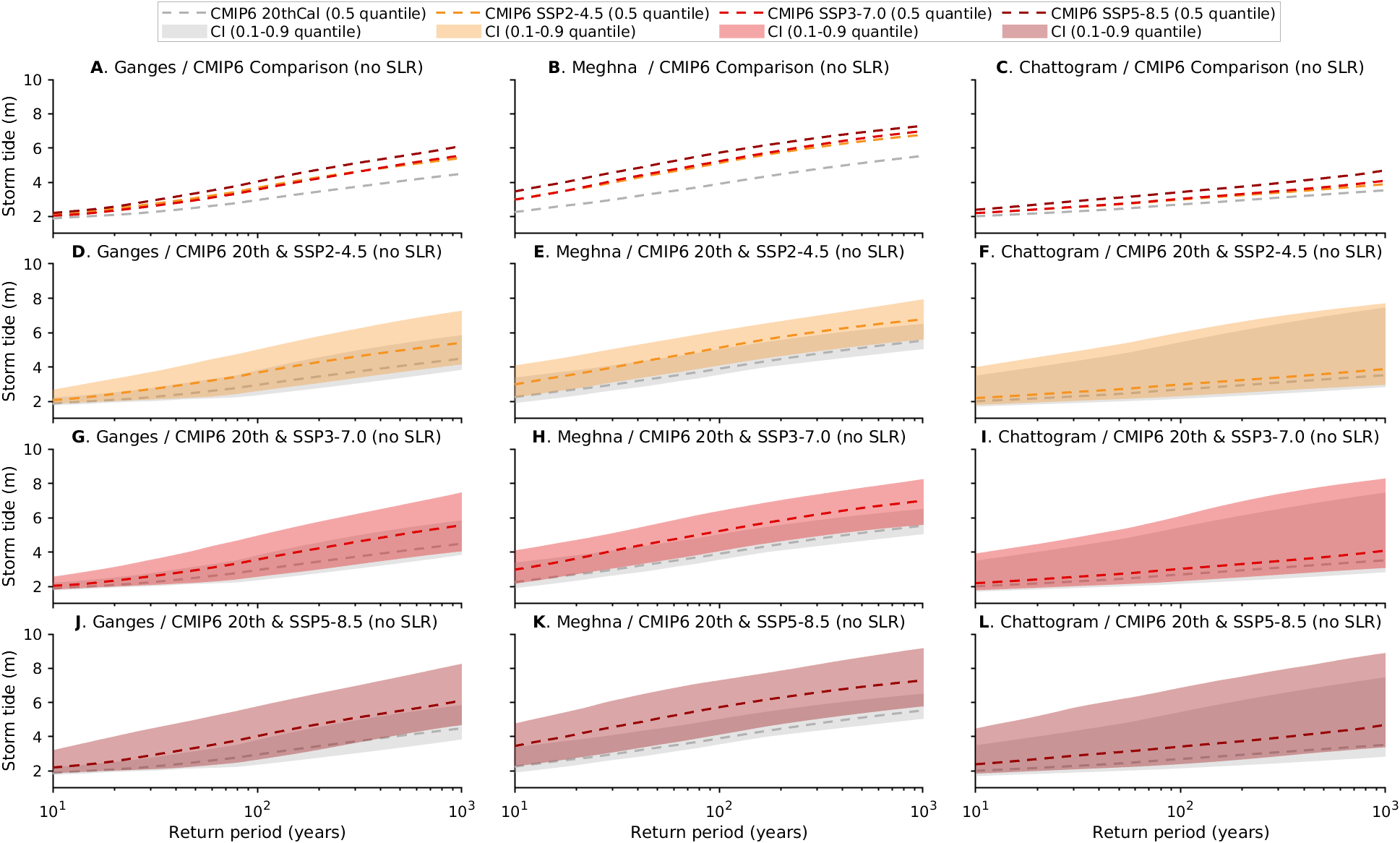}
\caption{\textbf{Bangladesh's storm tide versus return period curves projected by CMIP6 models at the regional scale.} Storm tides are the water level elevations (combining components of astronomic tide, storm surge, and mean sea-level state) relative to the mean sea-level of the 1995-2014 baseline. \textbf{A, D, G, J,} Projections for the Ganges. \textbf{B, E, H, K,} Projections for the Meghna. \textbf{C, F, I, L,} Projections for the Chattogram. \textbf{A, B, C,} Comparison between CMIP6 SSP2-4.5, SSP3-7.0 and SSP5-8.5 scenarios. \textbf{D, E, F,} CMIP6 model ensemble composite under the SSP2-4.5 scenario. \textbf{G, H, I,} CMIP6 model ensemble composite under the SSP3-7.0 scenario. \textbf{J, K, L,} CMIP6 model ensemble composite under the SSP5-8.5 scenario. Dashed lines indicate the ensemble median (0.5 quantiles) while shaded areas indicate each estimate's confidence interval (CI, 0.1-0.9 quantile). The current climate period spans from 1981 to 2000, while the future climate period spans from 2081 to 2100. The confidence intervals account for variability in Tide, SLR, TCs, climate models, multi-station data, and Kernel-GPD parameters.} 
\label{fig:figs2}
\end{figure}  

\begin{figure}[htb!]%
\centering
\includegraphics[width=1\textwidth]{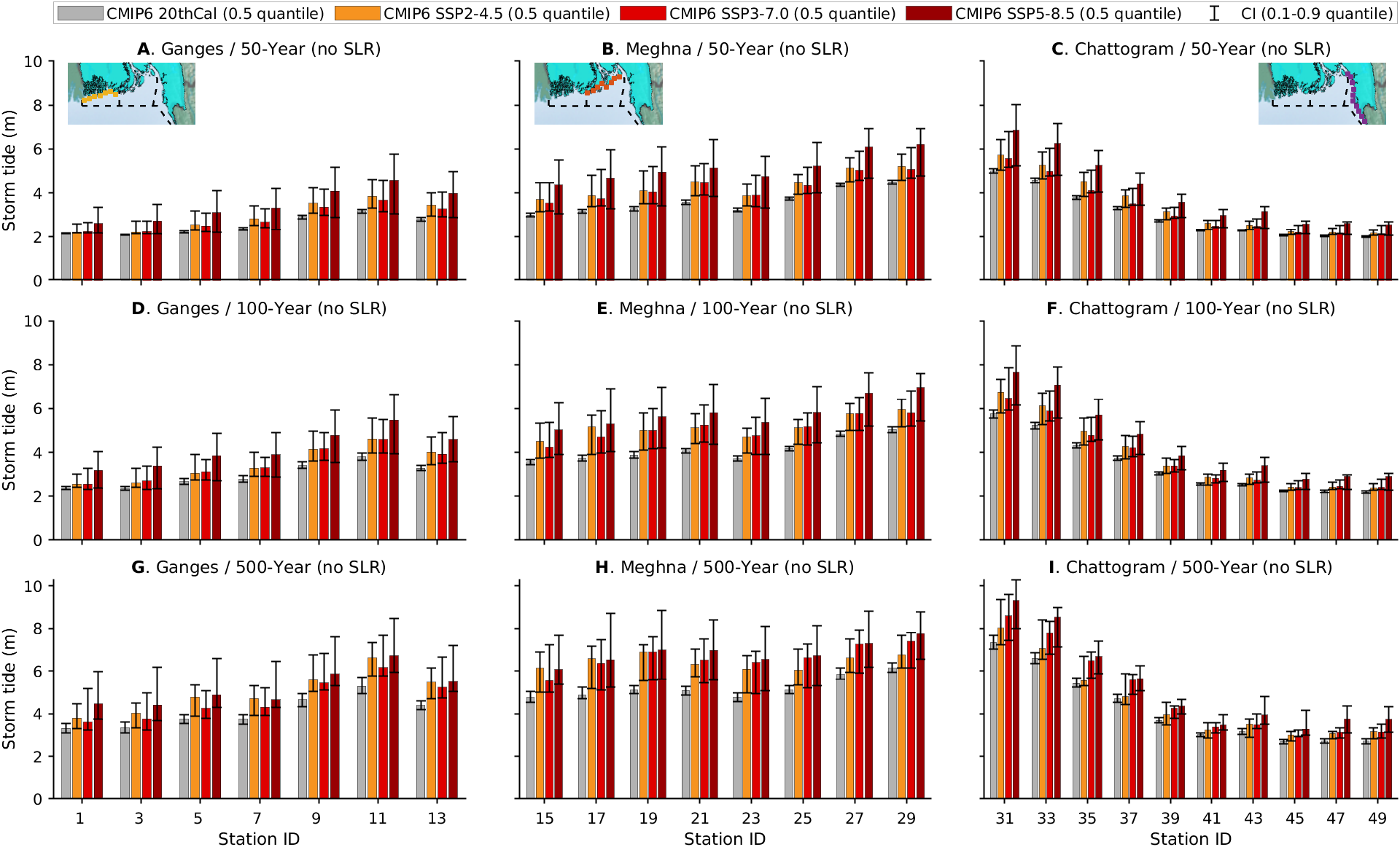}
\caption{\textbf{Bangladesh's 50-, 100-, 500-year storm tides projected by CMIP6 models at the station scale.} Storm tides are the water level elevations (combining components of astronomic tide, storm surge, and mean sea-level state) relative to the mean sea-level of the 1995-2014 baseline. \textbf{A, B, C,} 50-year return period. \textbf{D, E, F,} 100-year return period. \textbf{G, H, I,} 500-year return period. \textbf{A, D, G,} Stations located in the Ganges (southwest Bangladesh). \textbf{B, E, H,} Stations located in the Meghna (middle Bangladesh). \textbf{C, F, I,} Stations located in the Chattogram (east Bangladesh). Projections are conducted for all 58 stations, but the graph displays only every other VS. Histogram heights indicate the ensemble median (0.5 quantiles) for the CMIP6 current climate, future climate under the SSP2-4.5, SSP3-7.0 and SSP5-8.5 scenarios, respectively. Vertical error bars indicate each estimate's confidence interval (CI, 0.1-0.9 quantile). The current climate period spans from 1981 to 2000, while the future climate period spans from 2081 to 2100. The confidence intervals account for variability in Tide, SLR, TCs, climate models, and Kernel-GPD parameters. Base map sourced from the BDP 2100 (Baseline Volume 1, pg. 403), Humanitarian Data Exchange, World Bank, ESRI ArcGIS, Maxar, Earthstar Geographics, USDA FSA, USGS, Aerogrid, IGN, IGP, and the GIS User Community.}
\label{fig:figs3}
\end{figure}  

\begin{figure*}[htb!]%
\centering
\includegraphics[width=1\textwidth]{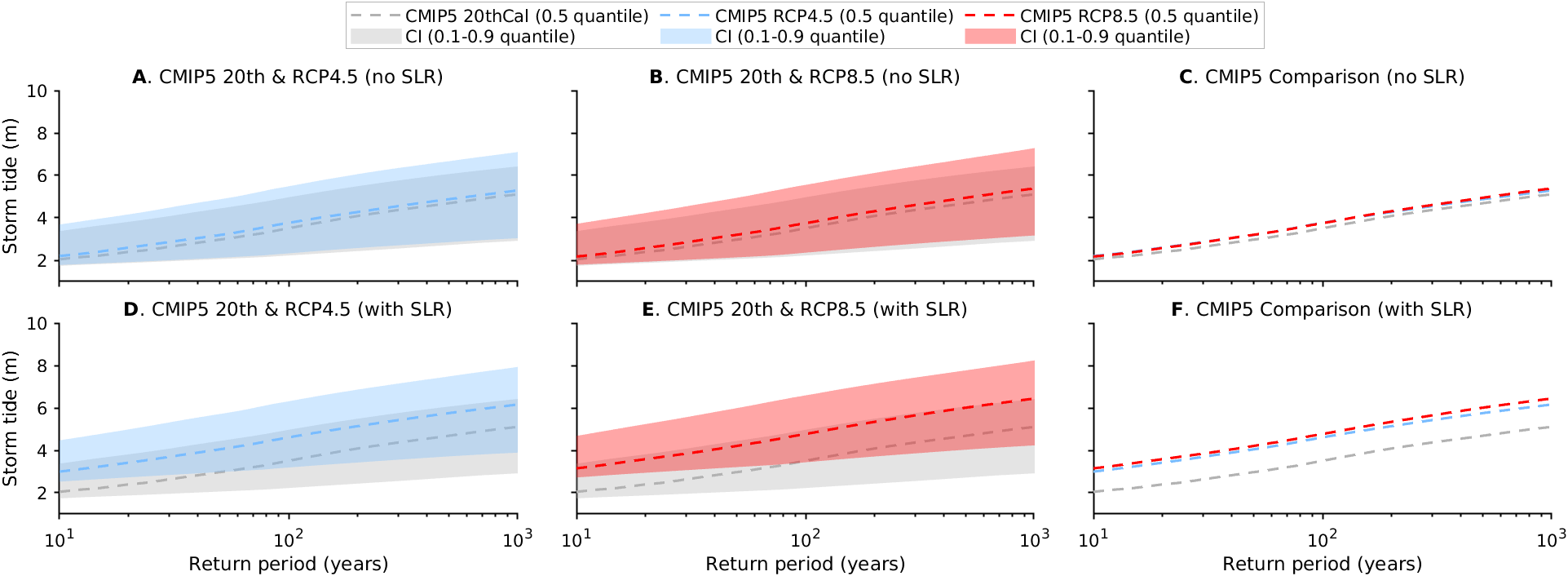}
\caption{\textbf{Bangladesh's storm tide versus return period curves projected by CMIP5 models at the national scale.} Storm tides are the water level elevations (combining components of astronomic tide, storm surge, and mean sea-level state) relative to the mean sea-level of the 1995-2014 baseline. \textbf{A, B, C,} Projections without considering SLR. \textbf{D, E, F,} Projections with incorporating SLR. \textbf{C, F,} Comparison between CMIP5 current, RCP4.5 and RCP8.5 climates, respectively. Dashed lines indicate the ensemble median (0.5 quantiles) while shaded areas indicate each estimate's confidence interval (CI, 0.1-0.9 quantile). The current climate period spans from 1981 to 2000, while the future climate period spans from 2081 to 2100. The confidence intervals account for variability in Tide, SLR, TCs, climate models, multi-station data, and Kernel-GPD parameters.}
\label{fig:figs4}
\end{figure*}  

\begin{figure*}[htb!]%
\centering
\includegraphics[width=1\textwidth]{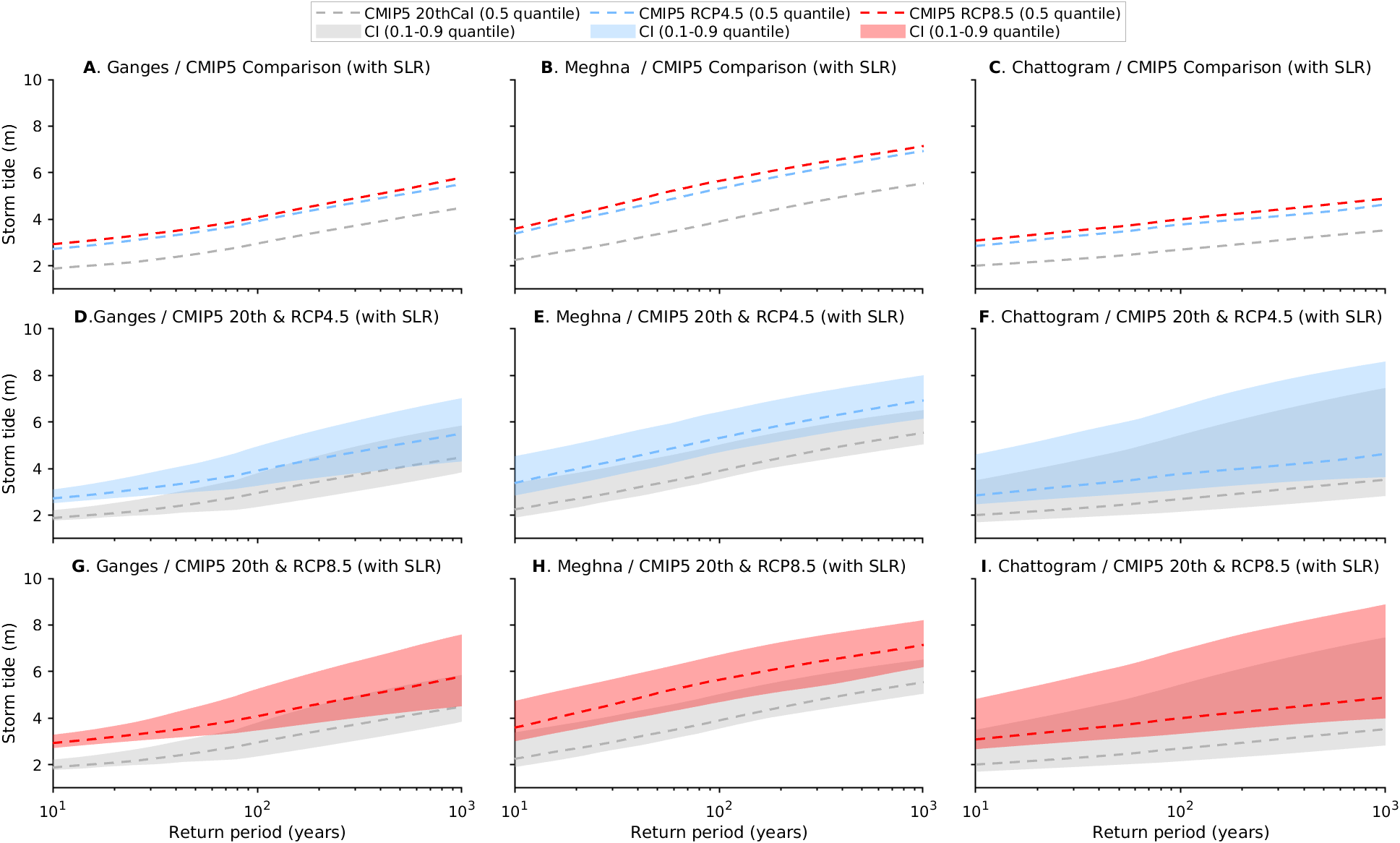}
\caption{\textbf{Bangladesh's storm tides versus return periods projected by CMIP5 models at the regional scale.} Storm tides are the water level elevations (combining components of astronomic tide, storm surge, and mean sea-level state) relative to the mean sea-level of the 1995-2014 baseline. \textbf{A, D, G,} Projections for the Ganges. \textbf{B, E, H,} Projections for the Meghna. \textbf{C, F, I,} Projections for the Chattogram. \textbf{A, B, C,} Comparison between CMIP5 current climate and future climate under the RCP4.5 and RCP8.5 scenarios. \textbf{D, E, F,} CMIP5 model ensemble composite under the RCP4.5 scenario. \textbf{G, H, I,} CMIP6 model ensemble composite under the RCP8.5 scenario. Dashed lines indicate the ensemble median (0.5 quantiles) while shaded areas indicate each estimate's confidence interval (CI, 0.1-0.9 quantile). The current climate period spans from 1981 to 2000, while the future climate period spans from 2081 to 2100. The confidence intervals account for variability in Tide, SLR, TCs, climate models, multi-station data, and Kernel-GPD parameters.}
\label{fig:figs5}
\end{figure*}  

\begin{figure}[htb!]%
\centering
\includegraphics[width=1\textwidth]{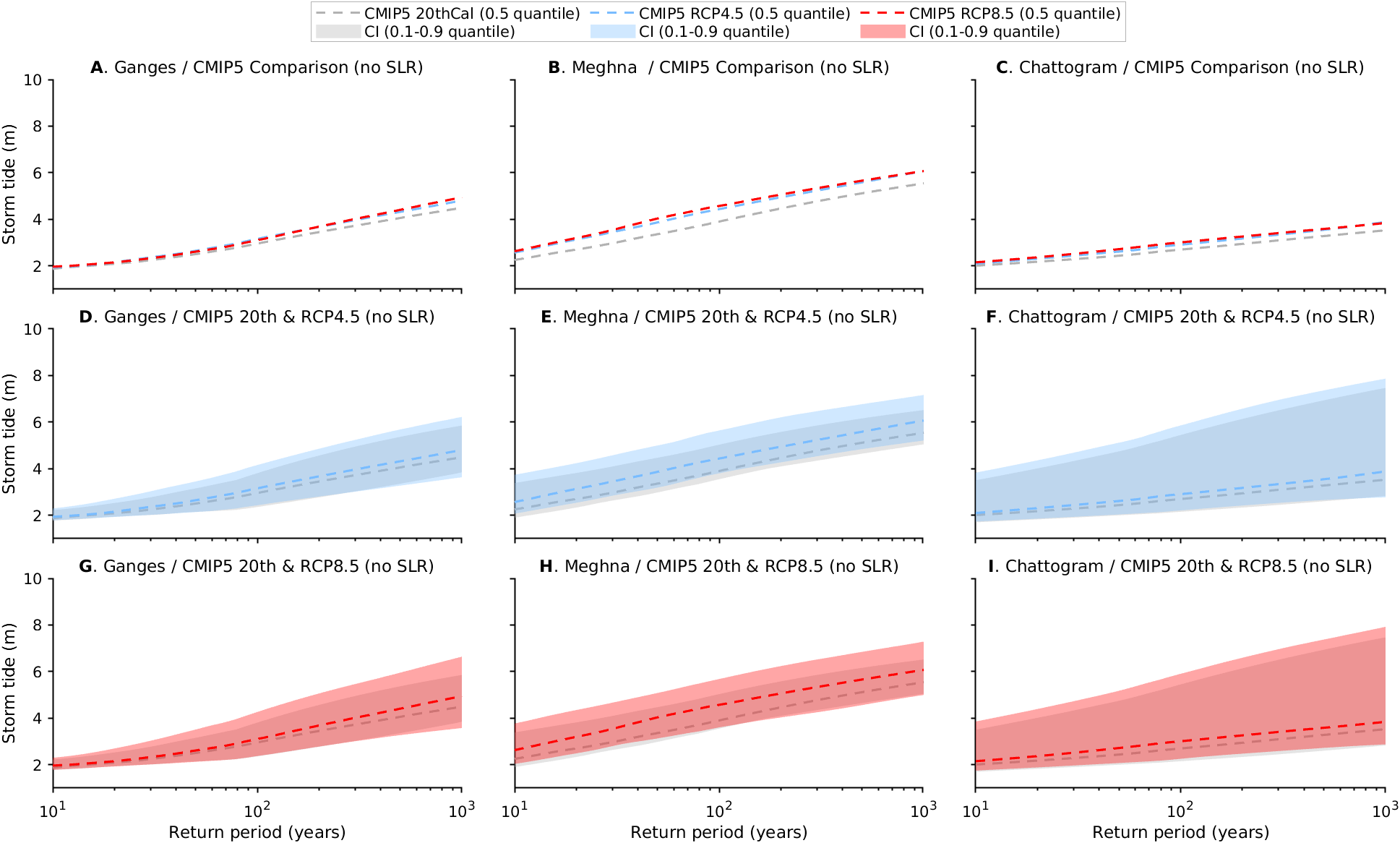}
\caption{\textbf{Bangladesh's storm tide versus return period curves projected by CMIP5 models at the regional scale.} Storm tides are the water level elevations (combining components of astronomic tide, storm surge, and mean sea-level state) relative to the mean sea-level of the 1995-2014 baseline. \textbf{A, D, G,} Projections for the Ganges. \textbf{B, E, H,} Projections for the Meghna. \textbf{C, F, I,} Projections for the Chattogram. \textbf{A, B, C,} Comparison between CMIP5 current climate and future climate under the RCP4.5 and RCP8.5 scenarios. \textbf{D, E, F,} CMIP5 model ensemble composite under the RCP4.5 scenario. \textbf{G, H, I,} CMIP6 model ensemble composite under the RCP8.5 scenario. Solid lines indicate the ensemble median (0.5 quantiles) while shaded areas indicate each estimate's confidence interval (CI, 0.1-0.9 quantile). The current climate period spans from 1981 to 2000, while the future climate period spans from 2081 to 2100. The confidence intervals account for variability in Tide, SLR, TCs, climate models, multi-station data, and Kernel-GPD parameters.}
\label{fig:figs6}
\end{figure}  

\begin{figure*}[htb!]%
\centering
\includegraphics[width=1\textwidth]{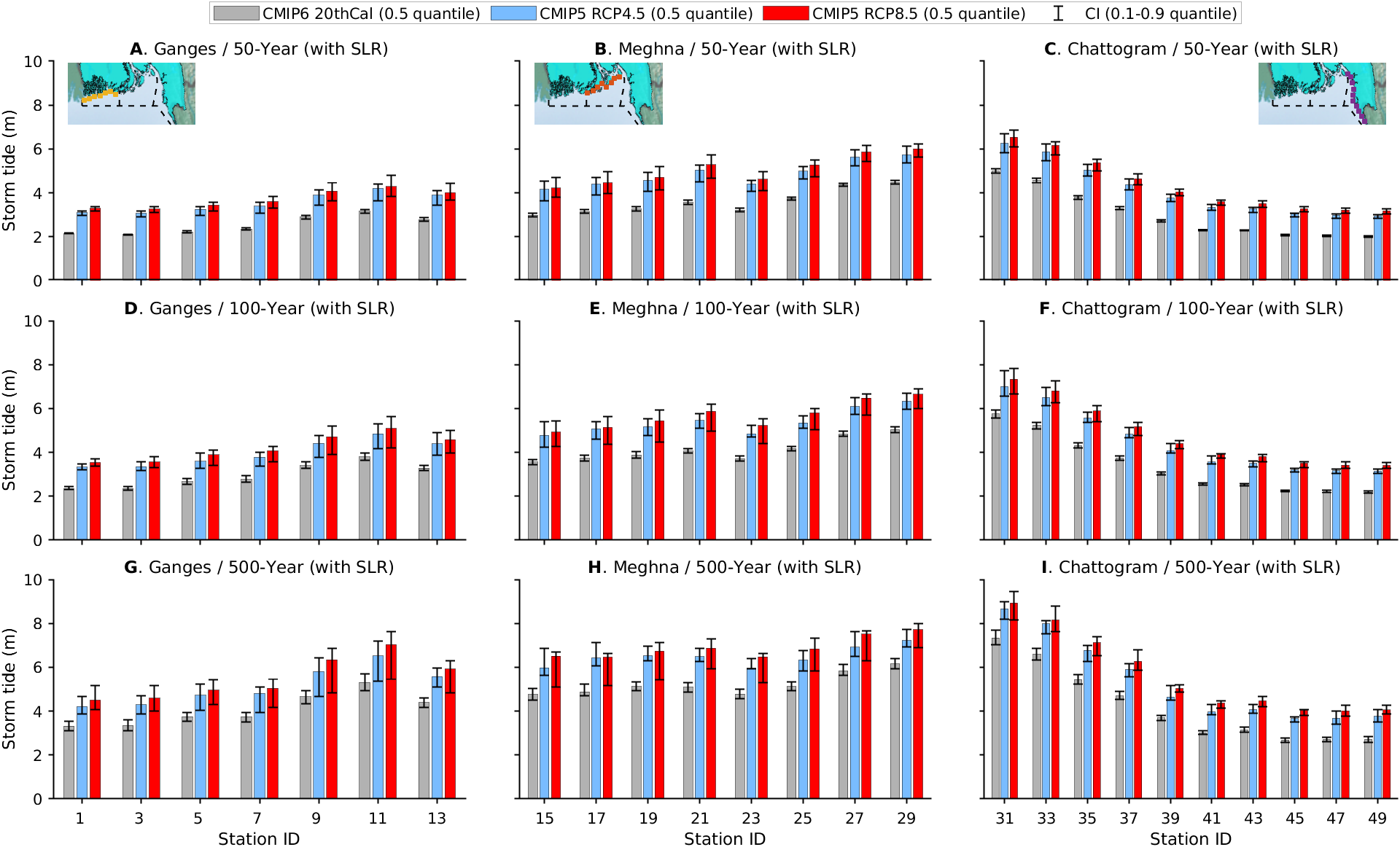}
\caption{\textbf{Bangladesh's 50-, 100-, 500-year storm tides projected by CMIP5 models at the station scale.} Storm tides are the water level elevations (combining components of astronomic tide, storm surge, and mean sea-level state) relative to the mean sea-level of the 1995-2014 baseline. \textbf{A, B, C,} 50-year return period. \textbf{D, E, F,} 100-year return period. \textbf{G, H, I,} 500-year return period. \textbf{A, D, G,} stations located in the Ganges. \textbf{B, E, H,} stations located in the Meghna. \textbf{C, F, I,} stations located in the Chattogram. Projections are conducted for all 58 stations, but the graph displays only every other VS. Histograms indicate the ensemble median (0.5 quantiles) for the current climate, CMIP5 RCP4.5 and RCP8.5 climate, respectively. Vertical error bars indicate each estimate's confidence interval (CI, 0.1-0.9 quantile). The current climate period spans from 1981 to 2000, while the future climate period spans from 2081 to 2100. The confidence intervals account for variability in Tide, SLR, TCs, climate models, and Kernel-GPD parameters. Base map sourced from the BDP 2100 (Baseline Volume 1, pg. 403), Humanitarian Data Exchange, World Bank, ESRI ArcGIS, Maxar, Earthstar Geographics, USDA FSA, USGS, Aerogrid, IGN, IGP, and the GIS User Community.}
\label{fig:figs7}
\end{figure*} 

\begin{figure}[htb!]%
\centering
\includegraphics[width=1\textwidth]{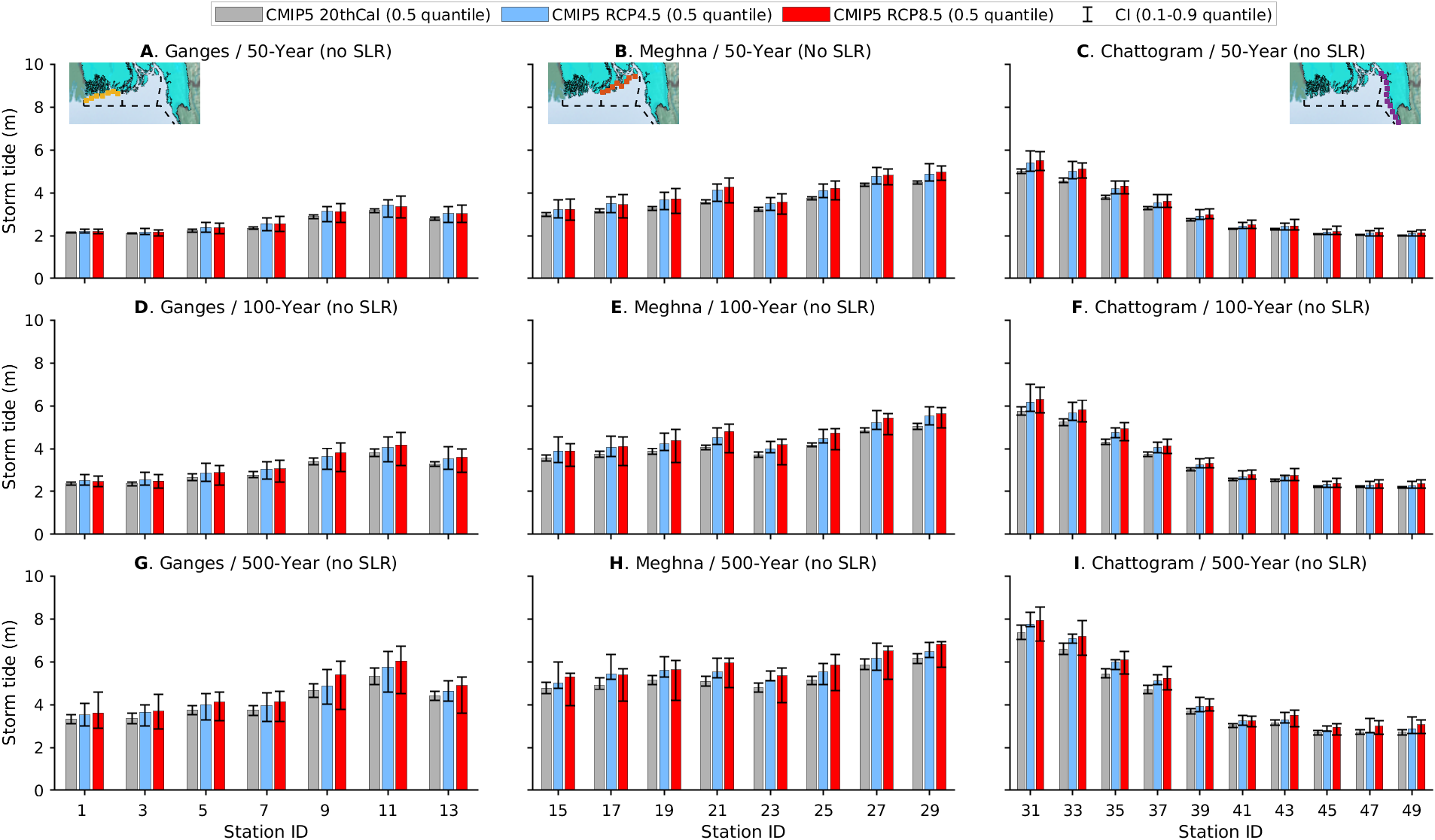}
\caption{\textbf{Bangladesh's 50-, 100-, 500-year storm tides projected by CMIP5 models at the station scale.} Storm tides are the water level elevations (combining components of astronomic tide, storm surge, and mean sea-level state) relative to the mean sea-level of the 1995-2014 baseline. \textbf{A, B, C,} 50-year return period. \textbf{D, E, F,} 100-year return period. \textbf{G, H, I,} 500-year return period. \textbf{A, D, G,} stations located in the Ganges. \textbf{B, E, H,} stations located in the Meghna. \textbf{C, F, I,} stations located in the Chattogram. Projections are conducted for all 58 stations, but the graph displays only every other VS. Blue, orange, and red histograms indicate the ensemble median (0.5 quantile) for the current climate, CMIP5 RCP4.5 and RCP8.5 climate, respectively. Vertical error bars indicate each estimate's confidence interval (CI, 0.1-0.9 quantile). The current climate period spans from 1981 to 2000, while the future climate period spans from 2081 to 2100. The confidence intervals account for variability in Tide, SLR, TCs, climate models, and Kernel-GPD parameters. Base map sourced from the BDP 2100 (Baseline Volume 1, pg. 403), Humanitarian Data Exchange, World Bank, ESRI ArcGIS, Maxar, Earthstar Geographics, USDA FSA, USGS, Aerogrid, IGN, IGP, and the GIS User Community.}
\label{fig:figs8}
\end{figure}  

\begin{figure}[htb!]%
\centering
\includegraphics[width=1\textwidth]{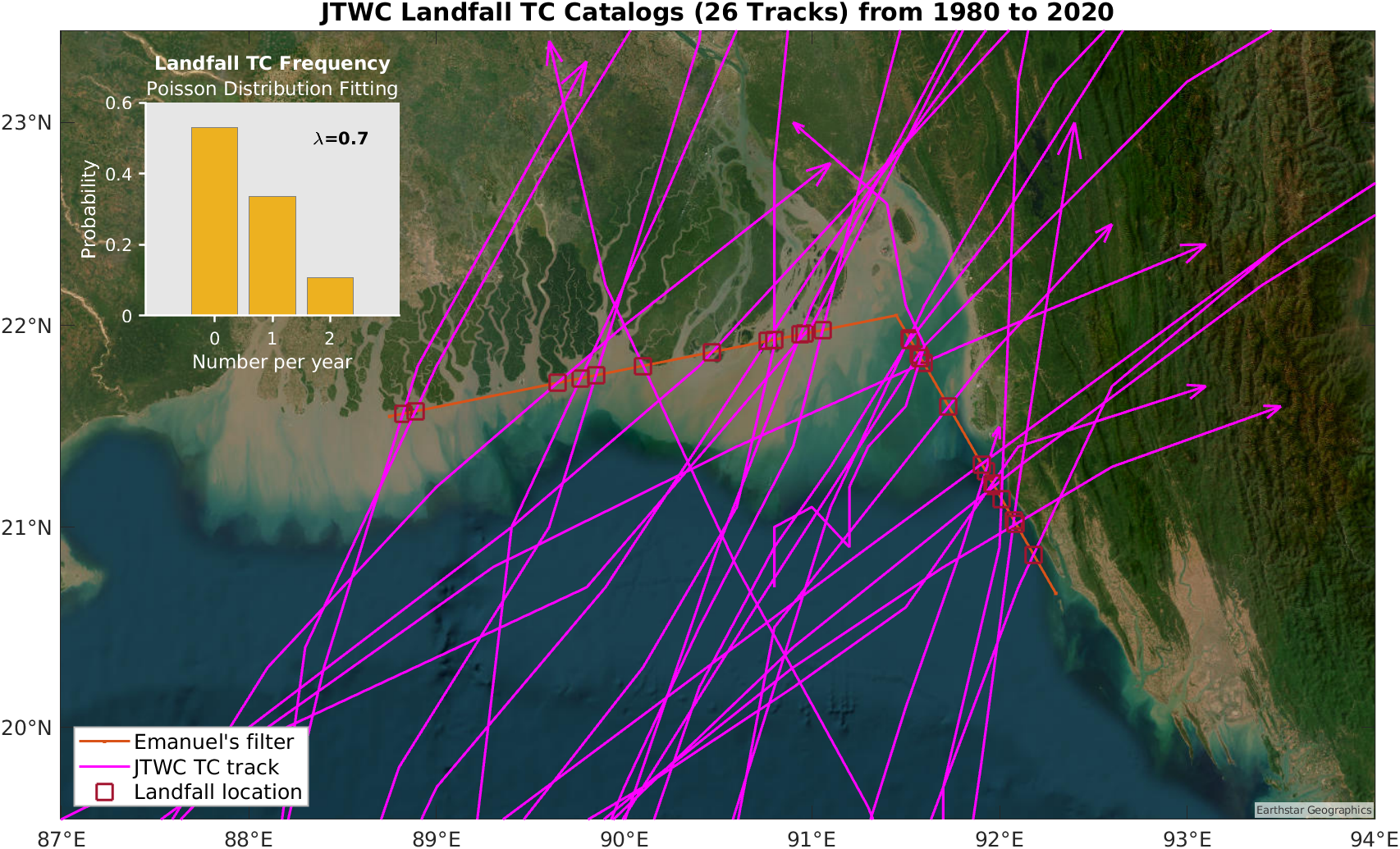}
\caption{\textbf{Annual TC frequency analyzed using 26 JTWC TCs that made landfall in Bangladesh from 1980 to 2020.} These magenta lines with arrows indicate the TC's track. The bold orange two-segment lines indicate Emanuel's filter used in generating synthetic TC tracks affecting Bangladesh. Red squares on the filter indicate TC landfall locations. The upper left panel displays the annual TC frequency estimation ($\lambda=0.7$) by fitting a Poisson distribution with the annual occurrence rate. Base map sourced from the ESRI ArcGIS, Maxar, Earthstar Geographics, USDA FSA, USGS, Aerogrid, IGN, IGP, and the GIS User Community.}
\label{fig:figs9}
\end{figure}  

\begin{figure}[htb!]%
\centering
\includegraphics[width=0.7\textwidth]{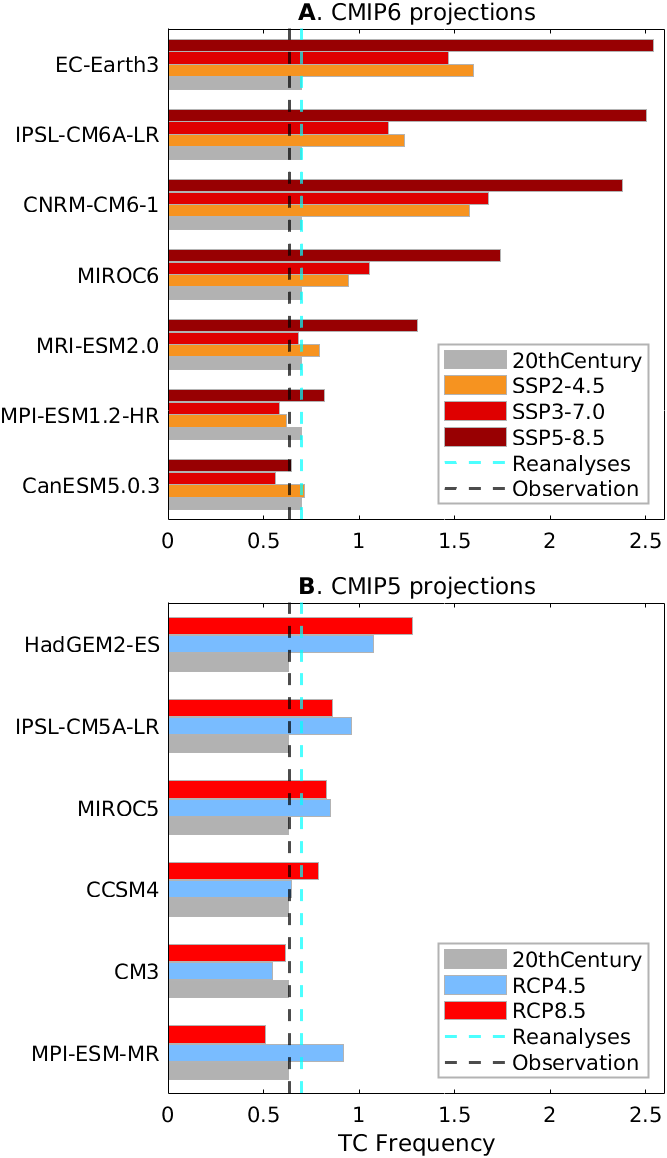}
\caption{\textbf{Projected annual TC frequency (without bias correction) from CMIP6 (A) and CMIP5 (B) models.} The horizontal bar lengths with different colors indicate projected annual TC frequency in current and future climates (RCP4.5 and RCP8.5 for CMIP5, SSP2-4.5, SSP3-7.0 and SSP5-8.5 for CMIP6), respectively. The vertical cyan and black dash lines indicate the annual TC frequency estimated based on climate reanalyses (including ECMWF/ERA5 and GMAO/MERRA2) and JTWC historical observation. The bias-corrected annual TC frequency is summarized in Table~\ref{tab:tabs3}.}
\label{fig:figs10}
\end{figure}  

\begin{figure}[htb!]%
\centering
\includegraphics[width=0.8\textwidth]{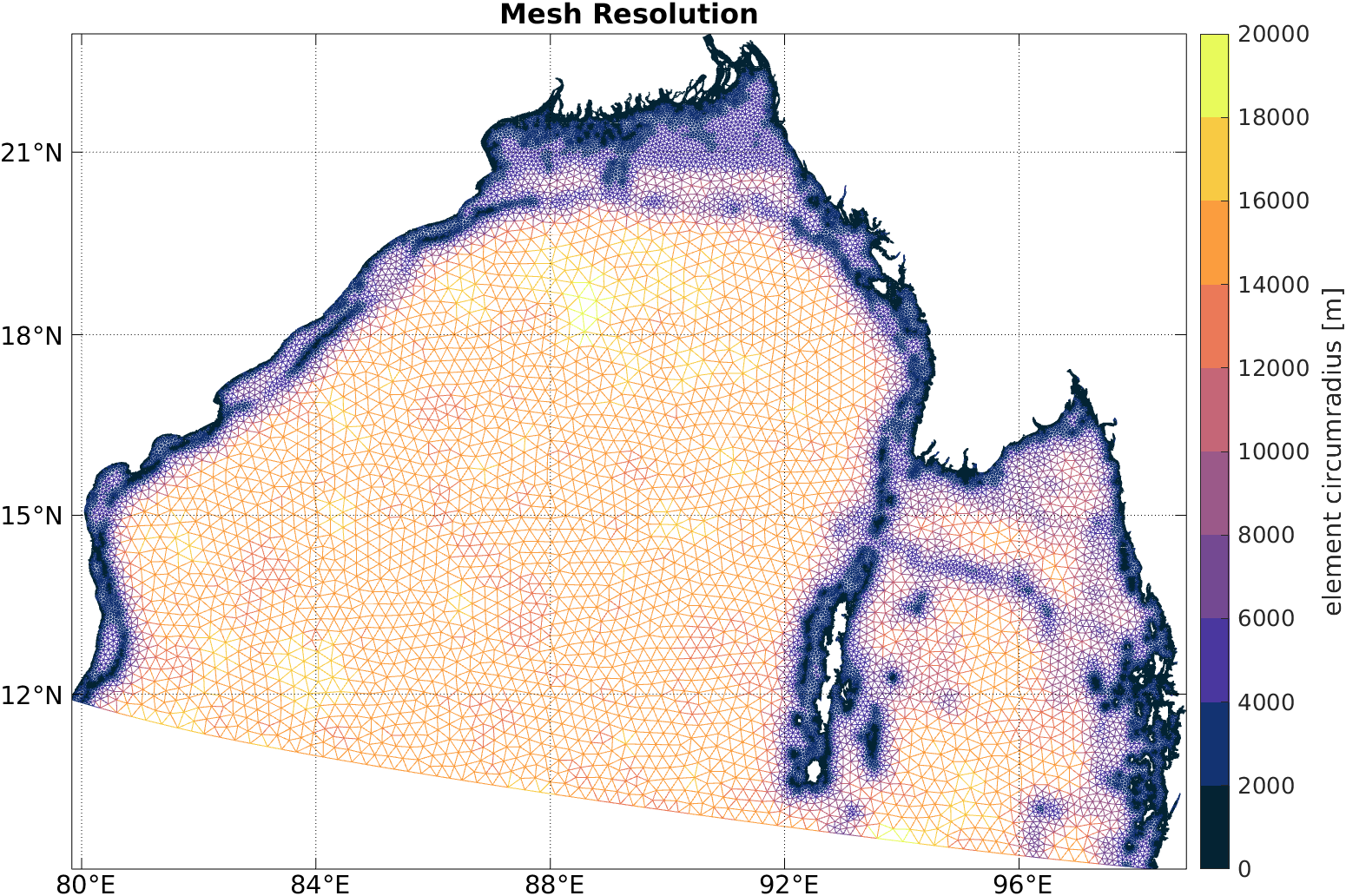}
\caption{\textbf{Mesh triangulation and resolution displayed on a Miller projection for the model domain.}}
\label{fig:figs11}
\end{figure}  

\begin{figure}[htb!]%
\centering
\includegraphics[width=0.8\textwidth]{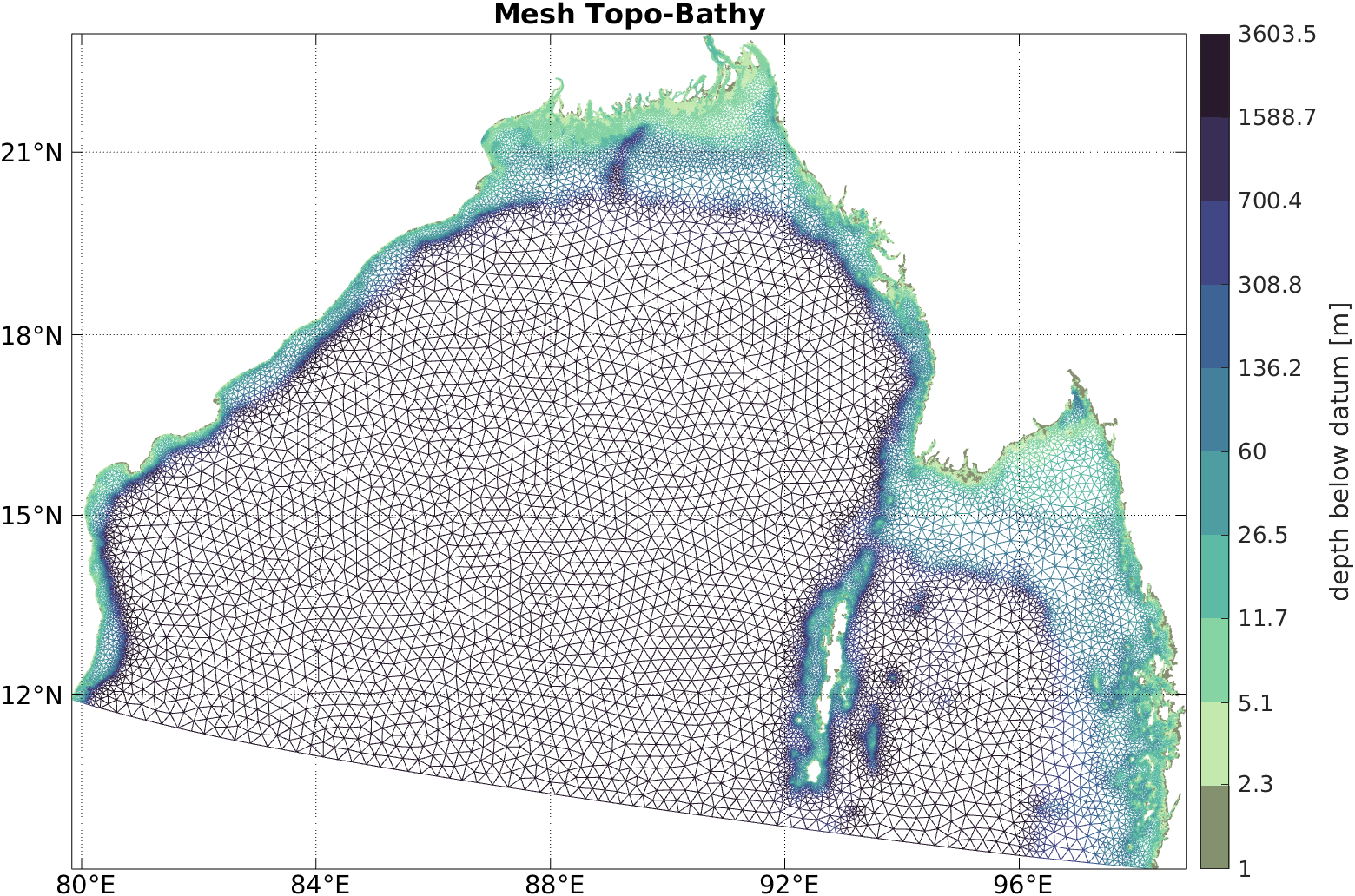}
\caption{\textbf{Mesh triangulation and depth (in logarithmic space) displayed on a Miller projection for the model domain.}}
\label{fig:figs12}
\end{figure}  

\begin{figure}[htb!]%
\centering
\includegraphics[width=0.8\textwidth]{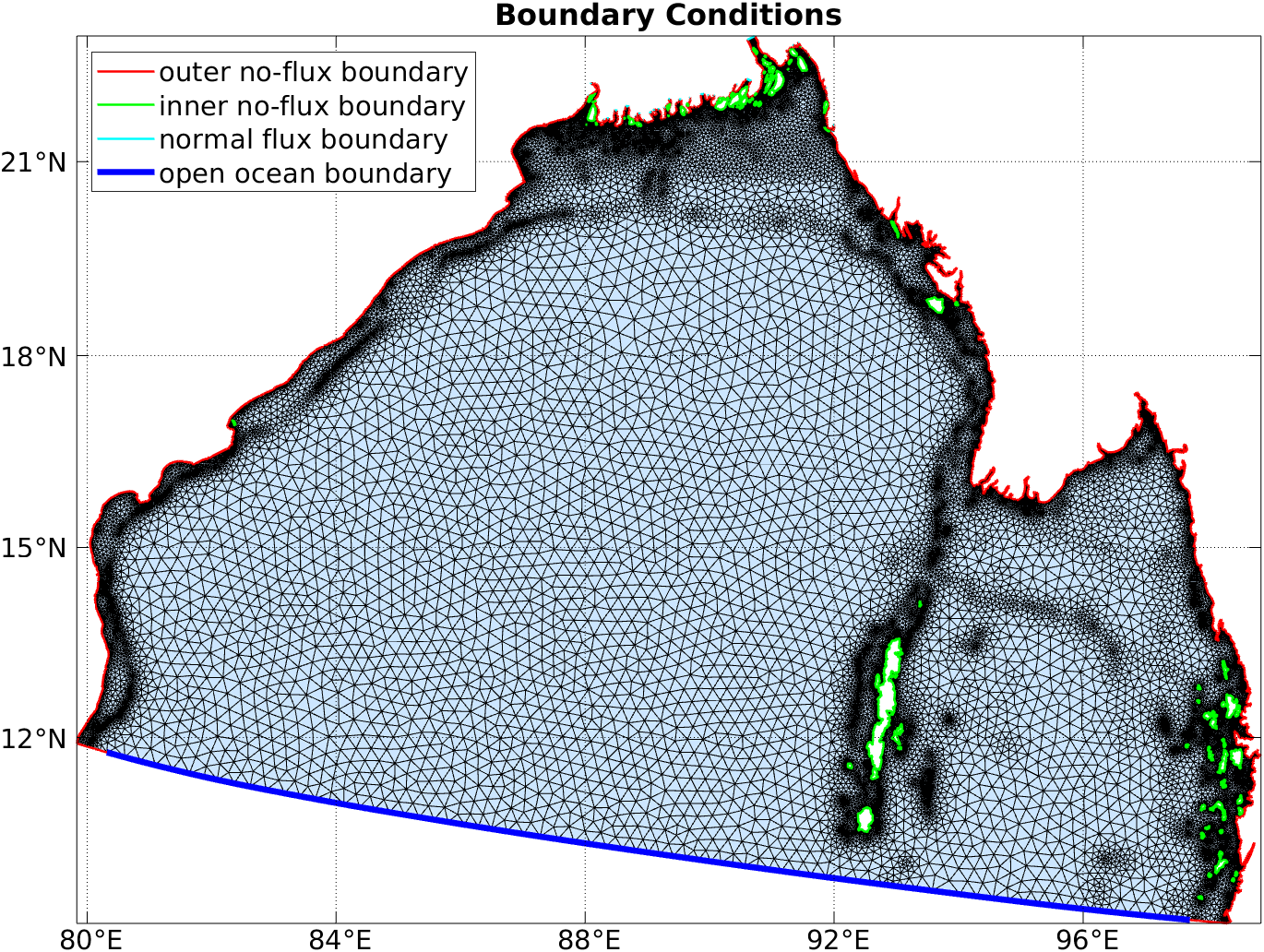}
\caption{\textbf{Mesh triangulation and open boundaries displayed on a Miller projection for the model domain.}}
\label{fig:figs13}
\end{figure}  

\begin{figure}[htb!]%
\centering
\includegraphics[width=0.8\textwidth]{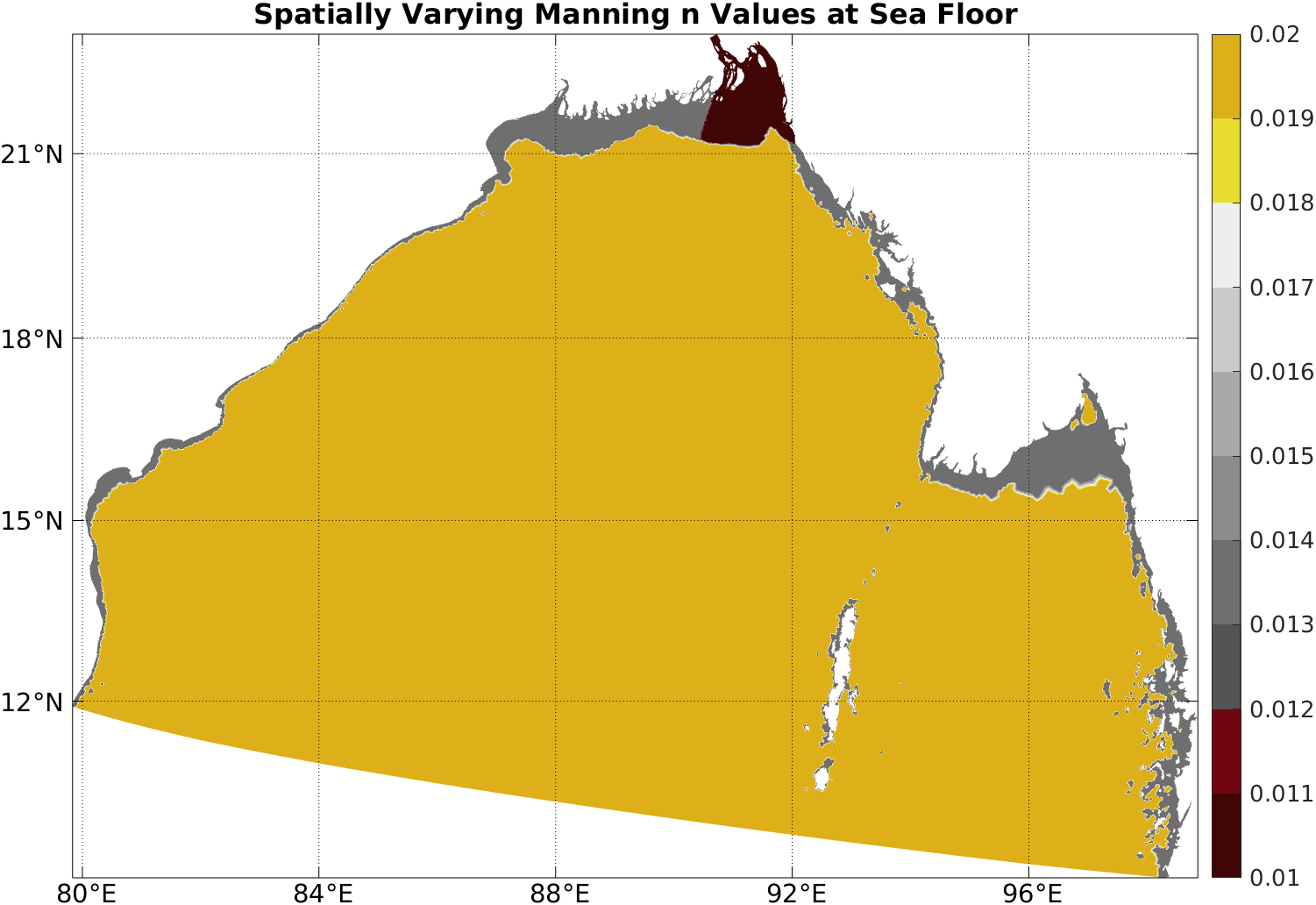}
\caption{\textbf{Spatially varying \textit{Manning's-N} values at sea floor displayed on a Miller projection for the model domain.}}
\label{fig:figs14}
\end{figure}  
\newpage

\begin{figure}[htb!]%
\centering
\includegraphics[width=0.8\textwidth]{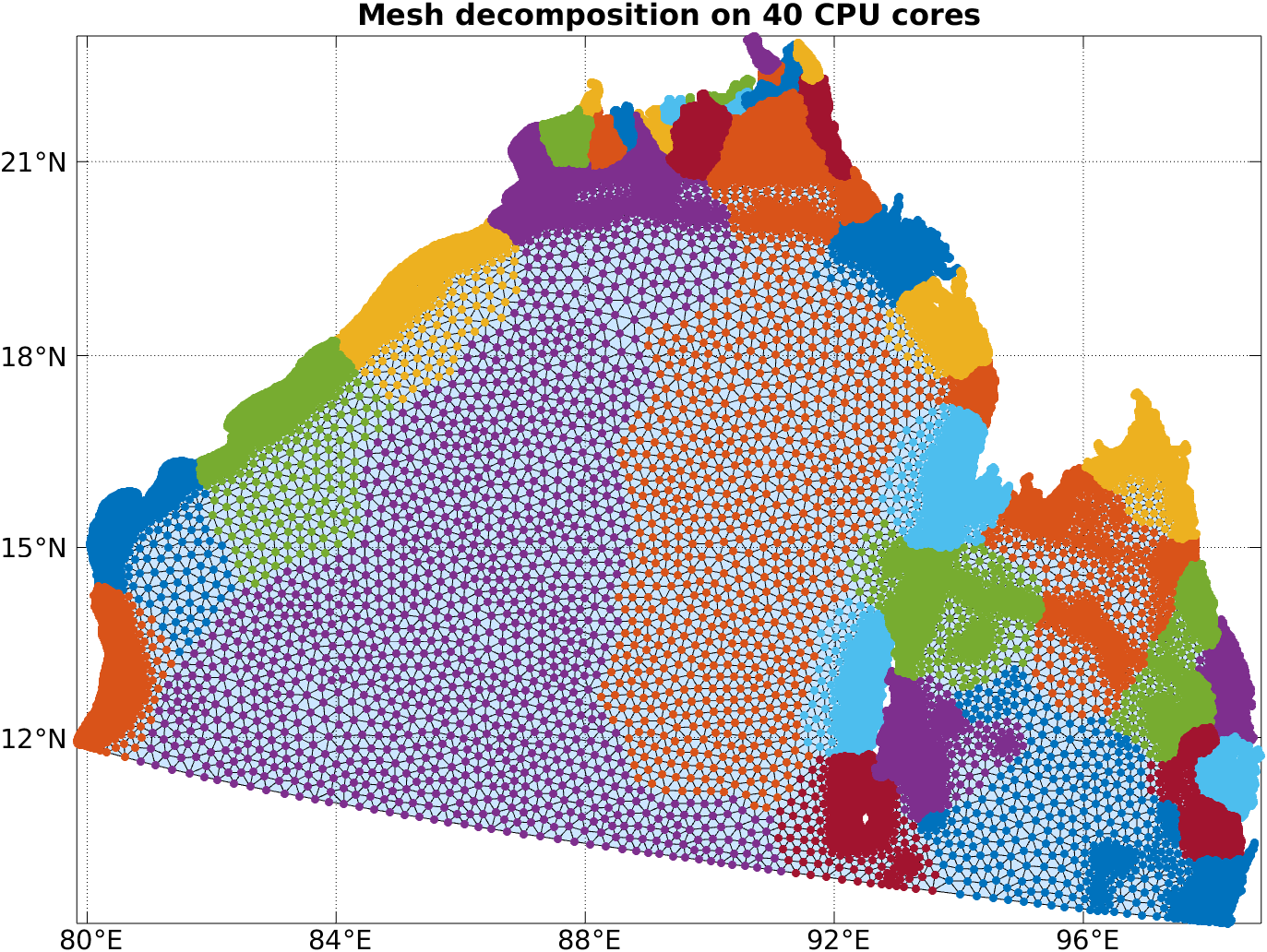}
\caption{\textbf{Mesh decomposition on 40 CPU cores displayed on a Miller projection for the model domain.}}
\label{fig:figs15}
\end{figure}  
\newpage

\begin{figure}[htb!]%
\centering
\includegraphics[width=0.6\textwidth]{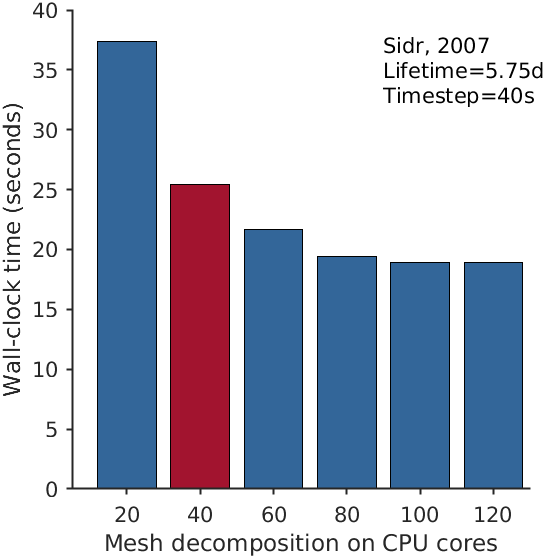}
\caption{\textbf{Computational performance evaluation of the model tested by historical TC Sidr (2007) for increasing CPU cores.} A parallel simulation using 40 CPU cores took only 25 seconds to complete (illustrated by the red histogram). All the historical TCs used for model verification and downscaling synthetic TCs are finally simulated using 40 CPU (marked red) cores for parallel.}
\label{fig:figs16}
\end{figure}

\begin{figure}[htb!]%
\centering
\includegraphics[width=1\textwidth]{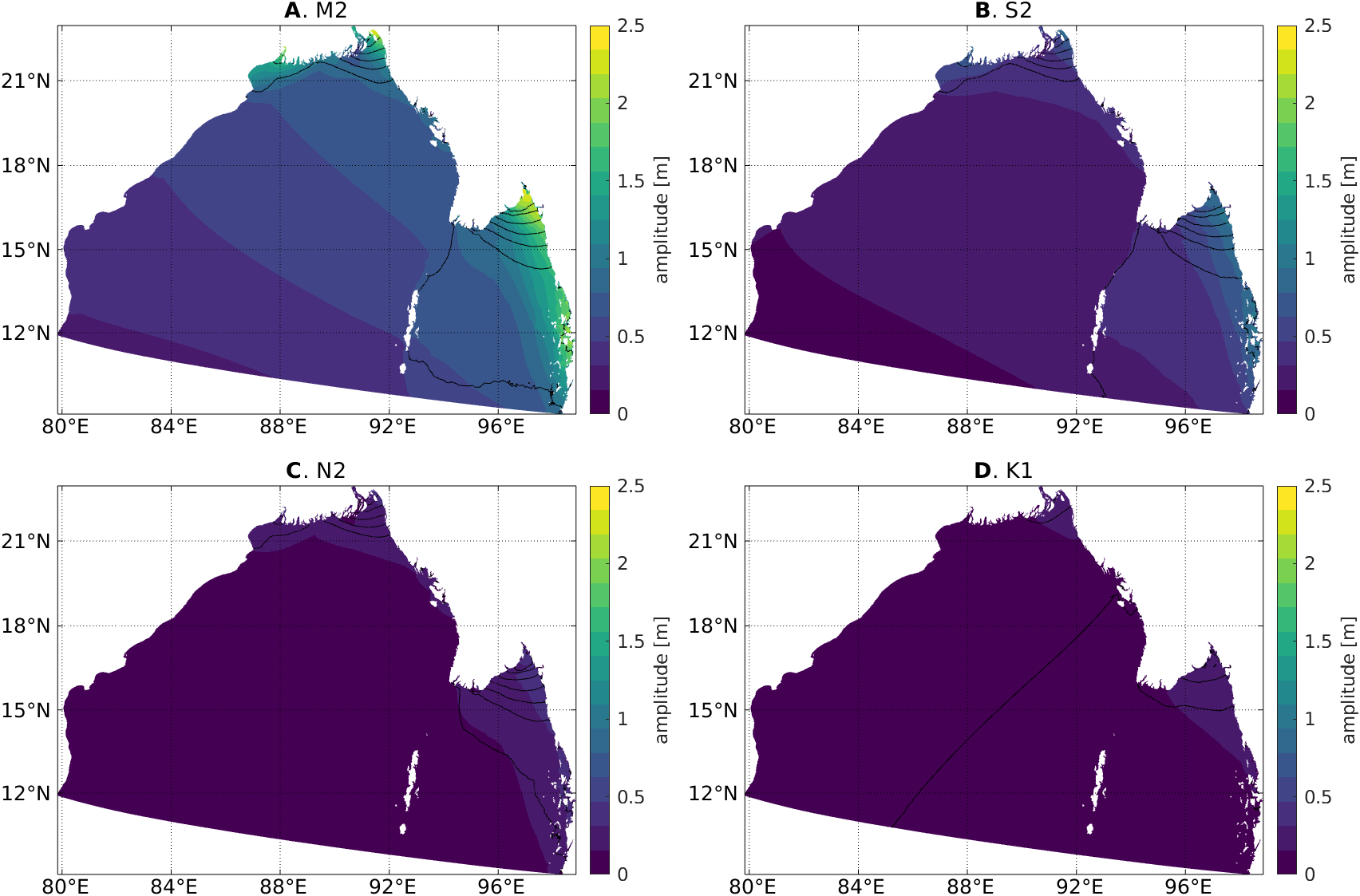}
\caption{\textbf{Amplitude (background color, unit: m) and phase (cotidal lines with 30-degree increments) responses of the $M_2$ (A), $S_2$ (B), $N_2$ (C), $K_1$ (D) tidal waves.} Five tidal constitutes are activated in the harmonic analysis for the astronomic tide verification; only the first four tidal constitutes' results are shown here.}
\label{fig:figs17}
\end{figure}  

\begin{figure}[htb!]%
\centering
\includegraphics[width=1\textwidth]{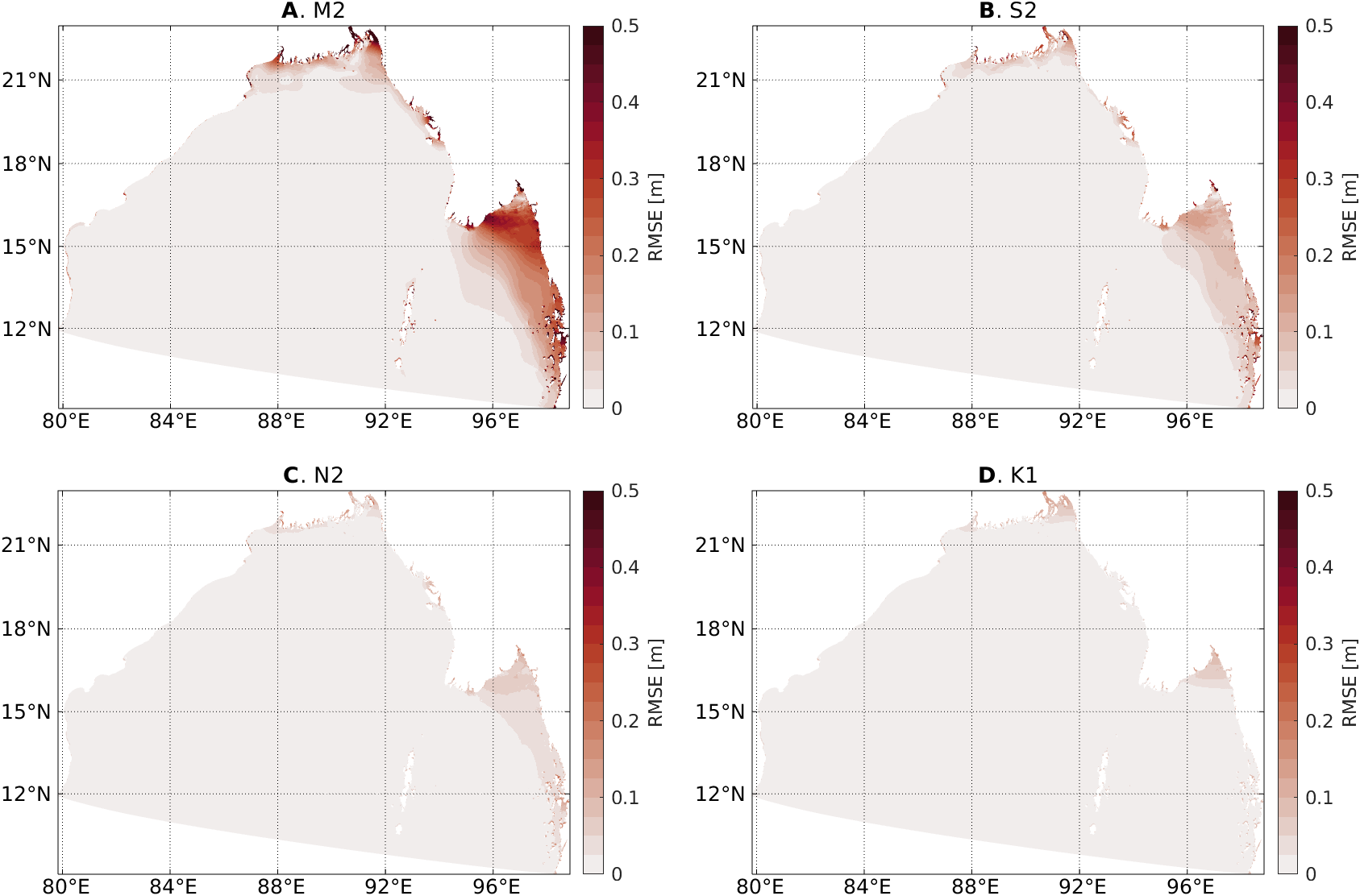}
\caption{\textbf{Root-mean-square error (m) of our model setup against the TPXO10-Altas-v2 for the $M_2$ (A), $S_2$ (B), $N_2$ (C) and $K_1$ (D) tidal waves.} Five tidal constituents are activated in the harmonic analysis for the astronomic tide verification; only the first four tidal constituents' results are shown here.}
\label{fig:figs18}
\end{figure}  

\begin{figure}[htb!] 
\centering
\includegraphics[width=0.6\textwidth]{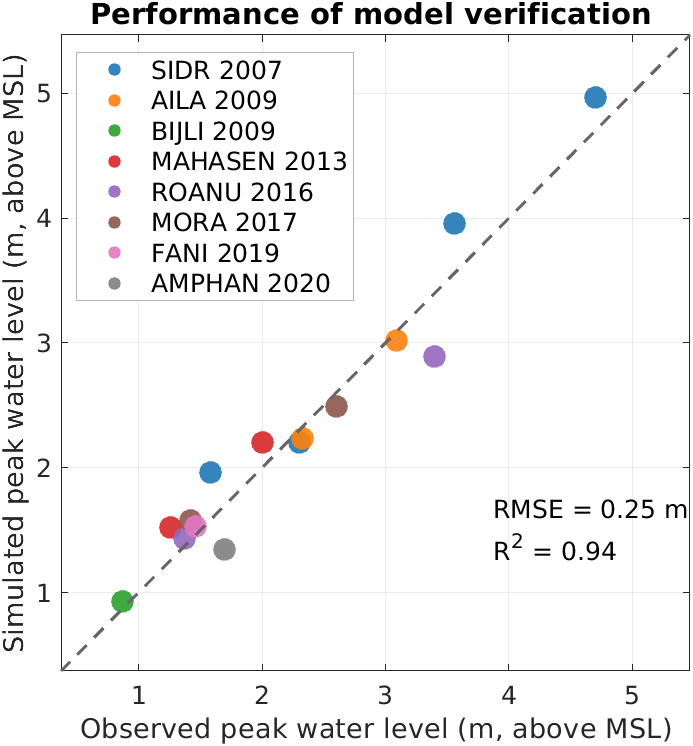}
\caption{\textbf{Performance of model verification.} The modeled and observed peak water levels for eight historical TC events among multiple tidal stations are used to calculate the metrics of RMSE and $R^2$. Multiple stations verified a single event, as indicated by the same color. The black dashed line is the 1:1 line.}
\label{fig:figs19}
\end{figure}  

\begin{figure}[htb!]
\centering
\includegraphics[width=1\textwidth]{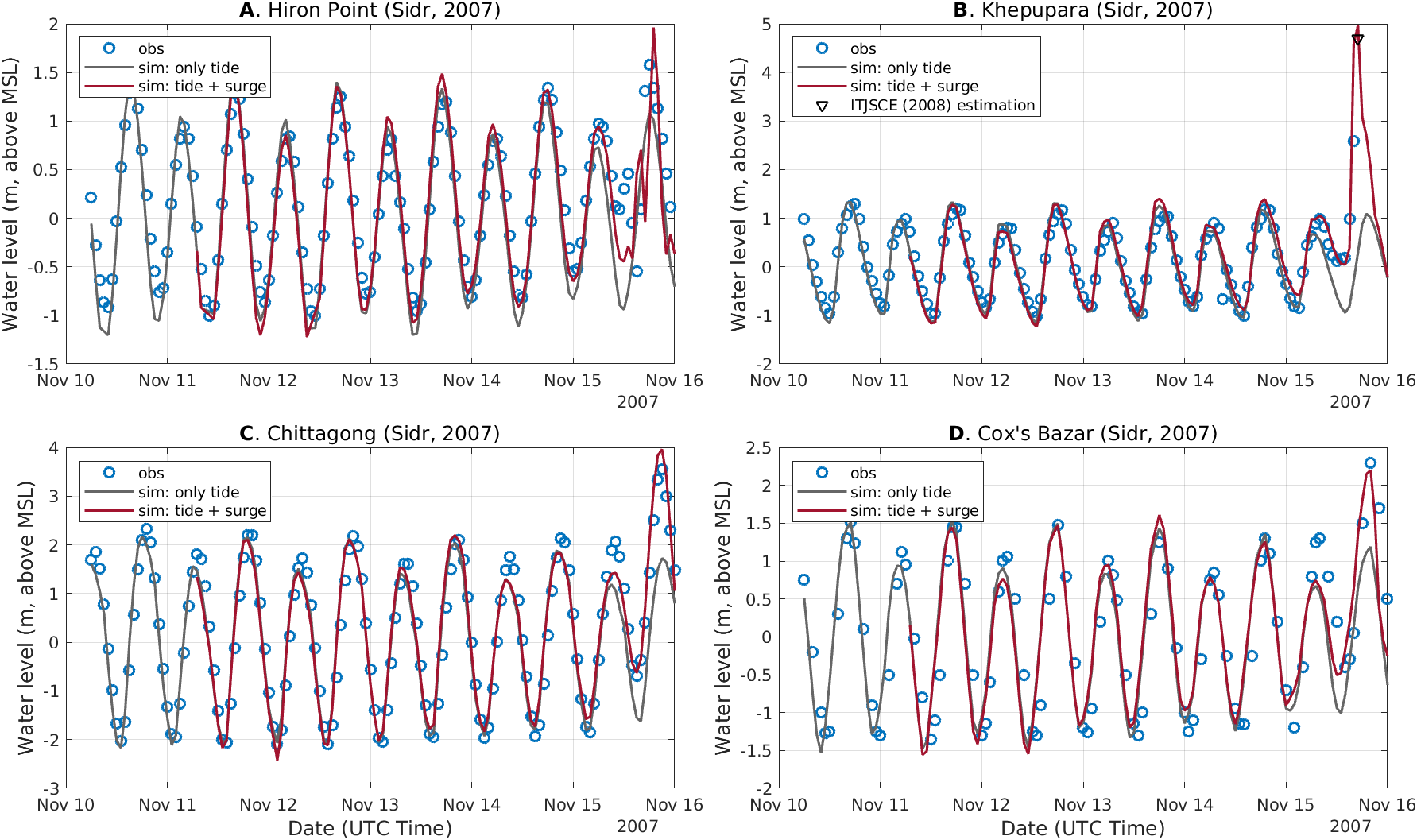}
\caption{\textbf{Comparison between modeled and observed total water levels at tidal station Hiron Point (A), Khepupara (B) Chittagong (C), and Cox's Bazaar (D) for the verification of Sidr (2007)-induced storm tide.} The model performance can be evaluated by comparing observations (blue circles) with the simulated storm tide (red line). Additionally, a separate simulation excluding meteorological forcing is conducted to display the astronomical tide (gray line) during Cyclone Sidr, providing better insights into the surge process. Detailed information about the source of the observations and the preprocessing procedures for the time series used in model comparison are documented in Section~\ref{acknowledgements}, Section~\ref{dataavailability}, and Section~\ref{model verification}, respectively.}
\label{fig:figs20}
\end{figure}  

\begin{figure}[htb!]%
\centering
\includegraphics[width=1\textwidth]{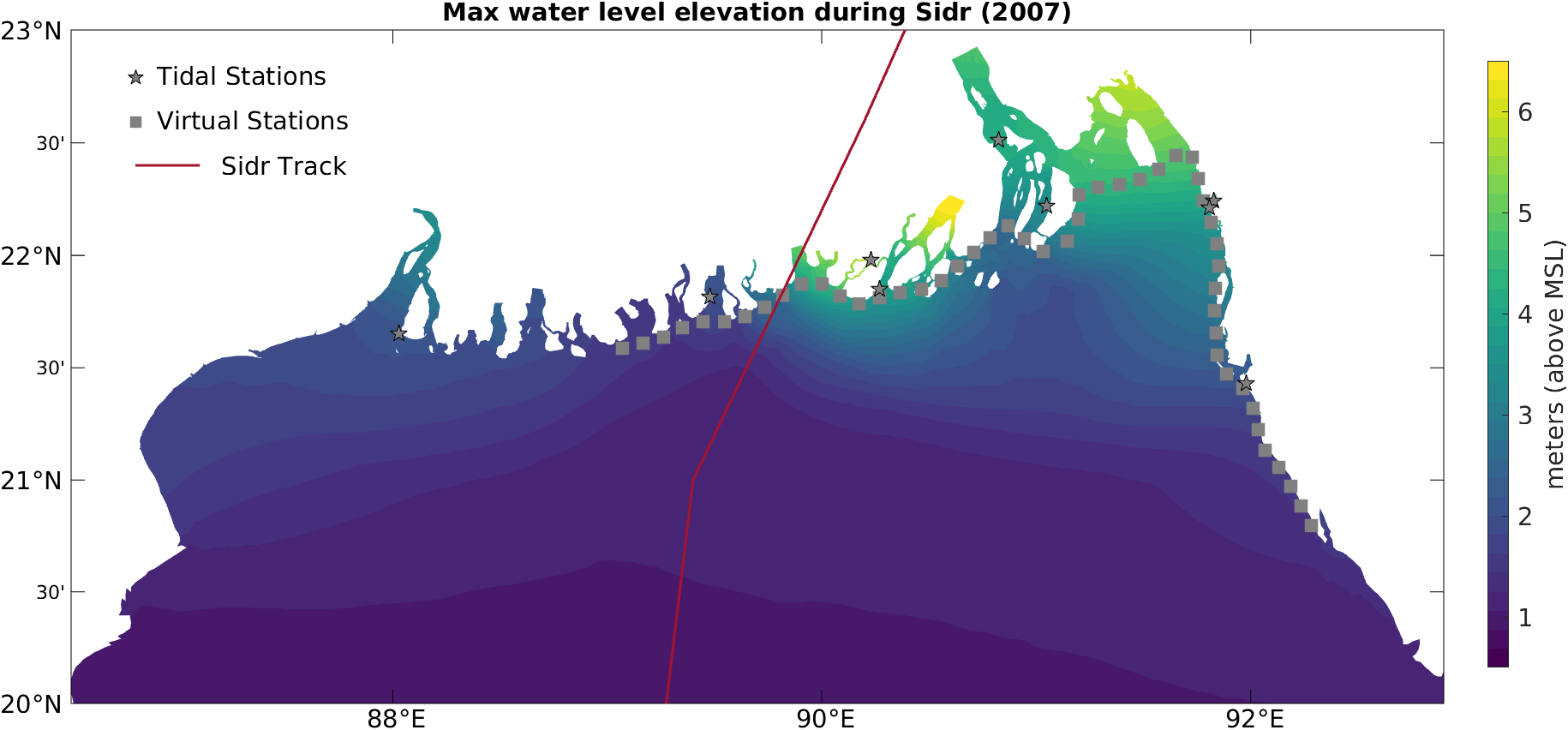}
\caption{\textbf{Maximum water level elevation during Cyclone Sidr (2007) make landfall to Bangladesh.}}
\label{fig:figs21}
\end{figure}  

\begin{figure}[htb!]%
\centering
\includegraphics[width=1\textwidth]{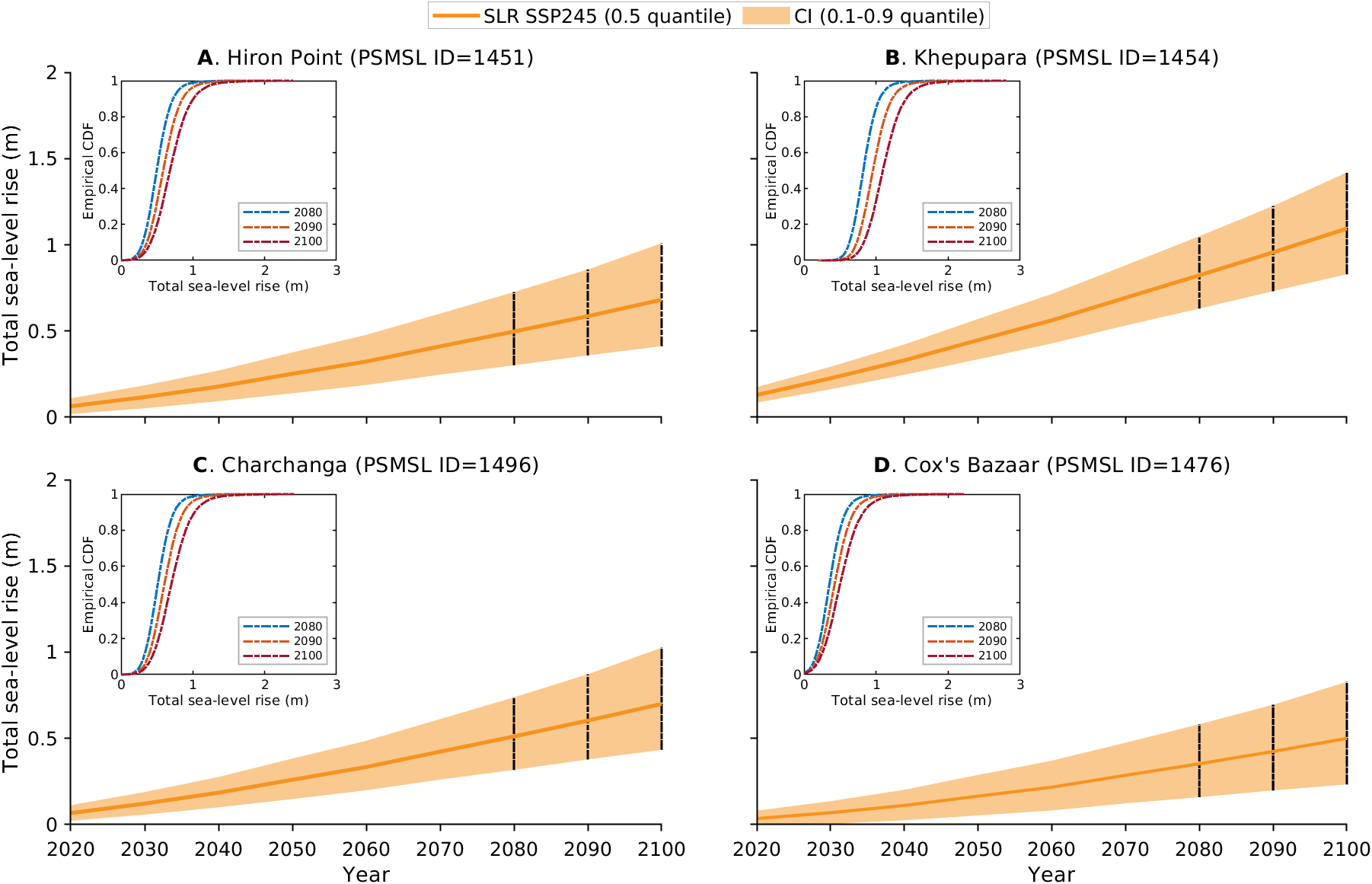}
\caption{\textbf{IPCC AR6 relative sea-level projections under CMIP6 SSP2-4.5 climate from 2020 to 2100 for station Hiron Point (A), Khepupara (B), Charchanga (C) and Cox's Bazaar (D).} Solid lines indicate the median value (0.5 quantile) of the projected total SLR, while shaded areas indicate the confidence interval (quantile 0.1-0.9). The upper left panel of each subplot displays the cumulative distribution function of the 20,000 Monte Carlo samples for the years 2080, 2090, and 2100, respectively.}
\label{fig:figs22}
\end{figure}  

\begin{figure}[htb!]%
\centering
\includegraphics[width=1\textwidth]{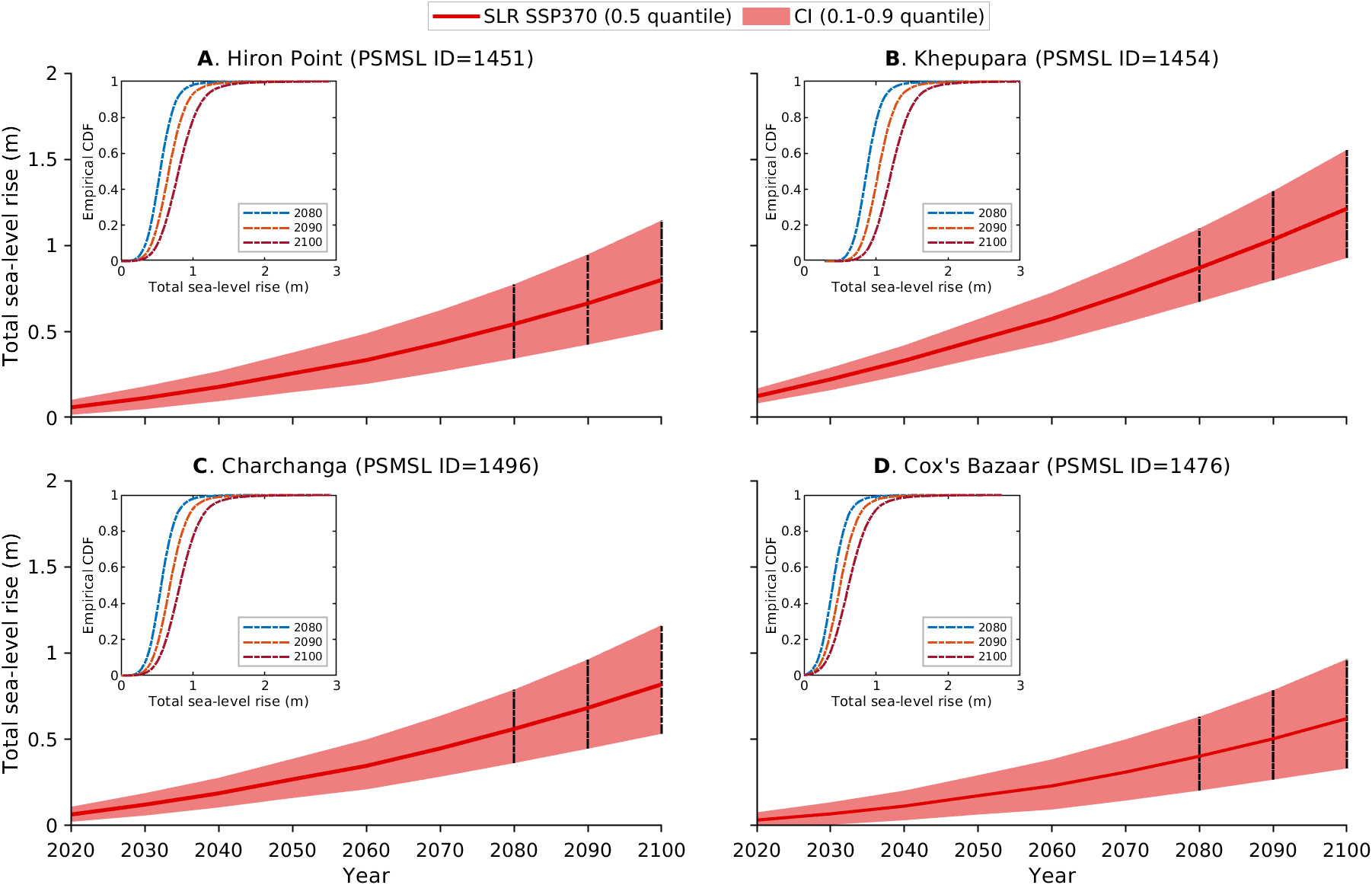}
\caption{\textbf{IPCC AR6 relative sea-level projections under CMIP6 SSP3-7.0 climate from 2020 to 2100 for station Hiron Point (A), Khepupara (B), Charchanga (C) and Cox's Bazaar (D).} Solid lines indicate the median value (0.5 quantile) of the projected total SLR, while shaded areas indicate the confidence interval (quantile 0.1-0.9). The upper left panel of each subplot displays the cumulative distribution function of the 20,000 Monte Carlo samples for years 2080, 2090, and 2100, respectively.}
\label{fig:figs23}
\end{figure}  

\begin{figure}[htb!]%
\centering
\includegraphics[width=1\textwidth]{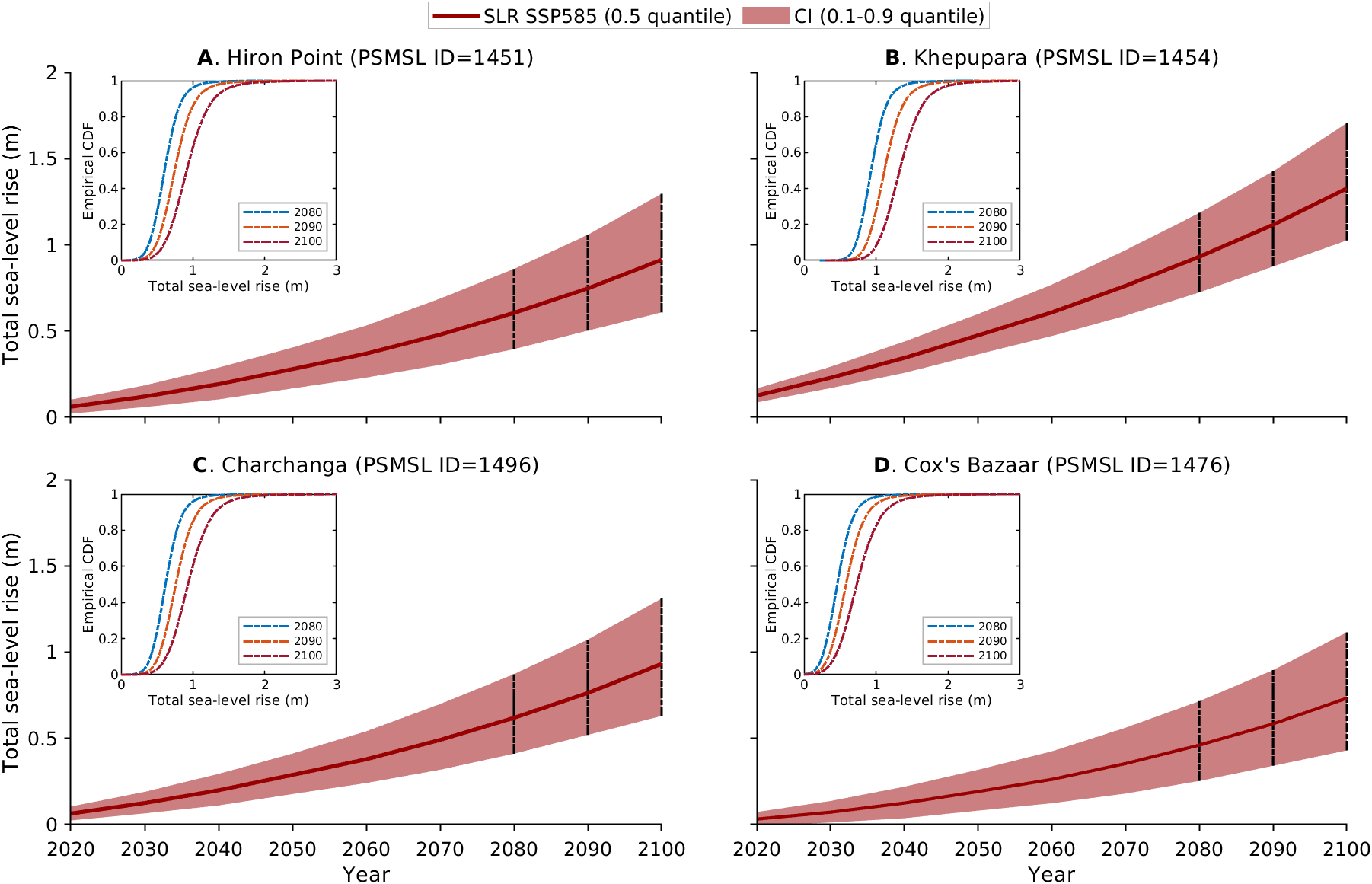}
\caption{\textbf{IPCC AR6 relative sea-level projections under CMIP6 SSP5-8.5 climate from 2020 to 2100 for station Hiron Point (A), Khepupara (B), Charchanga (C) and Cox's Bazaar (D).} Solid lines indicate the median value (0.5 quantile) of the projected total SLR, while shaded areas indicate the confidence interval (quantile 0.1-0.9). The upper left panel of each subplot displays the cumulative distribution function of the 20,000 Monte Carlo samples for the years 2080, 2090, and 2100, respectively.}
\label{fig:figs24}
\end{figure}  

\begin{figure}[htb!]%
\centering
\includegraphics[width=0.7\textwidth]{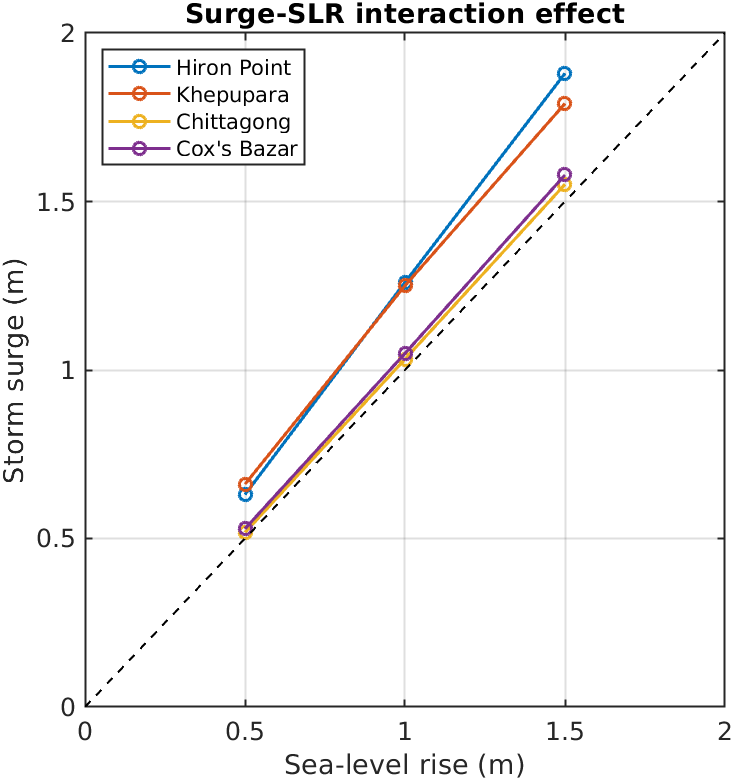}
\caption{\textbf{An example of the Surge-SLR interaction effect, illustrated using fixed SLR values of 0.5 m, 1.0 m, and 1.5 m, respectively.} The black dashed line is the 1:1 line.}
\label{fig:figs25}
\end{figure}  

\begin{figure}[htb!]%
\centering
\includegraphics[width=0.7\textwidth]{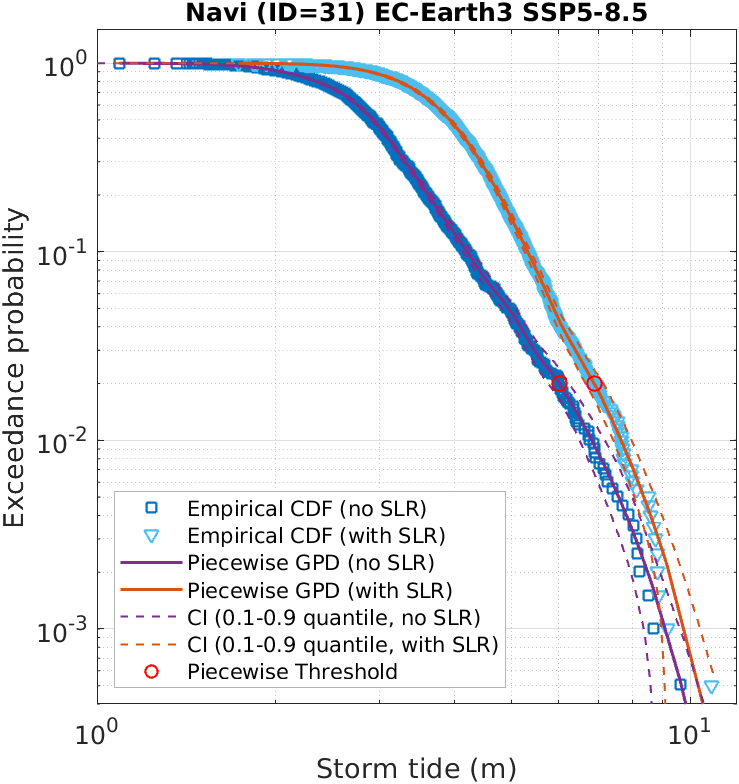}
\caption{\textbf{An example of exceedance probability fitting curve without (purple line) and with (orange line) considering the effect of SLR at VS Navi with ID=31 under EC-Earth3 SSP5-8.5 scenario.} The square and downward-pointing triangle dots represent the empirical exceedance probability for each storm tide, while the solid line indicates the fitted exceedance probability curve using the piecewise Kernel-GPD estimation method. The dashed line indicated the confidence interval (quantile 0.1-0.9), obtained through 1,000 bootstrap iterations. The red circle highlighted in bold indicates the threshold (quantile) used to segment the dataset for fitting the GPD model in the tail, while a kernel estimation is used in the remaining part. We performed the same fitting at all stations under all scenarios.}
\label{fig:figs26}
\end{figure}  

\begin{figure}[htb!]%
\centering
\includegraphics[width=0.7\textwidth]{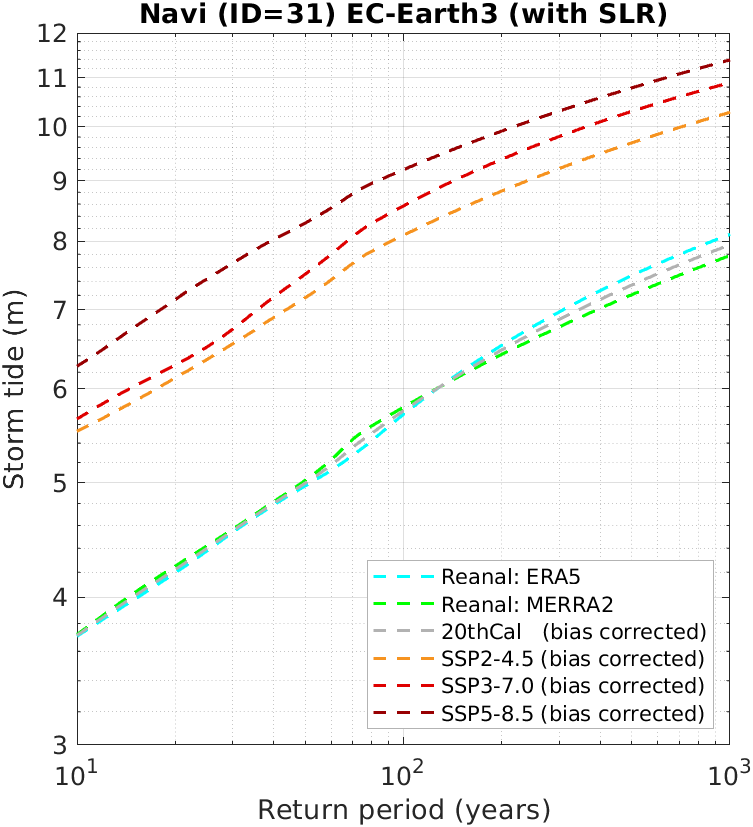}
\caption{\textbf{An example of the bias correction for the storm tide against return period curves with considering the effect of SLR at VS Navi}. This is based on the CMIP6 EC-Earth3 model, SSP2-4.5 (yellow dashed line), SSP3-7.0 (light red dashed line), and SSP5-8.5 (dark red dashed line) scenarios. The cyan, green, and gray dashed lines represent the storm tide and return period curves under the ERA5, MERRA2, and EC-Earth3 20th, respectively; these three lines overlap because we calibrated the EC-Earth3 20th to the averaged ERA5 and MERRA2 reanalyses. We performed the same bias correction procedure for all CMIP5 and CMIP6 models at all stations.}
\label{fig:figs27}
\end{figure}  

\begin{figure}[htb!]%
\centering
\includegraphics[width=0.7\textwidth]{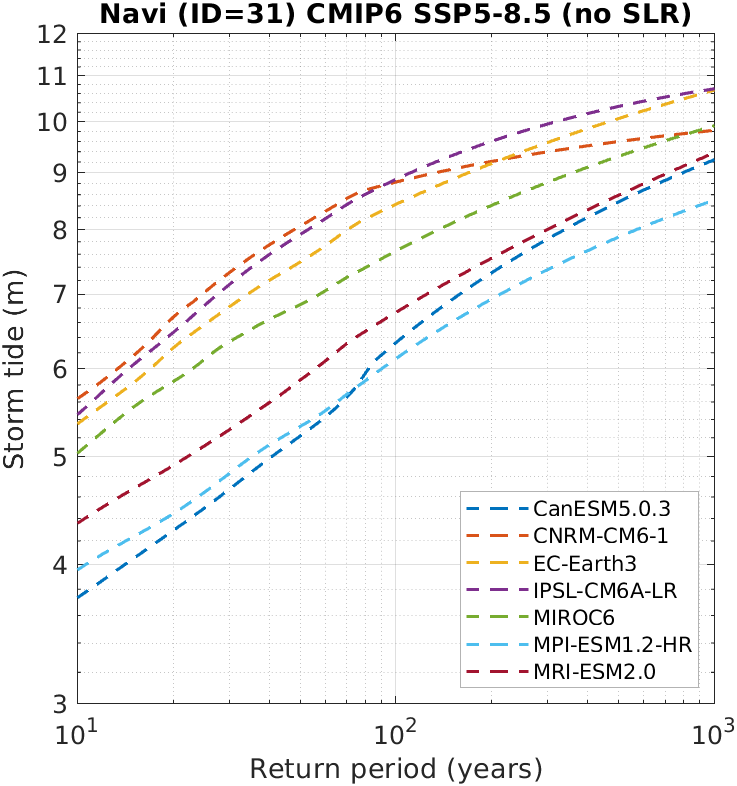}
\caption{\textbf{An example of bias-corrected storm tide against return period curves for seven climate models under SSP5-8.5, excluding the effect of SLR, at VS Navi}. The dashed lines represent the median (0.5 quantile) values derived from each of the 1,000 bootstrap ensembles (used to quantify the uncertainty of the Piecewise Kernel-GPD fitting) for each climate model. The station-scale uncertainty for each return period incorporates contributions from both the piecewise Kernel-GPD fitting process and the spread across the climate models used.}
\label{fig:figs28}
\end{figure}  

\begin{figure}[htb!]%
\centering
\includegraphics[width=0.7\textwidth]{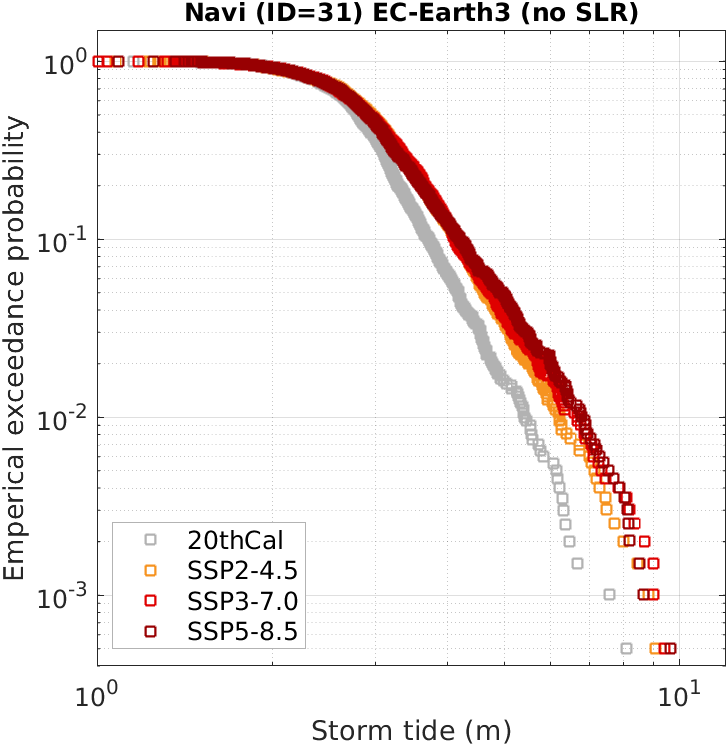}
\caption{\textbf{An example of the storm tide severity (empirical exceedance probability against storm tide) without considering the effect of SLR at VS Navi}. This is based on the CMIP6 EC-Earth3, SSP2-4.5 (yellow square), SSP3-7.0 (light red square), and SSP5-8.5 (dark red square) scenarios. Note that these dots represent the original simulated storm tides without bias correction using the ERA5 and MERRA2 reanalyses.}
\label{fig:figs29}
\end{figure}  

\begin{figure}[htb!]%
\centering
\includegraphics[width=1\textwidth]{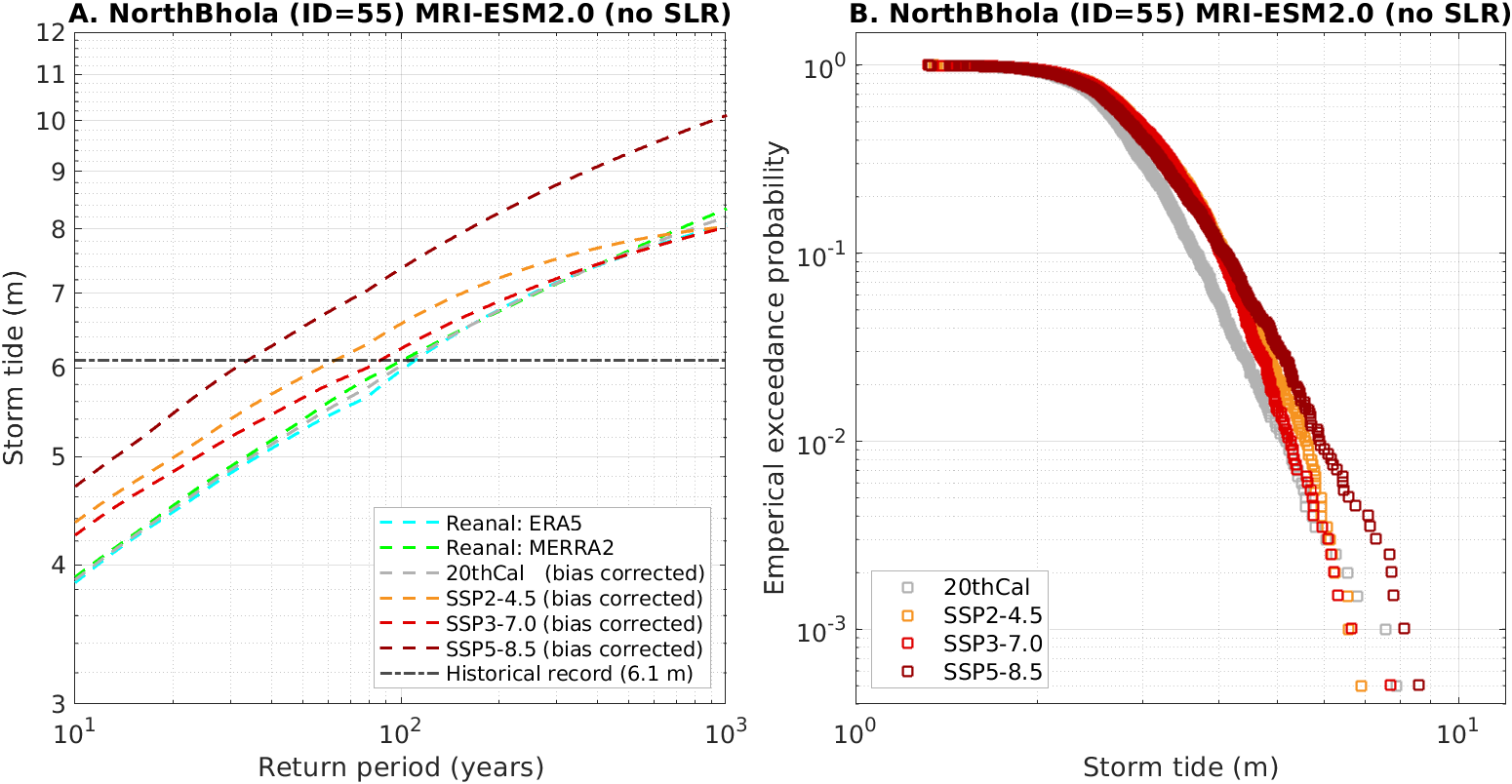}
\caption{\textbf{An example of the bias-corrected return level versus return period (A) and storm tide severity (B) without considering the effect of SLR at station Northern Bhola Island}. The horizontal dotted line indicates the historical maximum storm tide records observed at this location, source from previous study~\cite{frank1971deadliest}. This is based on the CMIP6 MRI-ESM2.0, SSP2-4.5 (yellow), SSP3-7.0 (light red), and SSP5-8.5 (dark red) scenarios. Note that these dots in subplot (B) represent the original simulated storm tides (storm tide severity) without bias correction using the ERA5 and MERRA2 reanalyses.}
\label{fig:figs30}
\end{figure}   
\newpage
\section{Supplementary2}\label{secS2}
\begin{sidewaystable}[htb!]
\caption{Detailed information about the tidal and ``virtual" stations along the Bangladesh coast.}
\label{tab:tabs1}
\begin{tabular*}{\textheight}{@{\extracolsep\fill}ccccccccccc}
\toprule%
\textbf{ID} & \textbf{Lon} & \textbf{Lat} & \textbf{Name} & \textbf{Location} & \textbf{ID} & \textbf{Lon} & \textbf{Lat} & \textbf{Name} & \textbf{Location} & \textbf{Source}\\
\midrule
1  & 89.0690124 & 21.5854591 & Alex       & Ganges           & 31 & 91.652238  & 22.4436514 & Navi       & Chattogram &\\
2  & 89.166293  & 21.6086212 & Ali        & Ganges           & 32 & 91.7268907 & 22.4347949 & Nick       & Chattogram &\\
3  & 89.2618475 & 21.6374707 & Amro       & Ganges           & 33 & 91.7564113 & 22.3395824 & Nishant    & Chattogram &\\
4  & 89.3519373 & 21.6802386 & Bakli      & Ganges           & 34 & 91.7800037 & 22.2425041 & Paco       & Chattogram &\\
5  & 89.4475076 & 21.7066318 & Barbehenn  & Ganges           & 35 & 91.8143059 & 22.148696  & Pitts      & Chattogram &\\
6  & 89.5471717 & 21.7052361 & Basha      & Ganges           & 36 & 91.8429101 & 22.0530798 & Rodi       & Chattogram &\\
7  & 89.644022  & 21.7271781 & Castillo   & Ganges           & 37 & 91.8496498 & 21.9537402 & Jiangchao  & Chattogram &\\
8  & 89.7331994 & 21.7709685 & Chen       & Ganges           & 38 & 91.83439   & 21.8551361 & Saha       & Chattogram &\\
9  & 89.8195397 & 21.8214198 & Chew       & Ganges           & 39 & 91.829805  & 21.7554847 & Sai        & Chattogram &\\
10 & 89.9057356 & 21.8721172 & Dada       & Ganges           & 40 & 91.8405295 & 21.6561052 & Salas      & Chattogram &\\
11 & 90.0015167 & 21.8720089 & Didi       & Ganges           & 41 & 91.8418013 & 21.5564125 & Seybold    & Chattogram &\\
12 & 90.0848174 & 21.8182692 & Duyck      & Ganges           & 42 & 91.8856959 & 21.4709733 & Sonia      & Chattogram &\\
13 & 90.1731304 & 21.7838998 & Edison     & Ganges           & 43 & 91.9618326 & 21.4064959 & Tagade     & Chattogram &\\
14 & 90.2687542 & 21.8099768 & Emanuel    & Ganges           & 44 & 92.0119427 & 21.3210051 & Thelonius  & Chattogram &\\
15 & 90.3658168 & 21.8339145 & Finn       & Meghna           & 45 & 92.0348013 & 21.2240663 & Trautner   & Chattogram &\\
16 & 90.4644153 & 21.8504308 & Fiyi       & Meghna           & 46 & 92.0657374 & 21.1308986 & Tswift     & Chattogram &\\
17 & 90.5564202 & 21.8883925 & Gabe       & Meghna           & 47 & 92.1303292 & 21.055069  & Yamaguchi  & Chattogram &\\
18 & 90.6336208 & 21.9512461 & Goran      & Meghna           & 48 & 92.1863032 & 20.9722584 & Yang       & Chattogram &\\
19 & 90.7103935 & 22.0153038 & Grace      & Meghna           & 49 & 92.2351253 & 20.8850721 & Zhuchang   & Chattogram &\\
20 & 90.7861956 & 22.0803504 & Hayek      & Meghna           & 50 & 92.2822905 & 20.7968937 & Ziwei      & Chattogram &\\ 
21 & 90.8696544 & 22.1299695 & Ilish      & Meghna           & 51 & 89.478     & 21.8169    & Hiron Point & Ganges & \cite{khan2022storm}\\
22 & 90.9461398 & 22.0760626 & Iris       & Meghna           & 52 & 90.23      & 21.98      & Khepupara  & Ganges & \cite{krien2017towards}\\
23 & 91.0310217 & 22.0190889 & Runge      & Meghna           & 53 & 90.27      & 21.85      & Dhulasar   & Ganges & \cite{khan2022storm}\\
24 & 91.1459135 & 22.0650456 & Kutta      & Meghna           & 54 & 91.05      & 22.2188    & Charchanga & Meghna & \cite{khan2022storm}\\
25 & 91.1964659 & 22.1630866 & Kyra       & Meghna           & 55 & 90.824     & 22.513     & Northern Bhola Island & Meghna & \cite{frank1971deadliest}\\
26 & 91.1983726 & 22.2695462 & Langlois   & Meghna           & 56 & 91.8274    & 22.2434    & Chittagong & Chattogram & \cite{khan2022storm}\\
27 & 91.2890317 & 22.3023262 & Lee        & Meghna           & 57 & 91.8054    & 22.2098    & Anwara     & Chattogram & \cite{as1998coastal}\\
28 & 91.3884806 & 22.312742  & Lubna      & Meghna           & 58 & 91.98      & 21.43      & Cox's Bazaar & Chattogram & \cite{bricheno2016tidal}\\
29 & 91.4855022 & 22.3358424 & Malai      & Meghna           & 59 & 88.03      & 21.65      & Sagar Roads &  Hooghly River & \cite{khan2022storm}\\
30 & 91.5727113 & 22.3830833 & Mitteldorf & Meghna           &    &            &            &            &            &\\
\botrule
\end{tabular*}
{\textit{Note}. Our storm tide risk assessments are conducted at all 58 stations except Sagar Roads, which is located at the mouth of the Hooghly River, far from Bangladesh, and is used solely for astronomical tide verification.}
\end{sidewaystable}

\begin{sidewaystable}[htb!]
\caption{List of climate reanalysis product, CMIP5, and CMIP6 model used in downscaling of tropical cyclones, including average horizontal resolution and principal reference.}
\label{tab:tabs2}
\begin{tabular*}{\textheight}{@{\extracolsep\fill}p{4cm}ccccc}
\toprule%
\textbf{Institution} & \textbf{Institute ID} & \textbf{Model name} & \textbf{Model type} & \textbf{Atmospheric resolution} & \textbf{Reference}  \\
\midrule
European Center for Medium-Range Weather Forecasts& ECMWF& ERA5&  Reanalyses&  0.25$^\circ$ × 0.25$^\circ$& \cite{hersbach2020era5} \\
NASA’s Global Modeling and Assimilation Office& GMAO& MERRA2&  Reanalyses&  0.5$^\circ$ × 0.625$^\circ$& \cite{gelaro2017modern} \\ 
National Center for Atmospheric Research& NCAR& CCSM4&  CMIP5&  1.25$^\circ$ × 0.94$^\circ$& \cite{lawrence2011parameterization} \\
NOAA Geophysical Fluid Dynamics Laboratory& GFDL& CM3&  CMIP5&  2.5$^\circ$ × 2.0$^\circ$& \cite{donner2011dynamical} \\
Met Office Hadley Center& MOHC& HadGEM2-ES&  CMIP5&  1.875$^\circ$ × 1.25$^\circ$& \cite{collins2011development} \\
Institute Pierre Simon Laplace&  IPSL& IPSL-CM5A-LR& CMIP5&  3.75$^\circ$ × 1.89$^\circ$& \cite{dufresne2013climate} \\
Atmosphere and Ocean Research Institute (The University of Tokyo), National Institute for Environmental Studies, and Japan Agency for Marine-Earth Science and Technology& MIROC& MIROC5&  CMIP5&  1.41$^\circ$ × 1.40$^\circ$& \cite{watanabe2010improved} \\
Max Planck Institute& MPI& MPI-ESM-MR&  CMIP5&  1.875$^\circ$ × 1.865$^\circ$& \cite{giorgetta2013climate} \\
Canadian Center for Climate Modeling and Analysis& CANESM& CanESM5.0.3&  CMIP6&  2.8$^\circ$ × 2.8$^\circ$& \cite{swart2019canadian} \\ 
Center National de Recherches Météorologiques& CNRM& CNRM-CM6-1& CMIP6& 1.4$^\circ$ × 1.4$^\circ$& \cite{voldoire2019evaluation} \\ 
EC-Earth consortium& ECEARTH& EC-Earth3& CMIP6& 0.7$^\circ$ × 0.7$^\circ$& \\ 
Institute Pierre Simon Laplace& IPSL& IPSL-CM6A-LR& CMIP6& 1.25$^\circ$ × 2.5$^\circ$& \cite{hourdin2020lmdz6a} \\
Center for Climate System Research; University of Tokyo; Japan Agency for Marine-Earth Science and Technology; National Institute for Environmental Studies& MIROC& MIROC6& CMIP6& 1.4$^\circ$ × 1.4$^\circ$& \cite{tatebe2019description} \\
Max Planck Institute& MPI& MPI-ESM1.2-HR& CMIP6& 0.94$^\circ$ × 0.94$^\circ$& \cite{muller2018higher} \\ 
Meteorological Research Institute (Japan)& MRI& MRI-ESM2.0& CMIP6& 1.12$^\circ$ × 1.125$^\circ$& \cite{yukimoto2019meteorological} \\ 
\botrule
\end{tabular*}
{\textit{Note}. The atmospheric resolution listed here refers to the resolution of the output used to drive the downscaling model; it may not directly correspond to the native resolution of the GCM.}
\end{sidewaystable}

\begin{table}[htb!]
\caption{Annual frequency bias corrections for six CMIP5 and seven CMIP6 models. }
\label{tab:tabs3}
\begin{tabular}{ccc}
\hline
\textbf{CMIP5}    & \textbf{RCP4.5}  &  \textbf{RCP8.5} \\
\hline
MPI-ESM-MR    & 0.8 (+33.3$\%$)   & 0.5 (-16.7$\%$)  \\
CM3           & 0.5 (-16.7$\%$)   & 0.6 (0$\%$)      \\
CCSM4         & 0.6 (0$\%$)       & 0.7 (+16.7$\%$) \\
MIROC5        & 0.8 (+33.3$\%$)   & 0.8 (+33.3$\%$)  \\
IPSL-CM5A-LR  & 0.9 (+50.0$\%$)   & 0.8 (+33.3$\%$)  \\
HadGEM2-ES    & 1.0 (+66.7$\%$)   & 1.2 (+100.0$\%$) \\
\end{tabular}
\begin{tabular}{cccc}
\hline
\textbf{CMIP6}   & \textbf{SSP2-4.5}   & \textbf{SSP3-7.0}   & \textbf{SSP5-8.5} \\
\hline
EC-Earth3       & 1.4 (+133.3$\%$) & 1.3 (+116.7$\%$) & 2.3 (+283.3$\%$) \\
IPSL-CM6A-LR    & 1.1 (+83.3$\%$)  & 1.0 (+66.7$\%$)  & 2.3 (+283.3$\%$) \\
CNRM-CM6-1      & 1.4 (+133.3$\%$) & 1.5 (+150.0$\%$) & 2.2 (+266.7$\%$) \\
MIROC6          & 0.9 (+50.0$\%$)  & 1.0 (+66.7$\%$)  & 1.6 (+166.7$\%$) \\
MRI-ESM2.0      & 0.7 (+16.7$\%$)  & 0.6 (0$\%$)      & 1.2 (+100.0$\%$) \\
MPI-ESM1.2-HR   & 0.6 (0$\%$)      & 0.5 (-16.7$\%$)  & 0.7 (+16.7$\%$)  \\
CanESM5.0.3     & 0.6 (0$\%$)      & 0.5 (-16.7$\%$)  & 0.6 (+0$\%$)     \\
\hline
\end{tabular}
{\textit{Note}. ``-" means decrease while ``+" means increase.}
\end{table}

\begin{table}[htb!]
\centering
\caption{Mesh size functions and the corresponding parameters used to spatially distribute element resolution~\cite{qiu2022quantitative}}
\label{tab:tabs4}
\renewcommand{\arraystretch}{1.5}
\begin{tabular}{ccccc}
\hline
\textbf{Code} & \textbf{Full name} & \textbf{Function expression} & \textbf{Bengal Delta} & \textbf{BoB} \\ \hline
MinEle & Minimum element size bound & $E_R \geq \alpha$ & 250 m & 6000 m \\ 
MaxEle & Maximum element size bound & $E_R \leq \alpha$ & 2000 m & 20,000 m \\ 
G & Element-to-element gradation limiter & $\Rightarrow |\nabla E_R| < \alpha$ & 0.25 & 0.35 \\ 
Fs & Feature width & $E_R = 2 \ast \frac{d_s + d_m}{x}$ & 6 & 3 \\ 
WL & Wavelength-to-element size ratio & $E_R = \frac{T_{M2}}{\alpha} \sqrt{\frac{gh}{f}}$ & / & 30 \\ 
TLS & Topographic-length-scale & $E_R = \frac{2\pi}{\alpha} \frac{h}{| \nabla h^* |}$ & / & 10 \\ 
FL & Low-pass filter length & $h^* = Fl_p(L) \ast h$ & / & 50 \\ \hline
\end{tabular}
\begin{flushleft}
\textit{Note.} $E_R$ is the mesh size function used to determine the spatially explicit resolution for each mesh element; $\alpha$ is a user-specified parameter; $d_s$ and $d_m$ are the absolute distances to the nearest shoreline and the medial axis, respectively; $T_{M2}$ is the period of the $M_2$ tidal wave; $g$ is the acceleration due to gravity; $h$ is the still-water depth; $h^*$ is the low-pass filtered water depth, in which $Fl_p(L)$ is the low-pass filter with cutoff length $L$.
\end{flushleft}
\end{table}

\begin{sidewaystable}[htb!]
\caption{Performance of tidal solutions among models at various tide-gauge locations.}
\label{tab:tabs5}
\centering
\small
\renewcommand{\arraystretch}{1}
\begin{tabular}{cccccccccccccccc}
\toprule
\textbf{Station} & \textbf{Const} & \multicolumn{2}{c}{\textbf{Observation}} & \multicolumn{3}{c}{\textbf{Krien et al. (2016)}} & \multicolumn{3}{c}{\textbf{Khan et al. (2021/22)}} & \multicolumn{3}{c}{\textbf{TPXO10-Atlas-V2}} & \multicolumn{3}{c}{\textbf{This Model}} \\
\cmidrule(lr){3-4} \cmidrule(lr){5-7} \cmidrule(lr){8-10} \cmidrule(lr){11-13} \cmidrule(lr){14-16}
  &  & $A_o$ & $\phi_o$ & $A_m$ & $\phi_m$ & Error & $A_m$ & $\phi_m$ & Error & $A_m$ & $\phi_m$ & Error & $A_m$ & $\phi_m$ & Error \\ 
\midrule
\textbf{Sagar Roads} 
& M2 & 140 & 116 & 143 & 116 & 3 & 144.5 & 114.9 & 5.3 & 120.5 & 121.4 & 23.4 & 153.9 & 121 & 19 \\
& S2 & 66 & 150 & 62 & 155 & 7 & 62.4 & 153.3 & 5.2 & 58.7 & 160.4 & 13.4 & 65.6 & 161.8 & 13.5 \\
& K1 & 15 & 262 & 17 & 265 & 2 & 15.6 & 265.4 & 1.1 & 13.6 & 279.4 & 4.5 & 15.3 & 262.3 & 0.3 \\
& O1 & 5 & 250 & 6 & 248 & 1 & 5.7 & 251.6 & 0.8 & 5.7 & 278.2 & 2.7 & 5 & 246.9 & 0.3 \\
& $\sigma_s$   &   &  &   &  & \textbf{6}  &   &  & \textbf{5.3} &   &   & \textbf{19.4} &   &  & \textbf{16.5}  \\
\midrule
\textbf{Hiron Point} 
& M2 & 81 & 127 & 81 & 115 & 17 & 99.9 & 115 & 26.7 & 102.1 & 118.3 & 25 & 95.6 & 115.7 & 22.7 \\
& S2 & 34 & 159 & 35 & 148 & 7 & 41.6 & 150.5 & 9.3 & 41.7 & 153.9 & 8.4 & 42.1 & 148.8 & 10.5 \\
& K1 & 13 & 268 & 15 & 265 & 2 & 15 & 265.7 & 1.7 & 14.4 & 273.4 & 2 & 15 & 253.5 & 4 \\
& O1 & 5 & 258 & 6 & 245 & 1 & 5.7 & 255 & 0.7 & 5.6 & 265.9 & 1 & 5.5 & 244.8 & 1.3 \\
& $\sigma_s$   &   &  &   &   & \textbf{13}  &  &   & \textbf{20}  &  &   & \textbf{18.7}    &   &  & \textbf{18} \\
\midrule
\textbf{Dhulasar} 
& M2 & 73 & 158 & 51 & 156 & 22 & 67.6 & 143.3 & 18.8 & 82.8 & 133.2 & 34.8 & 74.4 & 136.7 & 27 \\
& S2 & 35 & 193 & 20 & 194 & 15 & 28.5 & 179.6 & 9.8 & 42.3 & 164.9 & 20 & 31.2 & 177.6 & 9.6 \\
& K1 & 13 & 286 & 12 & 297 & 3 & 13.3 & 287.8 & 0.5 & 14 & 278.7 & 2 & 14.6 & 280 & 2 \\
& O1 & 4 & 278 & 5 & 280 & 1 & 5.6 & 273.8 & 1.6 & 5.1 & 273.5 & 1 & 4.7 & 260 & 1.5 \\
& $\sigma_s$   &   &  &  &  & \textbf{19}  &  &  & \textbf{15} &  & &\textbf{28.4}&  &  & \textbf{20} \\
\midrule
\textbf{CharChanga} 
& M2 & 96 & 234 & 67 & 208 & 46 & 95.8 & 216.9 & 28.5 & 69.1 & 232.3 & 27 & 76.5 & 235.6 & 19.5 \\
& S2 & 37.5 & 265 & 27 & 241 & 17 & 36.6 & 250.3 & 9.5 & 30.1 & 253.1 & 10 & 29.2 & 262 & 8.5 \\
& K1 & 13 & 304 & 14 & 309 & 2 & 16.8 & 308.7 & 4 & 9.5 & 320 & 4.7 & 21.7 & 300 & 8.7 \\
& O1 & 8 & 285 & 8 & 289 & 0 & 8.1 & 293.1 & 1.1 & 3.2 & 317.3 & 5.6 & 9.1 & 285 & 1.1 \\
& $\sigma_s$   &   &  &   &  & \textbf{35}  &   &  & \textbf{21.5} &  &  & \textbf{21}  &  & & \textbf{16.3} \\
\midrule
\textbf{Chittagong} 
& M2 & 173 & 196 & 156 & 198 & 18 & 149.2 & 194.8 & 24.1 & 152.3 & 203.7 & 30 & 170.2 & 183.6 & 19.4 \\
& S2 & 64 & 229 & 58 & 235 & 9 & 55 & 225.8 & 9.6 & 56.6 & 230.2 & 7.5 & 67.6 & 218.2 & 12.9 \\
& K1 & 19 & 278 & 20 & 289 & 4 & 19.1 & 284.9 & 2.3 & 14.1 & 290.3 & 6 & 22.9 & 267.4 & 5.5 \\
& O1 & 8 & 263 & 8 & 269 & 1 & 7.9 & 267.3 & 0.6 & 4.6 & 286.9 & 4.2 & 8.2 & 254.2 & 1.3 \\
& $\sigma_s$   &   &  &  &  & \textbf{14}  & &  & \textbf{18.4}   &  &  & \textbf{22.5}  &   &  & \textbf{17} \\
\bottomrule
\end{tabular}
{\textit{Note}. The observations of harmonics are sourced from previous studies\cite{krien2016improved,khan2021towards,khan2022storm} and are also available at \url{https://www.google.com/maps/d/u/0/viewer?hl=en&ll=4.962661968171313%2C120.96490361684279&z=4&mid=1yvnYoLUFS9kcB5LnJEdyxk2qz6g}. The methods used to calculate the complex error are described in detail in the above referenced papers. The geographical location of these tidal stations are illustrated in Figure~\ref{fig:fig1}.}
\end{sidewaystable}

\begin{table}[htbp]
\centering
\caption{Projected Return Periods (Years) for Bhola's and Gorky's Storm Tides Under Different Scenarios}
\begin{tabular}{ccccc}
\hline
\textbf{Scenario} & \textbf{Storm Tide} & \textbf{Condition} & \textbf{Return Period} & \textbf{Confidence Interval} \\ 
 &&&(Years)&(Years) \\\hline
\multirow{6}{*}{SSP2-4.5} & Bhola & Without SLR & 39 & 22--78 \\ 
                          &       & With SLR    & 14 & 8--23 \\ 
                          & Gorky & Without SLR & 300 & 99--562 \\ 
                          &       & With SLR    & 86  & 49--180 \\ \hline
\multirow{6}{*}{SSP3-7.0} & Bhola & Without SLR & 42 & 17--83 \\ 
                          &       & With SLR    & 13 & 8--24 \\ 
                          & Gorky & Without SLR & 201 & 96--461 \\ 
                          &       & With SLR    & 101 & 41--117 \\ \hline
\multirow{6}{*}{SSP5-8.5} & Bhola & Without SLR & 22 & 10--76 \\ 
                          &       & With SLR    & 6  & 5--24 \\ 
                          & Gorky & Without SLR & 79  & 34--370 \\ 
                          &       & With SLR    & 26  & 15--100 \\ \hline
\end{tabular}
\label{tab:return_periods}
\end{table}

\end{appendices}

\end{document}